\documentclass[useAMS,usenatbib]{mn2e}
\usepackage{graphicx}
\usepackage{times}
\usepackage{natbib}
\usepackage{subfigure}
\usepackage{url}
\usepackage{float,lscape}
\usepackage{amsmath}
\usepackage[dvipsnames]{xcolor}
\usepackage{amssymb}
\usepackage{multirow}
\usepackage{tikz}
\usepackage[titletoc,page]{appendix}

\def\checkmark{\tikz\fill[scale=0.4](0,.35) -- (.25,0) -- (1,.7) -- (.25,.15) -- cycle;} 
\newcommand*{\textquotedouble}[1]{\textquotedblleft #1\textquotedblright}
\def\hii{\mbox{H\,{\sc ii}}}

\def\sun{$_\odot$}

\title[Star formation in  IRAS~17256--3631]{Star formation towards the southern Cometary
H\,{\Large\bf II}~region IRAS~17256--3631}
\author[Veena et al.]{V. S. Veena$^{1}$\thanks{E-mail: veenavs.13@iist.ac.in},
S. Vig$^1$, A. Tej$^1$, W. P. Varricatt$^2$, S. K. Ghosh$^3$ 
\newauthor T. Chandrasekhar$^4$, N. M. Ashok$^4$ \\\\
$^1$Dept. of Earth and Space Science, Indian Institute of Space Science and Technology, Trivandrum, 695 547, India \\
$^2$United Kingdom Infrared Telescope, 660 N. Aohoku Place, Hilo, HI96720, USA\\
$^3$National Centre for Radio Astrophysics (NCRA-TIFR), Pune, 411 007, India\\
$^4$Physical Research Laboratory, Navrangpura, Ahmedabad, Gujarat, India}

\begin{document}

\date{}

\pagerange{\pageref{firstpage}--\pageref{lastpage}} \pubyear{}

\maketitle

\label{firstpage}

\begin{abstract}
 IRAS ~17256--3631 is a southern Galactic massive star forming region located at a distance of 2 kpc. In this paper, we present a multiwavelength investigation of the embedded cluster, the \hii~region, as well as the parent cloud. Radio images at 325, 610 and 1372 MHz were obtained using GMRT, India while the near-infrared imaging and spectroscopy were carried out using UKIRT and Mt. Abu Infrared Telescope, India. The near-infrared K-band image reveals the presence of a partially embedded infrared cluster. The spectral features of the brightest star in the cluster, IRS-1, spectroscopically agrees with a late O or early B star and could be the driving source of this region. Filamentary H$_2$ emission detected towards the outer envelope indicates presence of highly excited gas. The parent cloud is investigated at far-infrared to millimeter wavelengths and eighteen dust clumps have been identified. The spectral energy distributions (SEDs) of these clumps have been fitted as modified blackbodies and the best-fit peak temperatures are found to range from $14-33$ K, while the column densities vary from $0.7-8.5\times10^{22}$~cm$^{-2}$. The radio maps show a cometary morphology for the distribution of ionised gas that is density bounded towards the north-west and ionization bounded towards the south-east. This morphology is better explained with the champagne flow model as compared to the bow shock model. Using observations at near, mid and far-infrared, submillimeter and radio wavelengths, we examine the evolutionary stages of various clumps.
\end{abstract}


%%%%%%%%%%%%%%%%%%%%%%%%%%%%%%%%%

\section{INTRODUCTION}

Massive stars  (M $\gtrsim$ 8 M$_{\odot}$) play an important role in the evolution of our universe. They supply energy and chemically enrich the interstellar medium. Despite their importance, a proper understanding of the formation mechanism of these stars still remains a challenge \citep{{1987ARA&A..25...23S},{2002ApJ...569..846Y},{2014prpl.conf..149T}}. They are deeply embedded within their parent molecular clouds in the initial stages of their evolution, and enter the zero age main sequence (ZAMS) phase while still undergoing accretion. The resulting radiation pressure would inhibit accretion beyond $\sim$8~M\sun \citep{{1987ApJ...319..850W}, {1993ApJ...418..414P}}. Regardless of the theoretical limitations, stars with masses of 25~M$_{\odot}$ or greater have been observed \citep{{1997AJ....113.1733H}, {1995AJ....110.2235D}}. Several models have been proposed to understand how massive young stars overcome these issues \citep{2007ARA&A..45..481Z}. On the observational front, the challenges include high obscuration (that makes it difficult to examine the early phases of massive star formation), their rarity, relatively large distances and short evolutionary time scales. Observational studies of young massive star forming regions are therefore significant not only for increasing an observational database, but also for understanding different facets of massive star formation. 

\par A vital phase in the formation of a massive star is that of an \hii~region that is still embedded in the nascent molecular cloud (i.e. not optically visible) and undergoing expansion. An examination of the \hii~region and its parental cloud are crucial for the comprehension of the evolution of the \hii~region and the formation of the next generation of stars \citep{1999RMxAC...8..161B}. Although we have a better understanding of the later stages when the molecular clouds disperse and the nebula is visible in the optical bands, we know very little about the initial phases of massive star formation. 
While observations of molecular clouds at far-infrared and millimeter wavelengths give an insight into the evolution of precluster clouds and fragmentation process, the ionised gas distribution and young stellar objects help examine the physical conditions in the advanced stages of high mass star formation. One has to carefully look at multiple wavelengths in order to realise a collective picture of the massive star formation mechanism and how it acts as a feedback system for the surrounding interstellar medium. 
\par In this paper, we focus on IRAS~17256--3631, a southern Galactic \hii~region. In literature, the distance to this source is given as 2~kpc \citep{{2005A&A...432..921F},{2006A&A...447..221B}, {2013A&A...550A..21S}} and 14.7~kpc \citep{2006ApJ...653.1226Q}. In the present work, we adopt a distance of 2~kpc from \citet{2005A&A...432..921F} for the following reasons: (i) the morphologies of ionised gas emission from radio and near-infrared \textbf{Br$\gamma$} line match, and the latter would be negligible due to extinction if we assumed the far distance, (ii) a distance of 14.7 kpc would mean larger \textbf{diffuse} interstellar extinction, \textbf{excluding the effect of dense molecular cloud} \citep[A$_V\sim27$~mag;][]{1992JBAA..102..230W} towards the stellar objects in the cluster region that is not observed, and (iii) the neighbouring star forming region IRAS 17258-3637 (located $\sim5'$ away) is at a distance of 2 kpc from the Sun \citep{2014MNRAS.440.3078V}. 
\par  The infrared IRAS luminosity of this region is 6.4~$\times$~10$^4$ L$_{\odot}$ \citep{2005A&A...432..921F}. Our estimate of the luminosity (from mid-infrared to millimetre wavelengths, discussed in Section 3.3.4) gives a value of 1.6~$\times$~10$^5$ L$_{\odot}$. This region is classified as a low (L) type region. These L type regions are  based on the IRAS color indices [25$-$12] $<$~0.57 and ~[60$-$12]~$<$~1.30 and are believed to be optimum targets to search for high mass protostars \citep{1991A&A...246..249P}.  The molecular line and millimeter continuum studies by \citet{2005A&A...432..921F} have identified this region to be associated with massive protostellar candidates. 

\citet{2006A&A...447..221B} identified 11 massive dust clumps with masses ranging from 387~--~23~M$_{\odot}$ from their millimeter wave continuum emission studies. Radio continuum observations by \citet{2013A&A...550A..21S} detected emission at 18.0 and 22.8 GHz from this region (Beam size $\sim$ 30$\arcsec$). \citet{2003A&A...404..223B}  detected an infrared star cluster (No. 166 in their catalog) with an angular size of 1.6$\arcmin$, located 19$\arcsec$ away from the 
luminous IRAS source position.

\par In this work, we present a multiwavelength study of the region using infrared, submillimeter and radio continuum wavelengths. Radio observations using the GMRT array enable us to probe the small and large scale structures simultaneously. The highly sensitive data from UKIRT enable us to delve deeper into the infrared cluster detected by \citet{2003A&A...404..223B}. In order to locate the clumps of dense gas and dust and to understand the energetics, we use the Herschel Hi-GAL and ATLASGAL data at far-infrared and submillimeter wavelengths. The details of  observations, archival data and data reduction are given in Section 2. The results are discussed in Section 3. We explore the models of ionized gas distribution in Section 4 and the evolutionary stage of molecular clumps are analysed in Section 5. Finally, in Section 6, we summarize the results.

%%%%%%%%%%%%%%%%%%%%%%%%%%%%%%%%%%%%%%%%%%%%%%%%5

\section{OBSERVATIONS AND DATA REDUCTION} 

\subsection{Radio Continuum Observations}

The emission from ionized gas in the region around IRAS~17256--3631 is probed using low frequency radio continuum observations from Giant Metrewave Radio Telescope (GMRT) \citep{1991CuSc...60...95S}, India. GMRT comprises of 30 antennas, each having a diameter of 45~m, and arranged in a Y-shaped configuration spread over 25~km. Twelve antennas are randomly placed in a central array within an area 1 km$^2$  and the remaining 18 are stretched out along the three arms of length 14 km each. The minimum and maximum baselines are 
105~m and 25~km, respectively. Continuum observations were carried out in three frequency bands: 1372, 610 and 325~MHz. The radio sources 3C286 and 3C48 were used as the primary flux calibrators, while 1830-360, 1626-298 and 1622-297 were used as phase calibrators. The observational details are listed in Table~\ref{tb1}.

\begin{table}
\footnotesize
\caption{Details of GMRT radio continuum observations.}
\begin{center}

\hspace*{-0.5cm}
\begin{tabular}{l c c c} \hline \hline
 & & & \\
Details             & 1372 MHz &   610 MHz                   &  325 MHz   \\
\hline
Date of observation & 2004 Jan 25  &   2002 Jan 6                &  2003 Oct 10 \\
Primary beam        & 20.4$\arcmin$    &   46.6$\arcmin$                     &  85$\arcmin$   \\
Synthesized beam    & 7.3$\arcsec\times\rm5.2\arcsec$ &   11.5$\arcsec\times\rm7.9\arcsec$   &  21.7$\arcsec\times\rm8.5\arcsec$  \\
Peak flux (mJy/beam) & 60  &   135                       &  213  \\
RMS flux (mJy/beam)  & 1.0  &   1.5                       &  2.5  \\ 
Flux density$^*$ (Jy)&8.5 &13.2 &33.6 \\
Flux calibrator &3C286, 3C48 &3C286 &3C286 \\
Phase calibrator &1622-297 &1626-298 &1830-360 \\
\hline
\multicolumn{4}{l}{\textsuperscript{*}\footnotesize{Up to 3$\sigma$ contour level}}
\end{tabular}
\label{tb1}
\end{center}
\end{table}

\par The data reduction was carried out using the NRAO Astronomical Image Processing System (AIPS). Data sets were carefully checked for radio frequency interference and poor quality (due to non-working antennas, bad baselines, etc.) using tasks UVPLT and VPLOT. Subsequent editing was carried out using TVFLG and UVFLG. Calibrated data was cleaned and deconvolved using the task IMAGR and a map of the field was generated. Several iterations of self calibration process was applied in order to minimize amplitude and phase errors. A correction for the contribution of Galactic plane, to the system temperature \citep{2004MNRAS.349L..25R}, significant at low frequencies, has been applied to the radio maps. A correction factor equal to $\rm{(T_{gal}+T_{sys})/T_{sys}}$, has been used to properly scale the fluxes at these frequencies. Here $\rm{T_{sys}}$ is the system temperature corresponding to the flux calibrators that are located away from the Galactic plane. For estimating $\rm{T_{gal}}$, we used the map of \citet{1982A&AS...47....1H} at 408~MHz and a frequency spectral index of -2.6 for the estimation of the correction factor. The flux-scaled maps are flattened (using FLATN) and primary beam corrected using PBCOR. 
\par We constructed two spectral index maps corresponding to frequencies, 1372 - 610~MHz and 610 - 325~MHz, to learn about the mechanism responsible for the radio emission observed in this region. The spectral index $\alpha$ is defined as \textit{$S_\nu \propto \nu^\alpha$}, where \textit{S$_\nu$} is the flux density at frequency \textit{$\nu$}. For comparison, we used visibilities from baselines sensitive to emission at same angular scales and as a result also produced images at the three frequencies using the same UV range: 180 - 24000~$\lambda$. As each spectral index map requires the two images to be of the same resolution and pixel size, we convolved and regridded maps to a common pixel size and resolution, corresponding to the lower frequency image. This is achieved using the tasks CONVL and LGEOM.  For constructing the spectral index and error maps, we used the task COMB. For this, we considered pixels with flux larger than 5$\sigma$ in each images, where $\sigma$ is the rms of the image. The 1372 - 610 map has a resolution of $20''\times20''$ while the beam in the 610 - 325 map is $25''\times25''$. The error in spectral index is ~0.1 near the peak radio emission while it is higher ($\sim$0.6) near the edges where the signal-to-noise ratio is poor.  

\begin{table}
\footnotesize
\caption{Details of UKIRT near-IR observations.}
\begin{center}

\hspace*{-0.5cm}
\begin{tabular}{c c c c c} \hline \hline
 & & & \\
Observation             &   Filters            &   Exp. time (s)                &   Total int. time (s)    &FWHM (\arcsec)   \\
date          &      used         &           &    &   \\
\hline
2014 June 05 &        K       &  5               & 180    &0.9\\
2014 June 06      &   H$_2$                &  40                      &  1440  &0.87\\   
2014 June 20 &     J                    &     10                     & 720   &1.14\\
2014 June 20 &     H                    &     5                     & 360   &1.22\\
2014 June 20 &     K                    &     5                     & 360   &1.14\\  
\hline
\end{tabular}
\label{ukirt}
\end{center}
\end{table}

\subsection{Near-infrared observations}
Imaging and spectroscopic observations in the near-infrared were carried out with the 3.8-m United Kingdom Infrared 
Telescope (UKIRT), Hawaii and 1.2-m Mt. Abu Infrared Telescope, India.  

\subsubsection{UKIRT Data}

\begin{figure*}
\centering
\includegraphics[angle=270,scale=0.35]{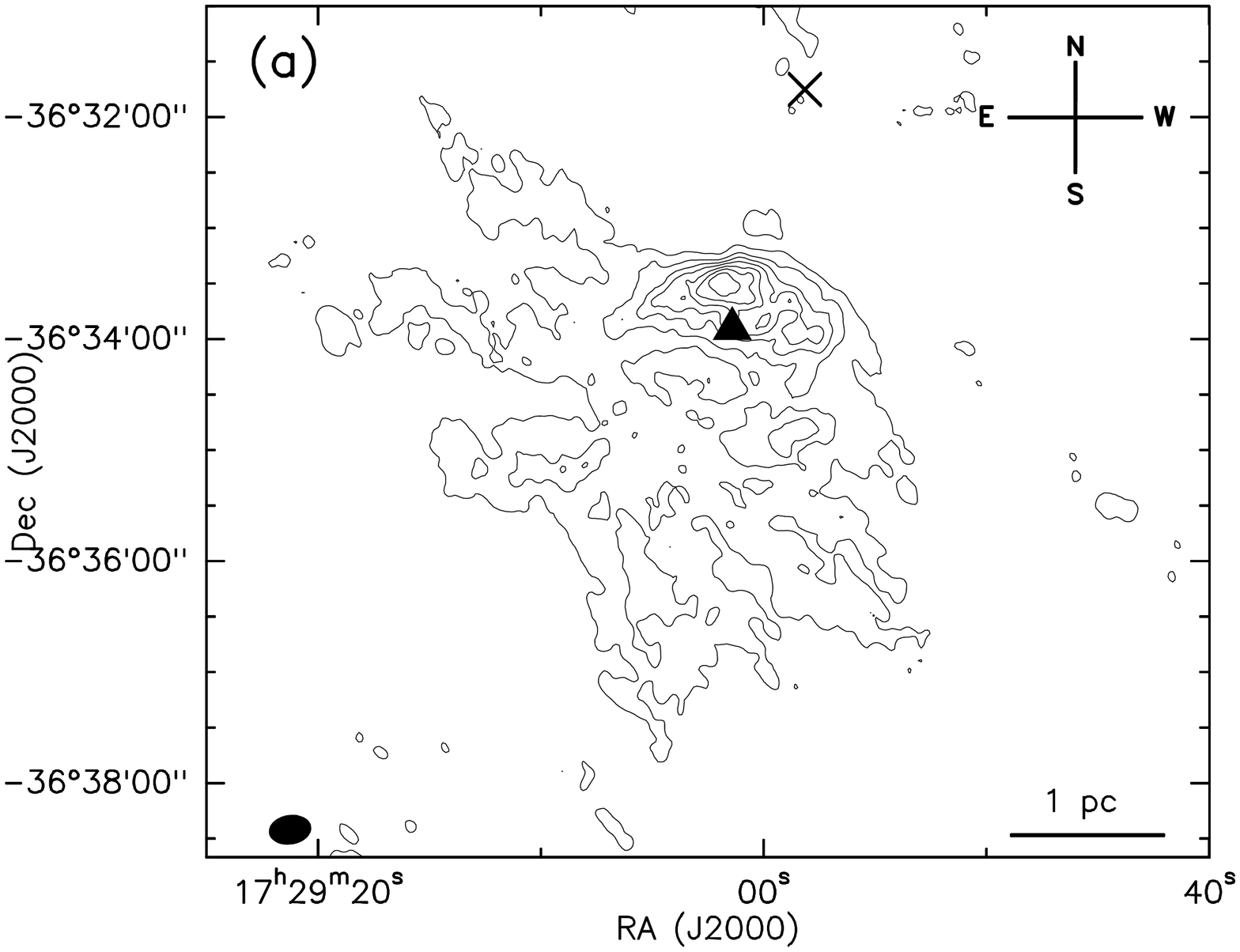} \quad \includegraphics[angle=270, scale=0.35]{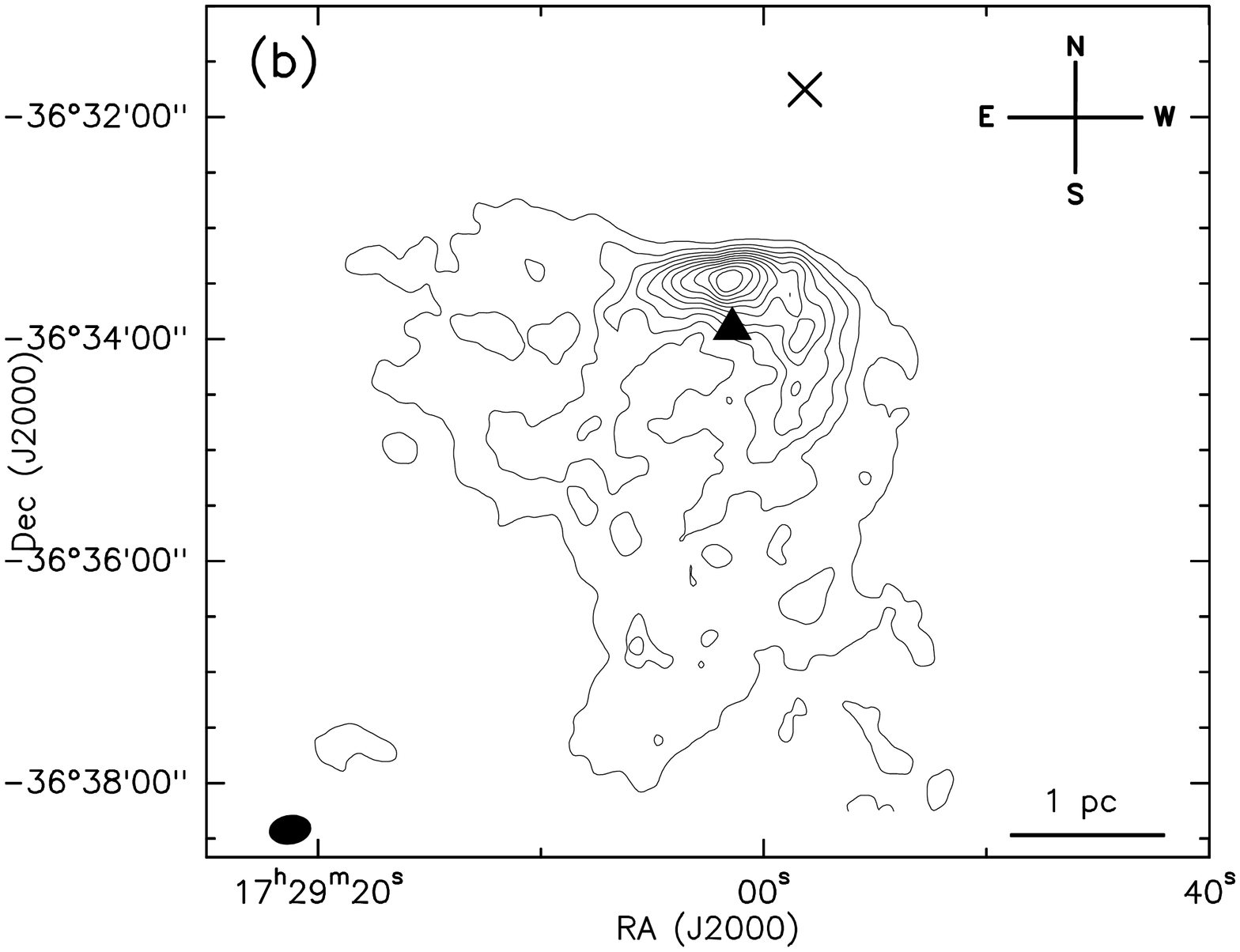} \quad \includegraphics[angle=270, scale=0.35]{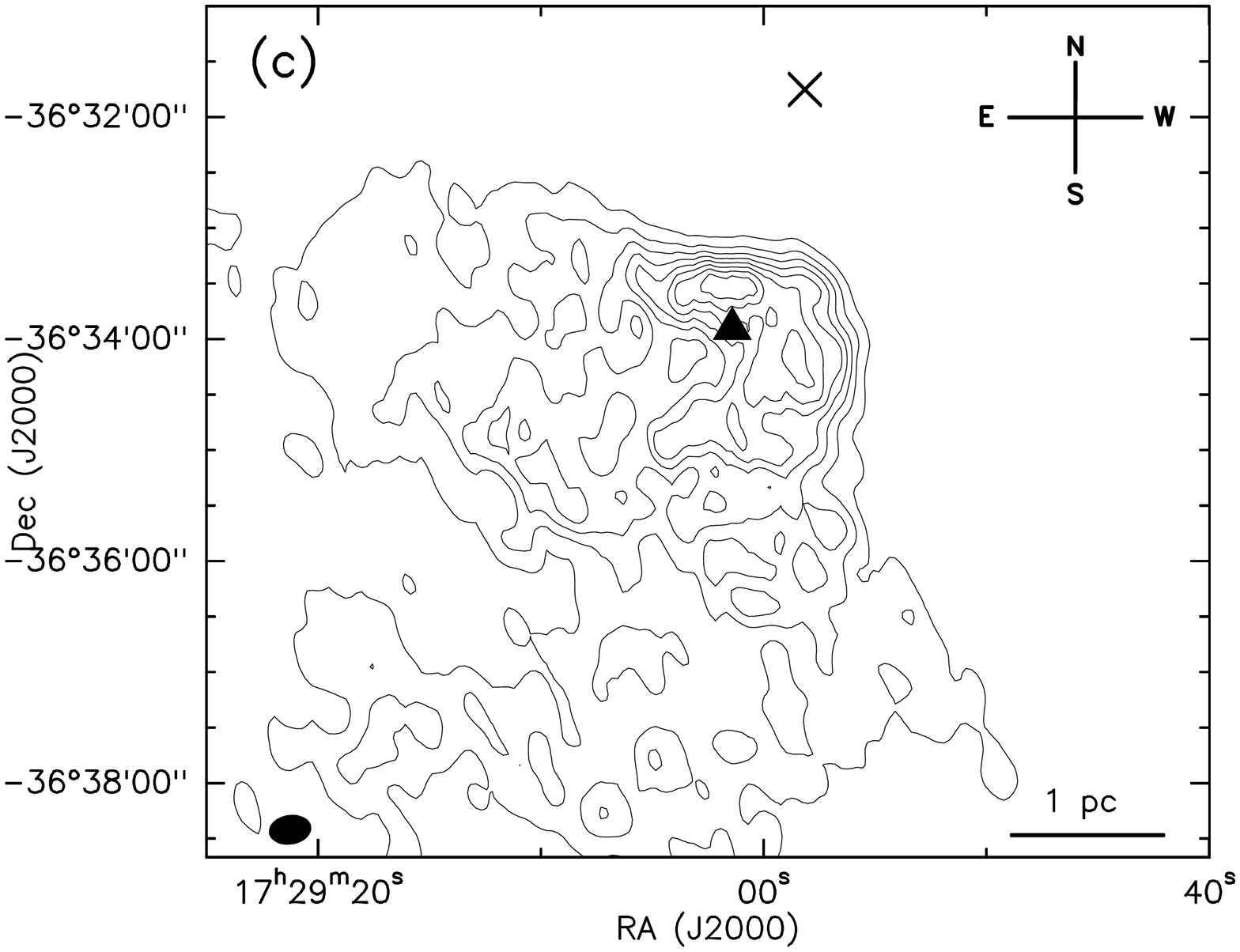}
\caption{ (a) Radio continuum map at 1372~MHz. The contour levels are at 3, 11, 19, 27, 36, 44, 52 and 60~mJy/beam with beam size of 7.3$\arcsec\times\rm5.2\arcsec$. (b) Radio continuum map at 610~MHz. The contour levels for 610 MHz map are at 8, 19.5, 31.0, 42.6, 54.2, 65.7, 77.2, 88.8, 100, 112, 123 and 135~mJy/beam where the beam size is 11.5$\arcsec\times\rm7.9\arcsec$. (c) Radio continuum map at 325~MHz. The contour levels for 325 MHz are 20, 44, 68, 92, 116, 140, 164, 188, and 212~mJy/beam where the beam size is 21.7$\arcsec\times\rm8.5\arcsec$. The corresponding beams are shown in the bottom left corner of the images in black. Cross and triangle correspond to EGO-1 and IRS-1 (discussed in text).}
\label{radio}
\end{figure*}

The region associated with IRAS 17256-3631 was imaged in the broad-band JHK 
filters and in a narrow-band H$_2$ filter centered at the 
wavelength of the H$_2$ (1-0) S1 line at 2.12~$\mu$m using the UKIRT Wide-Field Camera \citep[WFCAM,][]{2007A&A...467..777C}. WFCAM consists of four 2048$\times$2048 HgCdTe Hawaii-2 arrays. With
a pixel size of 0.4$\arcsec$, each array has a field of view 
of $13.5\arcmin\times13.5\arcmin$.  A $2\times2$ microstepping gives a final 
pixel scale of 0.2$\arcsec$ pixel$^{-1}$. Observations were obtained in the 9-point dithered mode with offsets less than 10$\arcsec$. 
The observing log for the WFCAM observations are given in Table~\ref{ukirt}.

\par The data reduction was carried out by the Cambridge Astronomical Survey Unit (CASU). Pipeline photometric calibrations were done using 
2MASS sources in the field, and the magnitudes derived were converted to the UKIRT system. The photometric system, calibration, pipeline processing and science archive are described in \citet{2009MNRAS.394..675H} and \citet{2008MNRAS.384..637H} and references therein. The pipeline reduction did not detect many of the sources surrounded by nebulosity in our field. For such sources, particularly towards the central region of IRAS 17256-3631, 
aperture photometry was carried out using the QPHOT task under the Image Reduction and Analysis Facility (IRAF) software \citep{1986SPIE..627..733T}. Photometric magnitudes of saturated sources were replaced by those from the 2MASS 
Point Source Catalog after converting to the UKIRT system \citep{hodgkin2009ukirt}.
Care was taken to identify and appropriately subtract out the components if an unresolved 2MASS single 
source was seen as a resolved binary or multiple sources, one component of which was saturated in the UKIRT-WFCAM images. 

Finally, a combined 
catalog was generated by converting to the Bessell \& Brett system \citep{1988PASP..100.1134B} using the transformation equations given in \citet{2001AJ....121.2851C} which was used for studying the stellar population associated with 
IRAS 17256-3631.

\par To obtain the emission in H$_2$ line, continuum emission was subtracted out from the H$_2$ images using scaled K-band images \citep{2005MNRAS.359....2V}. We have used the K-band image taken on 05~June~2014 for continuum subtraction as the seeing was comparable to that of H$_2$ image. Integrated counts using 
circular apertures of three times the average FWHM of isolated point sources in each mosaic were 
used to estimate the scaling factors. The K-band image was scaled using this factor and subtracted from the H$_2$ image to 
give the continuum subtracted H$_2$ image. This method removes the diffuse continuum emission efficiently but improper PSF 
matching between the narrow and broad band images results in residuals seen in stars.
\begin{figure*}
\hspace*{-0.5cm}
\begin{minipage}{0.497 \textwidth}
\centering \includegraphics[angle=270, scale=0.35]{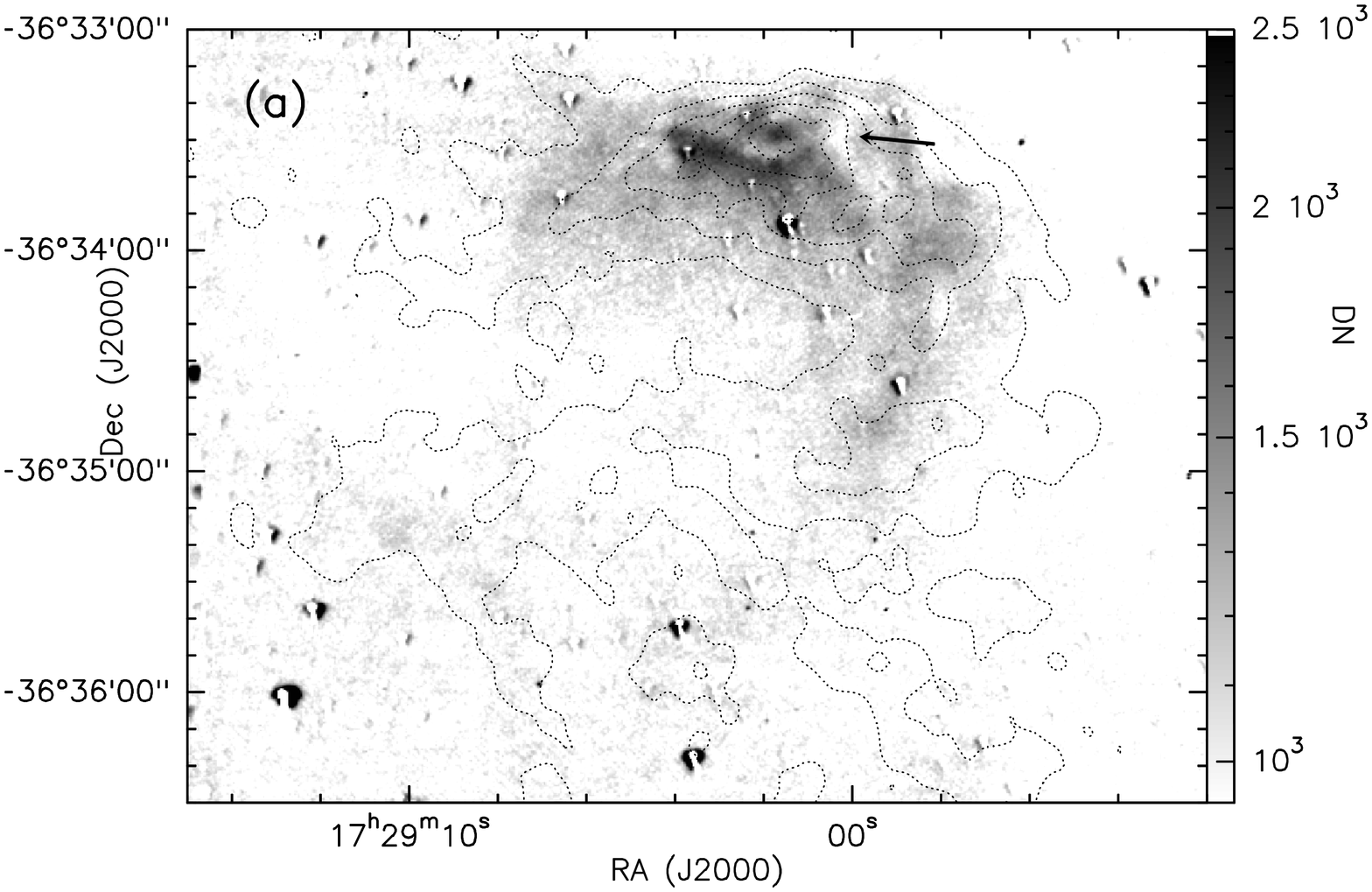} 
\end{minipage}
\hspace*{0.3cm}
\begin{minipage}{0.497 \textwidth}
\centering \includegraphics[scale=0.38]{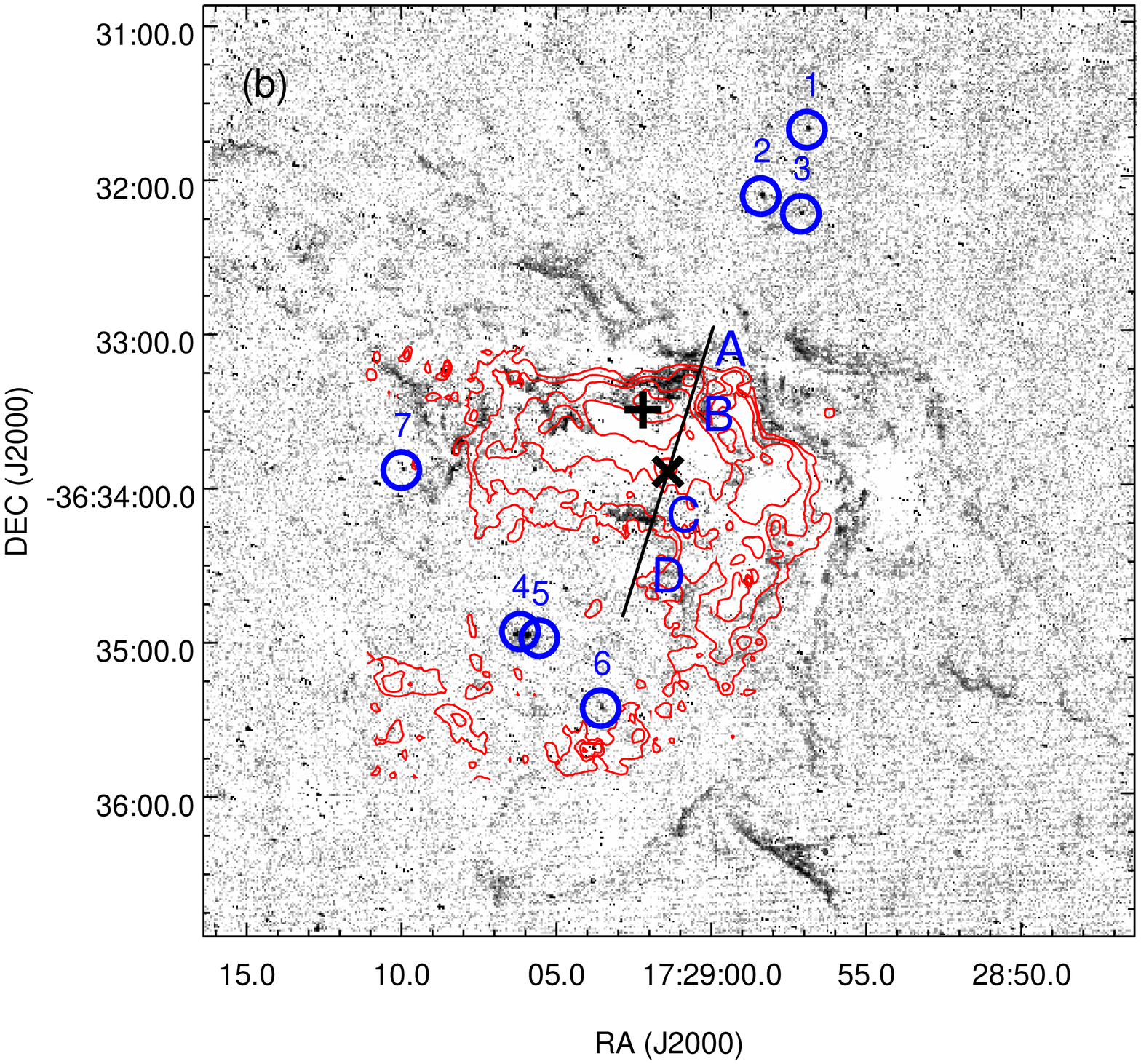}
\end{minipage}
\caption{(a) The continuum subtracted Br$\gamma$ image of the field around IRAS~17256--3631 overlaid with 1372~MHz radio contours. The contour levels are at 6, 15, 24, 33, 42, 51 and 60~mJy/beam. Residuals of continuum subtraction around the stars are seen in this image. The arrow point towards a high extinction filamentary structure. (b) The Narrow band H$_2$ image of the region IRAS~17256--3631 overlaid with Br$\gamma$ contours. The image has been smoothed to 3 pixels to improve the contrast against the background noise. Emission is seen as black filaments in the image. The detected H$_2$ knots are encircled and numbered. Radio peak is marked as +. The black line represents the slit used for obtaining the near-infrared spectrum, which was centered on IRS-1 (marked with $\times$). The four regions from which the nebular spectrum were taken are labelled as A, B, C and D in the image.}
\label{brg_H2}
\end{figure*}
\par Apart from imaging, spectroscopic observations of the near-infrared bright source G351.6921-01.1515 (hereafter IRS-1) was carried out 
using the UKIRT $1-5$ micron Imager Spectrometer (UIST). UIST employs a $1024\times1024$ InSb array, and has a pixel scale 
of 0.12$\arcsec$ pixel$^{-1}$ for spectroscopy.  The observations were performed using a 4-pixel-wide and 1.9$\arcmin$-long slit. The HK grism 
covers a wavelength regime of $1.395-2.506$~$\mu$m. The 4-pixel slit gives a spectral resolution of ~500. The slit was 
oriented 16.5$\degr$ west of north and centered on IRS-1 ($\rm{\alpha_{J2000}}$: 17$^h$29$^m$01.389$^s$, $\rm{\delta_{J2000}}$: -36$\degr$33$\arcmin$54.21$\arcsec$). This ensures sampling of the associated nebulosity 
in the NW direction. Given the spatial extent of the nebulosity, the observations were performed by nodding 
to blank sky ($\sim$ 3.5$\arcmin$ SE of the source). Sequence of object-sky pairs gave total on-
source integration time of 800 seconds, with an exposure time of 40~s per frame.

\par Preliminary data reduction was done using the 
UKIRT pipeline ORACDR. Flat fielding and wavelength calibrations were performed using the flat and arc spectra obtained prior to the target observations. Atmospheric air glow lines were removed by subtracting the sky frame from the 
target frame. Finally, the flat fielded, wavelength calibrated and sky subtracted frames were combined to give the 
resultant target spectral image. For telluric and instrumental corrections, a spectroscopic standard BS 6454, an F9V 
star, was observed. Subsequent reduction was performed using the starlink packages FIGARO and KAPPA. The spectrum of the 
standard star was extracted and the photospheric absorption lines were interpolated across. It was then corrected for the 
blackbody temperature of the star and used to divide the target spectrum to get the final spectral image. The wavelength calibration was further revised using the arc spectra, and the flux calibration was performed using the 2MASS magnitudes of the standard star, assuming that the seeing did not vary between the target and standard star observations. The flux calibration is expected to be accurate to $\sim$ 20\%. IRAF spectral extraction task APALL was also used to fine tune the presented spectra of the 
central star and the nebulosity.

\subsubsection{Mt Abu data}

\par IRAS~17256--3631 was imaged in the narrow-band Br$\gamma$ and broad-band K filters using the Near Infrared Camera and Spectrograph (NICS) on the Mt. Abu Infrared Telescope. 
NICS is equipped with a 1024~$\times$~1024 HgCdTe detector array giving a pixel scale of 0.5\arcsec pixel$^{-1}$ 
and the field of view is  $8\arcmin\times8\arcmin$. The observations were performed with a 5-position dither 
with offsets of 20\arcsec. The field is relatively crowded with faint extended nebulosity. Hence, for better sky 
subtraction, a nearby blank region of the sky was observed with the same pattern. Individual frame exposure times in the 
Br$\gamma$ and K bands were 90~s and 10~s, respectively. The sky subtracted frames were aligned and co-added to give 
resultant Br$\gamma$ and K band images with total on source integration times of 1350~s and 250~s, respectively. Continuum 
subtraction was carried out as discussed under WFCAM reduction.

%%%%%%%%%%%%%%%%%%%%%%%%%%%%%%%%%%%%%%%%%%%%%%%%%%%%%%%

\subsection{Archival Datasets}

In order to complement our radio and near-infrared observations, we used the available images and catalogs from the archives of {\it Spitzer Space Telescope}, ATLASGAL and \textit{Herschel Space Observatory}.  These have been used to study the dust emission from the nascent cloud associated with IRAS~17256-3631 and population of YSOs in this region.

\subsubsection{Spitzer Space Telescope Data}

We used the mid-infrared (MIR) images from the {\it Spitzer Space Telescope} to study the distribution of warm dust emission and young stellar objects (YSOs) towards IRAS~17256--3631.

The \textit{Spitzer Space Telescope} having a primary mirror of diameter 85-cm was launched in August 2003. One of the instruments on board the space telescope is the  InfraRed Array Camera (IRAC) that provides simultaneous images in four filters centred at 3.6, 4.5, 5.8 and 8.0 $\mu$m. The pixel size is $1.2\arcsec\times\rm1.2\arcsec$ and the angular resolutions achieved are $\rm1.66\arcsec, 1.72\arcsec, 1.88\arcsec$ and $1.98\arcsec$ at 3.6, 4.5, 5.8 and 8.0 $\mu$m, respectively. We used two sets of level 2 PBCD images from the Spitzer Heritage Archive for the analysis presented here. The first set of images are from GLIMPSE legacy project \citep{2003PASP..115..953B} in all the four IRAC bands. However, there is only a partial coverage at 3.6 and 5.8~$\mu$m, and the south-eastern portion towards our region of interest is not covered. The second set of IRAC images (AOR key:  40231680) are availbale at two bands only, 3.6 and 4.5 $\mu$m, but cover our region of interest fully.  

\par As the GLIMPSE point source catalogs do not completely cover our region of interest, we performed photometry for 3.6 and 4.5 $\mu$m bands using the second set of images and at 8.0~$\mu$m band using the GLIMPSE tile. We considered a $10\arcmin\times10\arcmin$ region centered on the IRAS peak ($\rm{\alpha_{J2000}}$: $\rm{17^h29^m01.1^s}$, $\rm{\delta_{J2000}}$: $\rm{-36\degr33\arcmin38.0\arcsec}$). Aperture photometry was carried out using the SExtractor software \citep{1996A&AS..117..393B}. In order to perform the background subtraction, a background map was constructed by estimating the background value in each mesh of a grid that covers the whole frame. The mesh size for each band were given as an input in the algorithm.  To check for detection efficiency and photometric accuracy, we took a 100$\arcsec$ region sufficiently away from the nebulosity, where there is GLIMPSE catalog coverage. We compared the magnitudes obtained using our photometry with values from the GLIMPSE catalogue. By comparing with the GLIMPSE catalog, we found $\sim$80\% common sources in all the bands. We also carefully checked for spurious sources and removed them from the catalog. By visual inspection, we see that in regions of strong nebulosity in the 8.0 $\mu$m band, the detection rate is lower compared to regions with little or no nebulosity. In these regions, we manually performed photometry for the sources using the aperture photometry task $\lq$qphot' of IRAF. The aperture size was chosen as 6$''$ and the background estimation has been carried out with a 2$\arcsec$ sky annulus. After performing manual photometry and removing the spurious sources, we combined the individual band catalogs to produce a final catalog. We detected 4155, 4163 and 572 sources in the 3.6, 4.5 and 8.0 $\mu$m bands, respectively. Of these, 275 sources were detected in all the three bands.

\subsubsection{ATLASGAL Survey} 

In order to investigate the cold dust emission at (sub)millimeter (submm) wavelengths, we used images from the ATLASGAL survey. The APEX Telescope Large Area Survey of the Galaxy (ATLASGAL) is a systematic survey of the Galaxy  carried out to observe the continuum emission from interstellar dust at submm wavelengths. Observations were performed with the Large APEX Bolometer Camera (LABOCA), a 295-element bolometer array observing at 870~$\mu$m \citep{2009A&A...504..415S}. The pixel size and angular resolution are 6$\arcsec$ and 18$\arcsec$, respectively. The archive provides image tiles of size 5$\arcmin$. As the emission from this region is larger than the tiles, we mosaiced the tiles to make a bigger image. Seven image tiles were obtained from the archive. These tiles were mosaiced by projecting them to larger sized images and then combined together by adding them. In the regions of overlap, near the edges, we averaged the pixel values. 

\subsubsection{Herschel Hi-GAL Survey}

We use the far-infrared (FIR) images from \textit{Herschel Space Observatory} to examine the cold dust clumps in this region.
The \textit{Herschel Space Observatory} is a 3.5-meter telescope capable of observing in the far-infrared and submm spectral range 55-671 $\mu$m \citep{2010A&A...518L...1P}. The images obtained from the Herschel Archive are part of the Herschel Hi-GAL survey \citep{2010PASP..122..314M} carried out using the Photodetector Array Camera and Spectrometer \citep[PACS,][]{2010A&A...518L...2P} and the Spectral and Photometric Imaging Reciever \citep[SPIRE,][]{2010A&A...518L...3G}. The Hi-GAL observations were carried out in $\lq$parallel mode' covering 70, 160 $\mu$m (PACS) as well as 250, 350 and 500~$\mu$m (SPIRE). The pixel sizes are 2$\arcsec$, 3$\arcsec$, 6$\arcsec$, 10$\arcsec$ and 14$\arcsec$  at 70, 160, 250, 350 and 500 $\mu$m and the angular resolutions achieved are 5$\arcsec$, 13$\arcsec$, 18.1$\arcsec$, 24.9$\arcsec$ and 36.4$\arcsec$, respectively. We used PACS level 2.5 and SPIRE level 3 data products for our analysis. Our region of interest has partial coverage in the PACS bands. However, we used these images to understand the dust emission from our region of interest where PACS coverage is available.

%%%%%%%%%%%%%%%%%%%%%%%%%%%%%%%%%%%%%%%%%%%%%%%%%%%%%%
\section{RESULTS AND DISCUSSION}

\subsection{Emission from Ionized Gas}

The radio continuum emission  from IRAS~17256--3631 at 1372, 610 and 325 MHz are shown in Figs.~\ref{radio}a, \ref{radio}b and \ref{radio}c respectively. The morphology of the \hii ~region is strikingly similar to that of the cometary-shaped ultracompact \hii~region described by \citet{1989ApJS...69..831W}. There is a  steep density gradient towards north-west characterized by a bright, arc-like leading edge (or head) and a low surface brightness tail of emission. The angular size of the radio emitting region is $\sim 5\arcmin\times\rm4\arcmin$ ($2.9\times2.3\ \rm{pc}$). The total flux densities (upto 3$\sigma$ contour level of the peak flux density) at  1372, 610 and~325 MHz are 8.5, 13.2 and 33.6~Jy, respectively. The radio emission at 1372 MHz peaks at $\rm{\alpha_{J2000}}$: $\rm{17^h29^m02.2^s}$, $\rm{\delta_{J2000}}$: $\rm{-36\degr33\arcmin30.2\arcsec}$. The location of radio peak positions at all three frequencies match well (within 3$\arcsec$). The peak flux densities, synthesized beams, flux densities and rms flux density values are given in Table~\ref{tb1}.  
 
\begin{figure*}
\includegraphics[angle=270, scale=0.35]{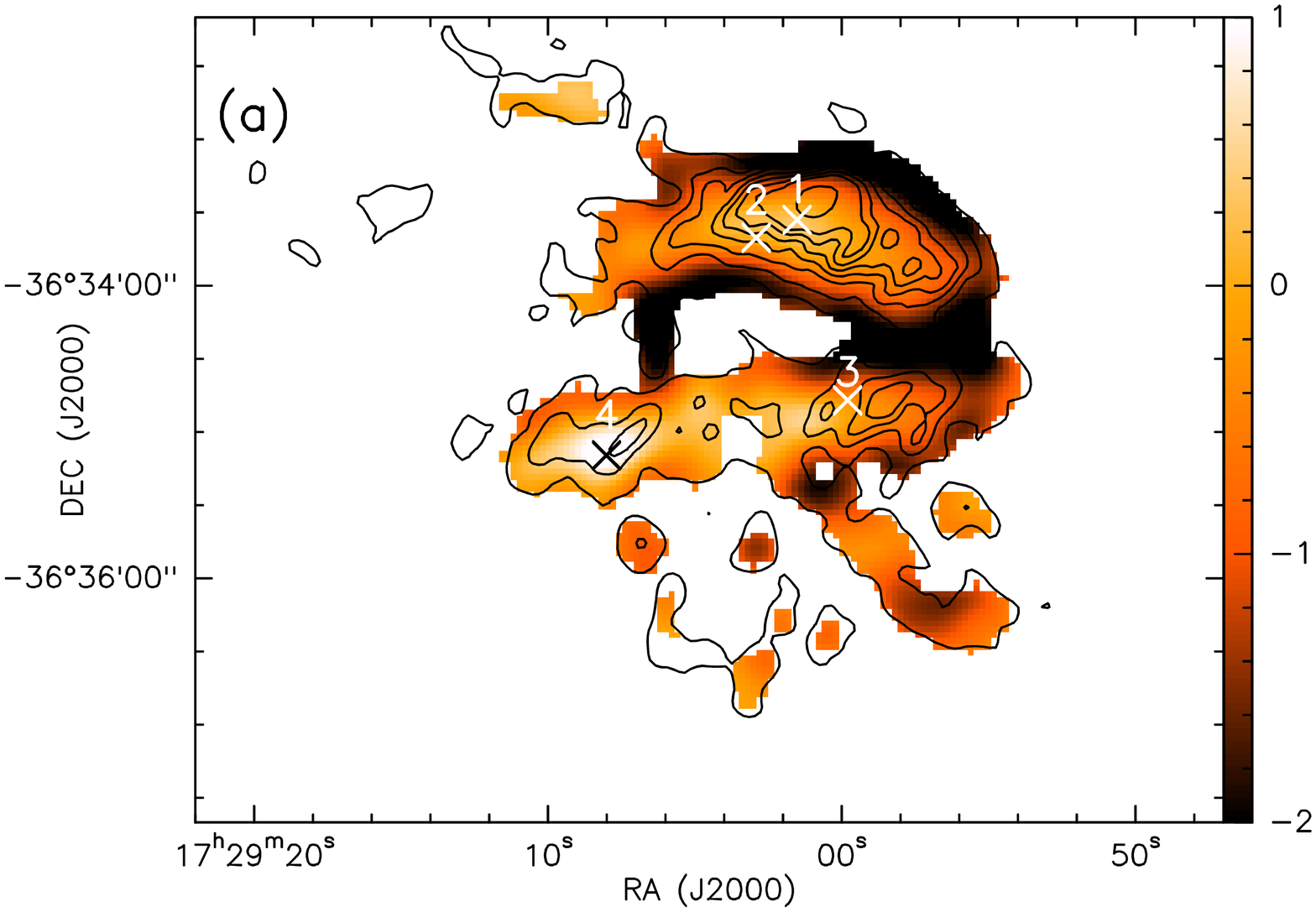} \quad \includegraphics[angle=270, scale=0.35]{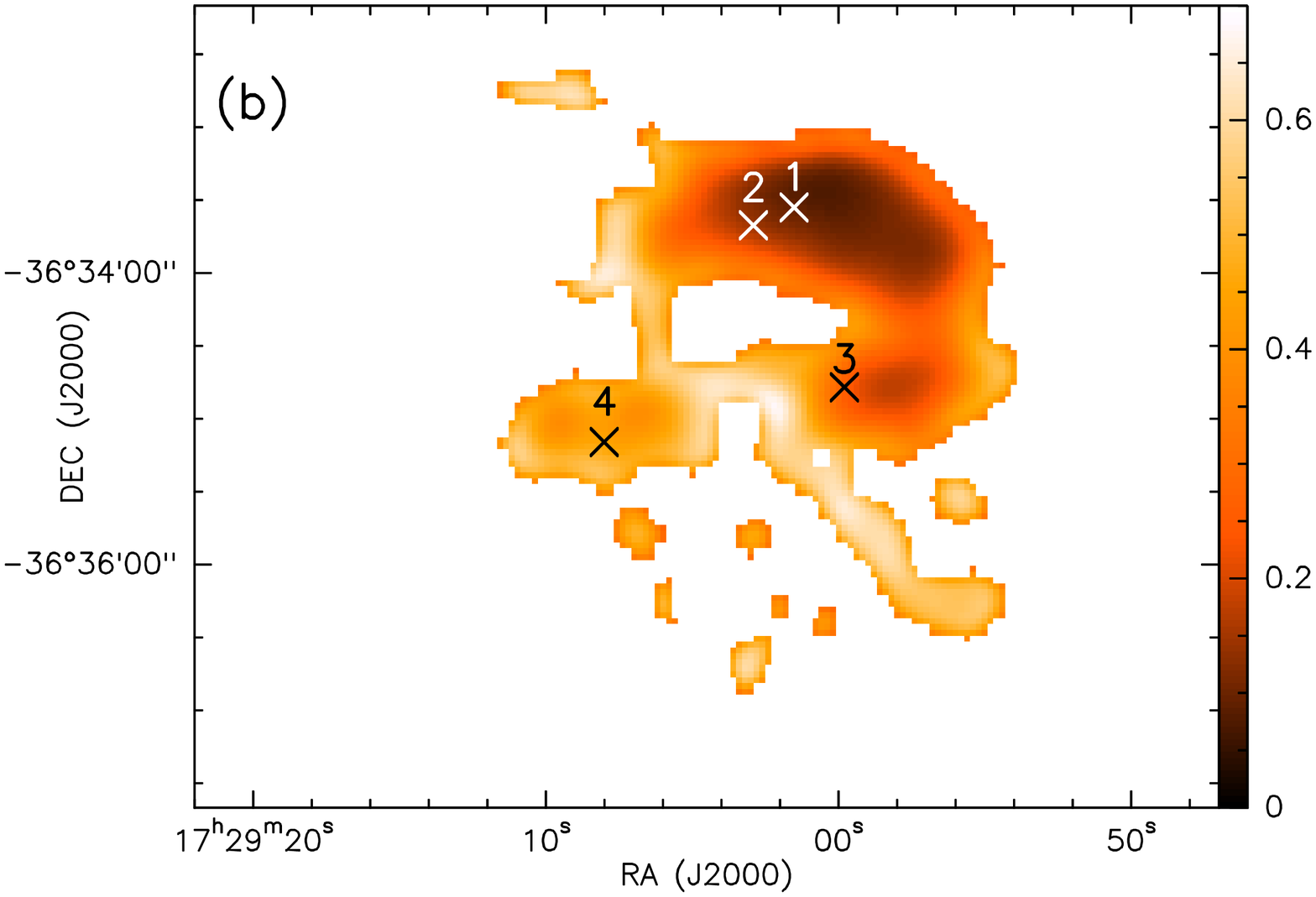}
\quad \includegraphics[angle=270, scale=0.35]{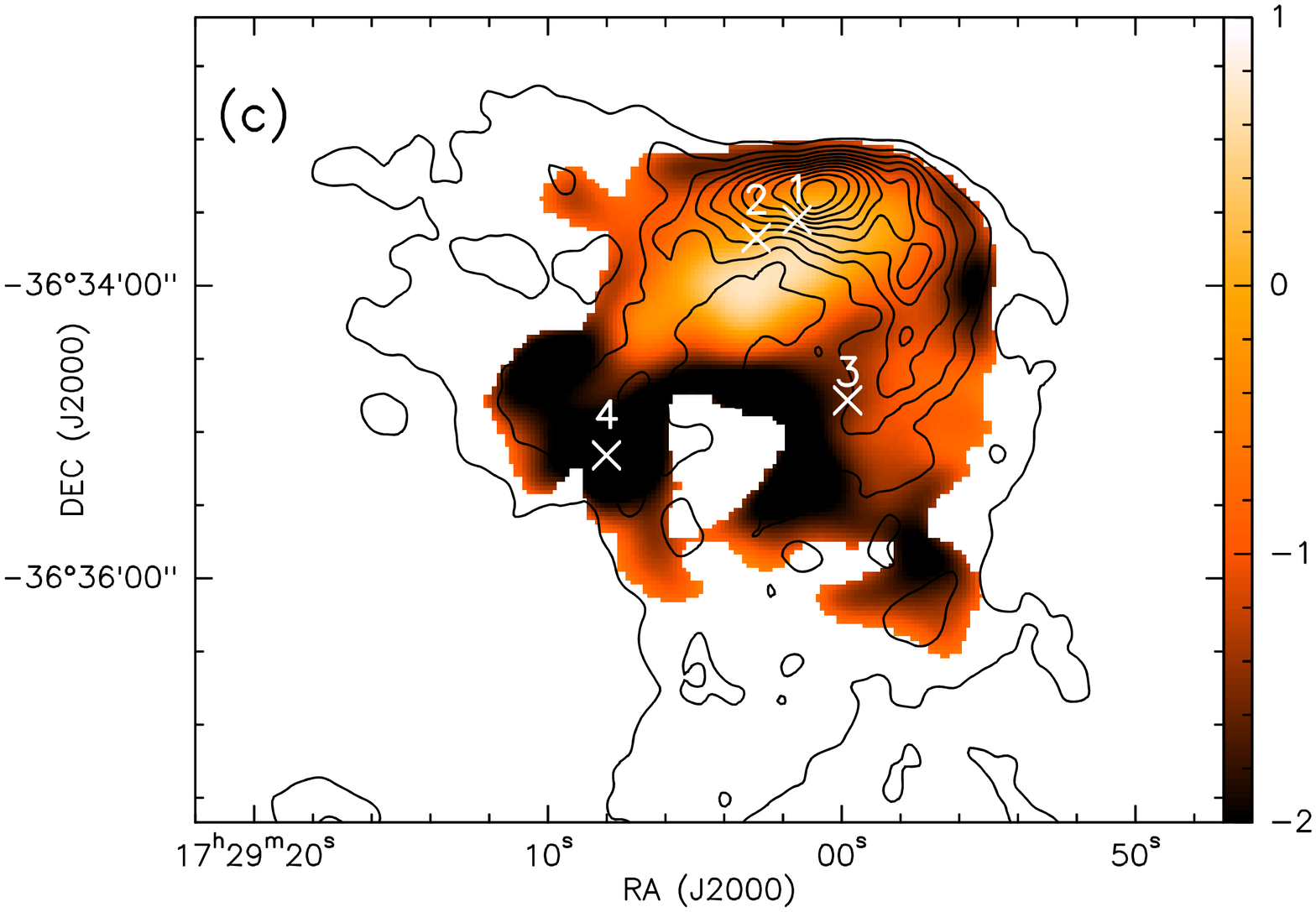} \quad \includegraphics[angle=270, scale=0.35]{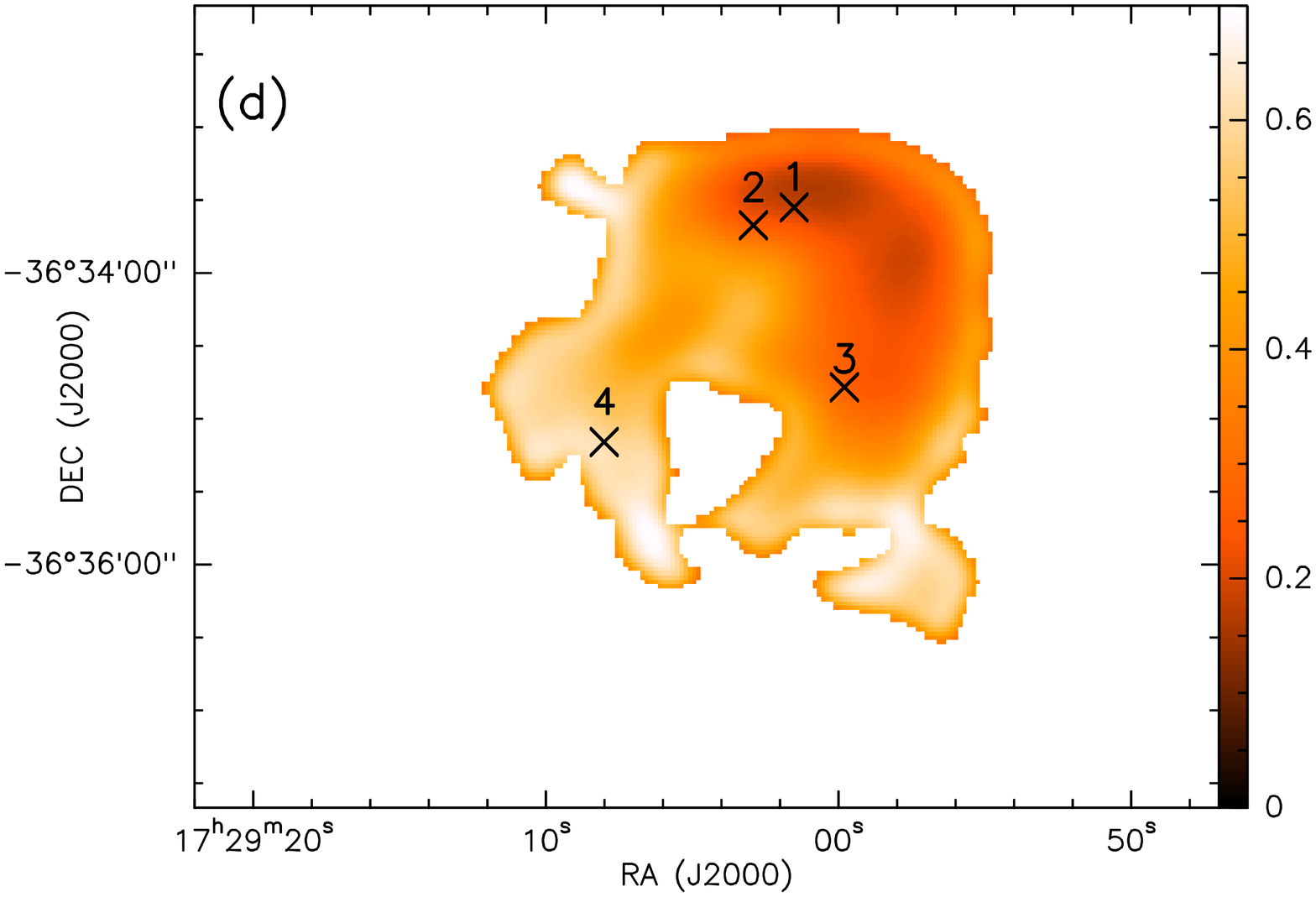}
 \caption{(a) The 1372-610~spectral index map overlaid with 1372~MHz radio continuum contours. The four locations at which the spectral indices are computed are also shown in the figure. (b) Error map of 1372-610~spectral index map. (c) 610-325~spectral index map overlaid with 610~MHz radio continuum contours. The four locations shown in (a) are also shown here. (d) Error map corresponding to the 610-325~spectral index map.}
\label{specin} 
\end{figure*}

\begin{figure*}
\includegraphics[scale=0.2]{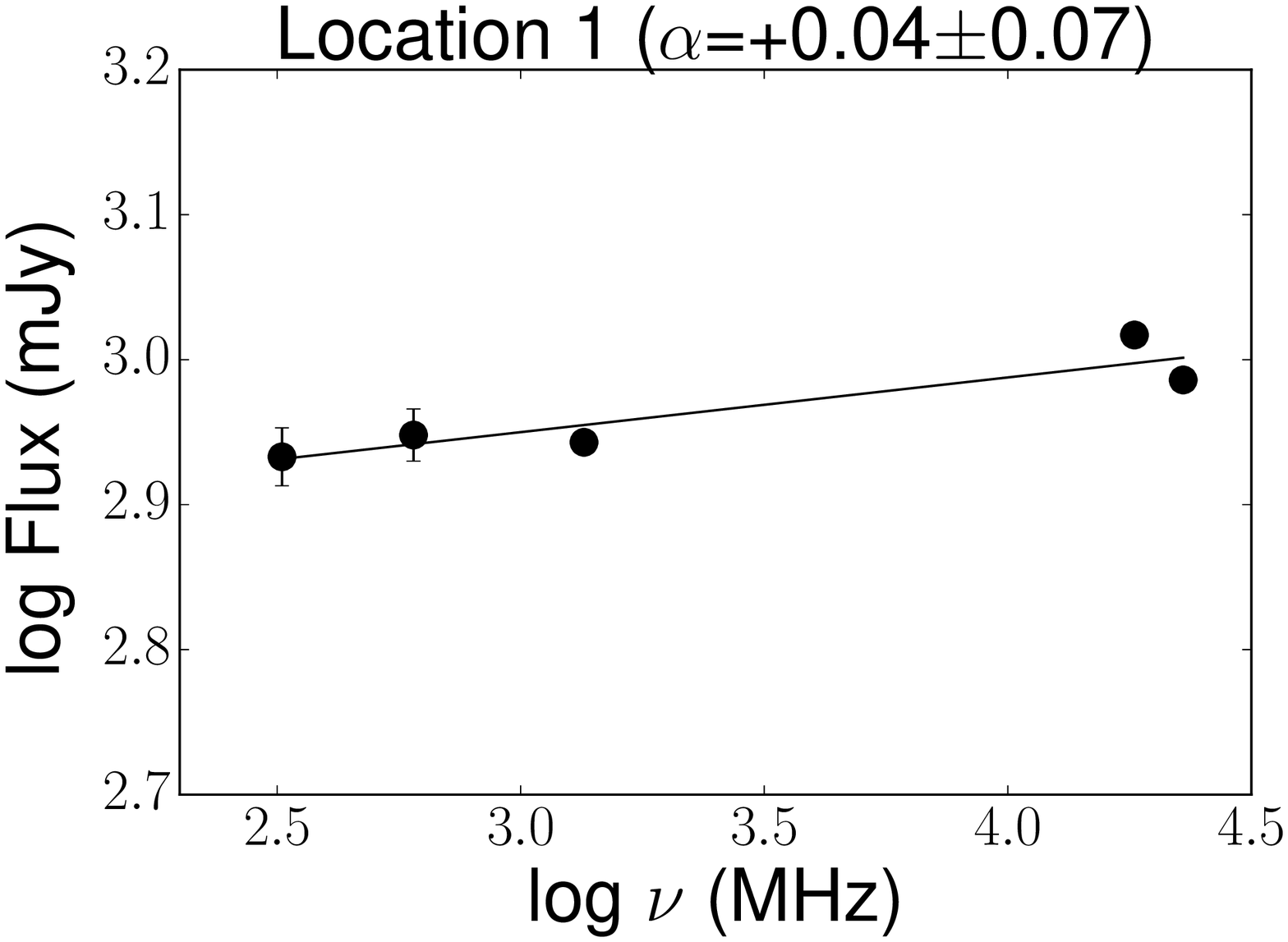} \quad \includegraphics[scale=0.2]{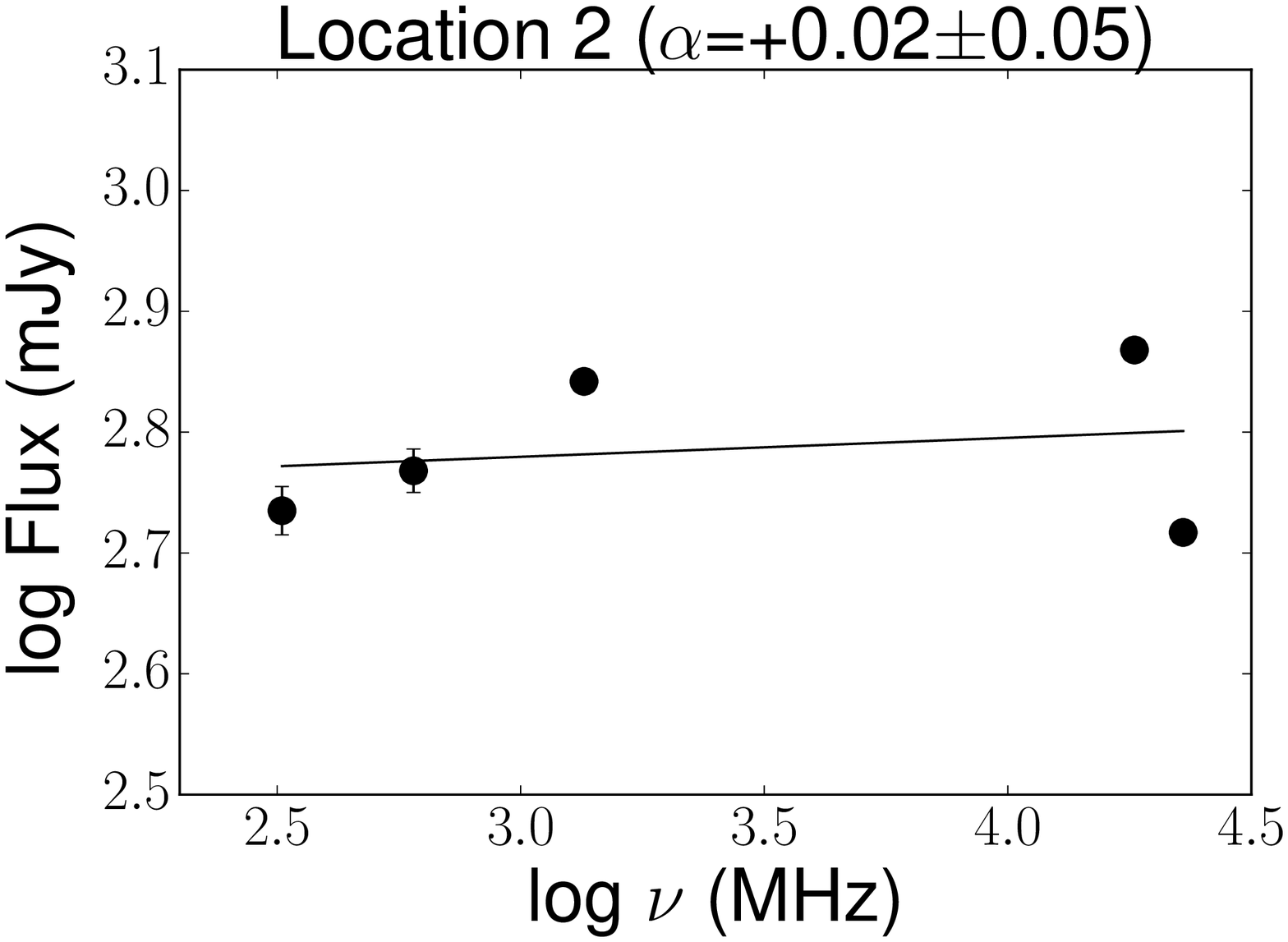} \quad \includegraphics[scale=0.2]{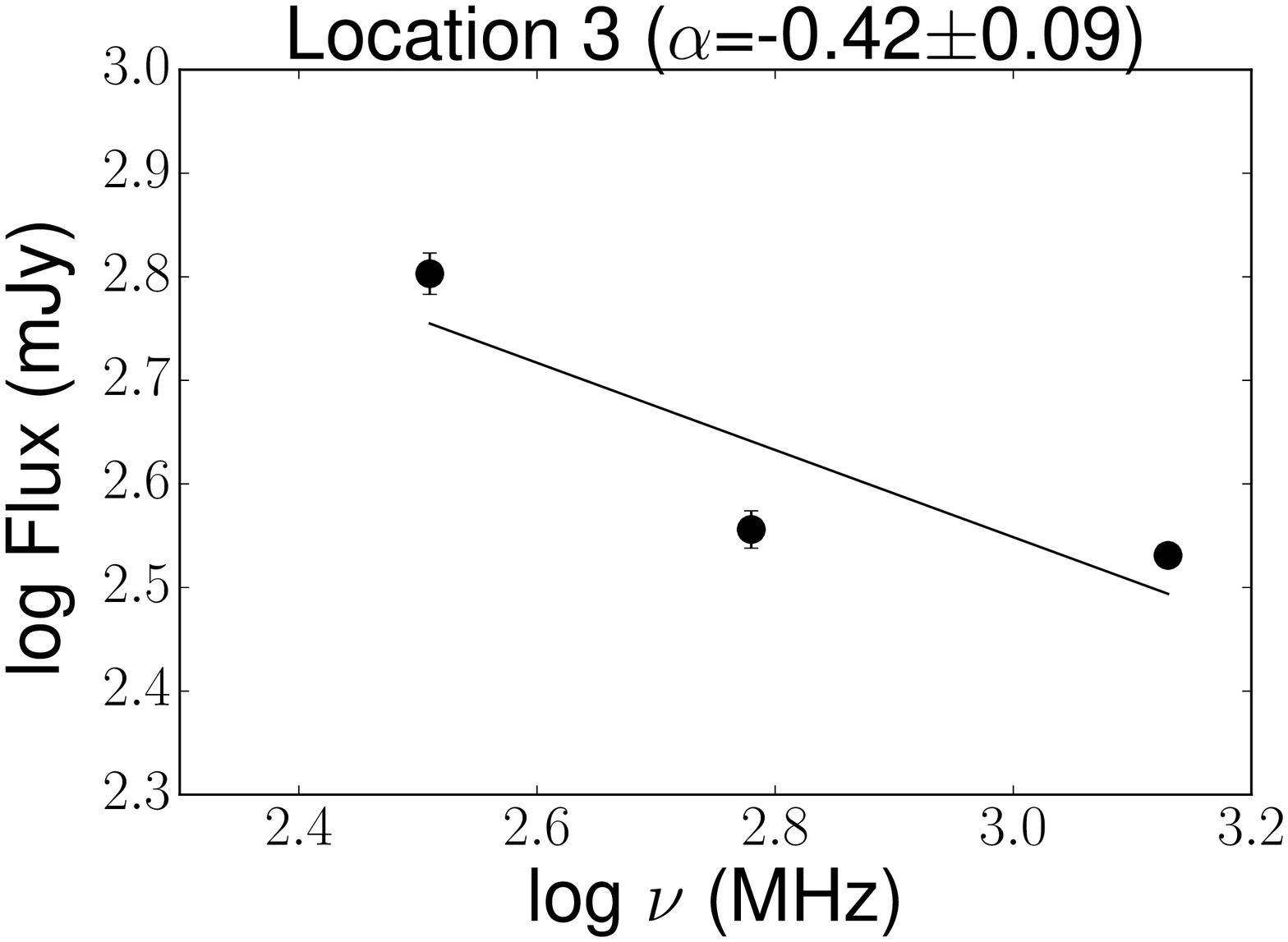} \quad \includegraphics[scale=0.2]{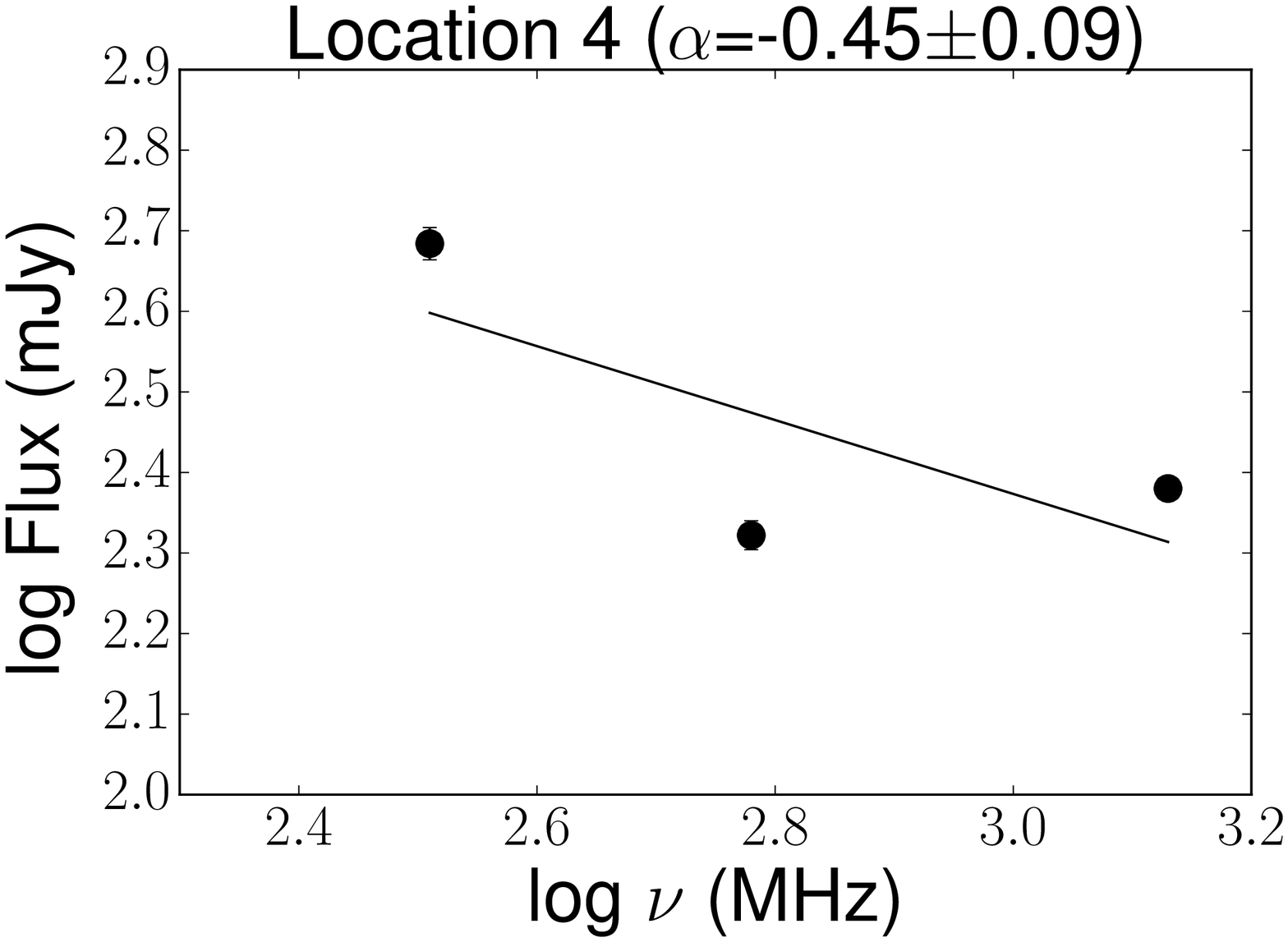}
\caption{The spectral index plots of the four locations shown in Fig.~\ref{specin}.}
\label{specinpl} 
\end{figure*}

\par The Lyman continuum photon flux at 1372~MHz was estimated using the equation
\citep{{1967ApJ...147..471M}, {2013A&A...550A..21S}},

\begin{equation}
\left[\frac{N_{Ly}}{\rm{s}^{-1}}\right]=8.9\times\rm10^{40}\left[\frac{S_\nu}{\rm{Jy}}\right]\left[\frac{\nu}{\rm{GHz}}\right]^{0.1}\left[\frac{T_e}{10^4\rm{K}}\right]^{-0.45}\left[\frac{d}{\rm{pc}}\right]^2
\end{equation} 

where, $S_{\nu}$ is the flux density at frequency $\nu$, $T_e$ is the electron temperature, and $d$ is the distance to the source. Based on the electron temperature gradient \textbf{across} the Galactocentric distance \citep{1978A&A....70..719C}, the electron temperature for this region is found to be 7500~K. Using this value, the Lyman continuum flux is found to be 3.5~$\times$~10$^{48}$~s$^{-1}$. This is a lower limit since the emission may be optically thick at this frequency. If a single ZAMS star is responsible for the ionization of the entire region, from our flux density measurements, we find that the ZAMS spectral type of this star to be earlier than $\rm{O7-O7.5}$ \citep{1973AJ.....78..929P}. From the high frequency radio continuum observations, \citet{2013A&A...550A..21S} estimated the spectral type of the ionizing source as O9 that is of later type when compared to our estimate. This is probably because
they may not be sampling the diffuse emission completely. An additional reason for their later spectral type is their assumption of electron temperature of 10$^4$~K compared to our 7500~K. Using the latter temperature would lead to a spectral type that is $\sim0.5$ earlier.

\par Another gauge of the ionised gas is the Br$\gamma$ recombination line in the near-infrared.  The continuum subtracted Br$\gamma$ image of the field around IRAS~17256--3631 is shown in Fig.~\ref{brg_H2}a. The image displays a similar cometary morphology, with the head towards the north-west direction that matches the radio continuum emission. However, being a near-infrared line, this suffers from extinction. This is evident from the high extinction regions seen in the Br$\gamma$ map.

\subsubsection{Spectral Index Maps}

The spectral index map is a useful tool to study the variation of spectral index across the \hii~region and to get an idea about the mechanisms responsible for radio emission.  We created two spectral indices maps, corresponding to 1372 - 610~MHz and 610 - 325~MHz. The spectral index maps along with their error maps are shown in Fig.~\ref{specin}. Both the spectral index maps are similar in morphology. From these maps, we see that the spectral indices in this region span a wide range, between -2.0 to 1.0. Regions with spectral index larger than $-0.1$ are believed to arise from thermal emission \citep{1975A&A....39..217O} while spectral index $< -0.5$ is believed to be mostly due to non-thermal mechanisms \citep{1999ApJ...527..154K}. Both the spectral index maps show evidences of non-thermal emission near the head towards the north and north-west, but the uncertainties are also larger here in both the maps. In the 610 - 325 map, the tail region shows steeper negative indices. One has to bear in mind that spectral indices include contributions from both thermal and non-thermal mechanisms, and spectral indices in both maps are likely to be different due to different mechanisms dominating at different frequencies.

\par We have compared the spectral indices at different locations using the three GMRT and two ATCA images (Sanchez-Monge et al. 2013), from images convolved to a common resolution of 31$\arcsec\times$31$\arcsec$. The radio SED at 4 representative positions, marked as 1, 2, 3 and 4 in Fig.~\ref{specin} are shown in Fig.~\ref{specinpl}. The spectral indices of these four locations are listed in Table~\ref{spectb}. Column 1 lists their locations, Column 2 lists the spectral index obtained from fitting to fluxes at five frequencies including ATCA, Column 3 lists the spectral index fitting to fluxes using three GMRT frequencies while Columns 4 and 5 list the spectral indices from~610 - 325 and 1372 - 610~MHz maps, respectively. The positions 1 and 2 have ATCA emission counterparts while positions 3 and 4 have only GMRT fluxes. This is because the ATCA maps are snapshot images of $\sim$10 min integration and we do not see the diffuse emission contribution, particularly towards the south-east. If the ATCA maps do not sample the diffuse emission, then the fluxes at some regions covered by ATCA maps are also likely to be lower than the actual values. While positions 1 and 2 near the radio peak show optically thin thermal emission with positive spectral indices less than 0.1, the other positions (3 and 4)  towards the outer envelope show evidences of non-thermal emission. Towards the south-east, near the tail region, the lower frequency (610 - 325) spectral indices are more non-thermal in nature (i.e. steeper negative index) when compared to the higher frequencies, ie. 1372 - 610. This is likely to be because thermal contribution dominates at higher frequencies. IRAS~17256--3631 exhibits a morphology where the spectral indices are nearly flat towards the core and relatively negative towards the diffuse envelope. A similar morphology is observed in other Galactic star forming regions \citep{{1993ApJ...415..191C},{felli1993radio},{1996ApJ...459..193G}}. In particular, \citet{2002ApJ...571..366M} have observed non-thermal emission towards the tail of two cometary \hii~regions. They attribute this to synchrotron emission produced by shocks or magnetic reconnection out of a pool of thermal particles. The combined thermal and non-thermal emission can be present in regions where there is a shock moving through a magnetized medium \citep{{1990ApJ...361L..49C}, {1991A&A...248..221H}}. \citet{1996ApJ...459..193G} proposed that the diffuse non-thermal emission corresponds to synchrotron radiation from electrons that are accelerated in the region of interaction between stellar wind and ambient cloud material. Hence it is likely that shock is responsible for the presence of non-thermal emission around the envelope.

\tabcolsep=0.10cm
\begin{table}
\footnotesize
\caption{Spectral indices corresponding to the locations shown in Fig.~\ref{specin}.}
\begin{center}
%\vskip 0.4cm
%\label{spec}
\hspace*{-0.5cm}
\begin{tabular}{c c c c c} \hline \hline
Location & $\alpha_{\text{\tiny{ATCA+GMRT}}}$  &$\alpha_{\text{\tiny{GMRT}}}$  &$\alpha_{610-325}$  &$\alpha_{1372-610}$  \\
\hline\\
1 &0.04$\pm$0.07&0.02$\pm$0.06&0.13$\pm$0.06 &-0.01$\pm$0.08 \\
2 &0.02$\pm$0.05 &0.18$\pm$0.03&0.24$\pm$0.08 &0.10$\pm$0.13 \\
3 & - &-0.42$\pm$0.09 &-0.91$\pm$0.15 &-0.07$\pm$0.16 \\
4 & - &-0.43$\pm$0.09 &-1.25$\pm$0.23 &0.09$\pm$0.25 \\

\hline
\end{tabular}
\label{spectb}
\end{center}
\end{table}

\tabcolsep=0.08cm
\begin{table}
\footnotesize
\caption{The positions of identified H$_2$ knots.}
\begin{center}

\hspace*{-0.5cm}
\begin{tabular}{c c c } \hline \hline
H$_2$ knot no. & $\rm{\alpha_{J2000}}$ & $\rm{\delta_{J2000}}$ \\
& ($^{h~m~s}$) & ($^{\degr~\arcmin~\arcsec}$)\\
\hline\\
1 &17:28:56.916 &-36:31:41.48 \\
2 &17:28:58.405 &-36:32:07.28 \\
3 &17:28:57.109 &-36:32:14.24 \\
4 &17:29:06.348 &-36:34:57.65 \\
5 &17:29:05.965 &-36:34:58.08 \\
6 &17:29:03.565 &-36:35:26.19 \\
7 &17:29:09.998 &-36:33:53.21 \\
\hline
\end{tabular}
\label{H2knot}
\end{center}
\end{table}

\subsection{H$_2$ Emission from Molecular Gas}
The high-excitation molecular gas can be probed using the near-infrared H$_2$ line. Our continuum-subtracted H$_2$ image (Fig.~\ref{brg_H2}b) shows faint diffuse filamentary emission. A comparison of the H$_2$ and Br$\gamma$ emission reveals that the morphology of H$_2$ is different compared to the ionized gas emission. H$_2$ is seen at the edges of the ionized gas emission while there is little or no emission in the central region towards the radio peak. This indicates the presence of highly excited molecular gas around the \hii~region. In addition to the filamentary structures, several H$_2$ knots are also seen in this region. These are marked in Fig.~\ref{brg_H2}b and listed in Table~\ref{H2knot}. All the knots other than knot 7 seem to be aligned towards NW-SE direction with respect to IRS-1 (position angle $\sim$24\degr). Similar knots are also detected in other star forming regions that are interpreted as shocked emission along the jet axis \citep{2010MNRAS.404..661V}. One of the H$_2$ knots (Knot No. 2) is located near the object EGO-1 (discussed later). This knot is likely to be associated with EGO-1.

\subsection{Dust Emission}

\subsubsection{Emission from Cold Dust}

Emission at seven wavelength bands have been used to study the physical properties of the cold dust clumps. These include five bands from $Herschel$ Hi-GAL (70, 160, 250, 350 and 500~$\mu$m), 870~$\mu$m emission from the ATLASGAL survey, and 1.2~mm emission from \citet{2006A&A...447..221B}. In order to estimate the dust clump properties, we first identified clumps in 350~$\mu$m image as the resolution is optimum, close to the lowest available resolution at 500~$\mu$m. Identifying clumps in the higher resolution images at 70 or 160~$\mu$m would lead to inappropriate allocation of fluxes in the low resolution images if we use the same apertures in all the wavelength bands. A flip side of identifying clumps using 350~$\mu$m image is that more than one clump in a higher resolution image (say 870~$\mu$m) could be associated with a single 350~$\mu$m image. The clumps are identified using the $\it{2{\rm D}-Clumpfind}$ algorithm, a two-dimensional variation of the $\it{Clumpfind}$ algorithm developed by \citet{1994ApJ...428..693W}. The algorithm works by contouring the data at multiples of rms noise in the map and searches for peaks of emission that are identified as clumps. The advantage of $\it{2{\rm D}-Clumpfind}$ is that it does not $\it{a priori}$ assume any clump profile, and the total flux is conserved.

\begin{figure}
\hspace*{-0.2cm}
\centering \includegraphics[scale=0.35]{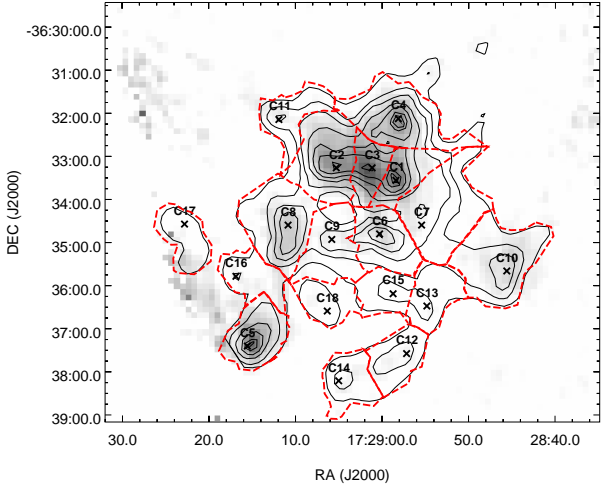}
\caption{1.2~mm SIMBA map overlaid with clump apertures used for identifying clumps. The peak positions of clumps are indicated. The contours represent SPIRE 350~$\mu$m emission with levels at 830, 1260, 1650, 2040, 2430, 2820, 3210, 3600, 3990~MJy/Sr.}
\label{350_aper}
\end{figure}

\tabcolsep=0.08cm
\begin{table*}
\footnote-size
\caption{Molecular clumps identified in this region.}
\begin{center}
%\label{spec}
\hspace*{-0.5cm}
\begin{tabular}{ c c c c c c c c c c}\hline \hline
\setlength{\tabcolsep}{1pt}
Clump & $\rm{\alpha_{J2000}}$ & $\rm{\delta_{J2000}}$ &Area &$\beta$ &Temperature &Column density&Mass &L &$\Sigma$\\
& $(^{h~m~s})$ &$(^{\degr~\arcmin~\arcsec})$&(pc$^{-2}$) &&(K)&(10$^{21}$ cm$^{-2}$)&($\times$10$^2$~M$_\odot$) &($\times$10$^2$~L$_\odot$) &(g cm$^{-2}$)\\
\hline\\
C1 &17:28:58.399 &-36:33:33.57 &0.5 &2.0 $\pm$ 0.1 & 28.6 $\pm$ 1.4 & 20.4 $\pm$ 3.2 &2&102&0.08\\
C2 &17:29:05.286 &-36:33:16.16 &0.8 &2.0 $\pm$ 0.2 & 27.1 $\pm$ 1.9 & 16.7 $\pm$ 3.8 &4&144&0.09\\
C3 &17:29:01.149 &-36:33:19.41 &0.3 &1.8 $\pm$ 0.1 & 33.2 $\pm$ 1.5 & 16.9 $\pm$ 2.2 &2&138&0.10\\
C4 &17:28:58.118 &-36:32:07.71 &1.1 &2.3 $\pm$ 0.3 & 20.2 $\pm$ 1.7 & 23.9 $\pm$ 8.5 &6&49&0.11\\
C5 &17:29:15.564 &-36:37:24.28 &0.8 &2.6 $\pm$ 0.2 & 14.2 $\pm$ 1.1 & 85.3 $\pm$ 23.8 &12&11&0.31\\
C6 &17:29:00.282 &-36:34:48.17 &0.8 &1.9 $\pm$ 0.6 & 27.1 $\pm$ 11.9 & 15.7 $\pm$ 15.2 &2&75&0.06\\
C7 &17:28:55.431 &-36:34:09.92 &1.1 &2.6 $\pm$ 0.4 & 24.9 $\pm$ 3.3 & 12.8 $\pm$ 5.0 &3&143&0.06\\
C8 &17:29:10.854 &-36:34:35.85 &1.1 &2.1 $\pm$ 0.6 & 19.6 $\pm$ 6.0 & 26.7 $\pm$ 22.6 &6&37&0.12\\
C9 &17:29:05.807 &-36:34:55.84 &0.5 &3.1 $\pm$ 0.6 &13.9 $\pm$ 2.7&75.8 $\pm$ 55&10&1&0.43\\
C10 &17:28:45.591 &-36:35:39.65 &1.1 &2.8 $\pm$ 0.4 &20.6 $\pm$ 2.6&14.7 $\pm$ 7.9&4&9&0.16\\
C11 &17:29:11.910 &-36:32:08.88 &0.5 &2.5 $\pm$ 0.4 &22.4 $\pm$ 2.7&12.4 $\pm$ 5.6&2&4&0.07\\
C13 &17:28:54.820 &-36:36:28.52 &0.5 &1.3 $\pm$ 0.2 &32.7 $\pm$ 2.8&6.6 $\pm$ 1.3&2&7&0.08\\
\hline
&&&&SED fit with $\beta$=2&&\\
\hline
C12 &17:28:57.160 &-36:37:34.76 &0.5 &2&21.9 $\pm$ 0.2&16.4 $\pm$ 0.03&1&3&0.03\\
C14 &17:29:05.008 &-36:38:12.65 &0.5 &2&20.4 $\pm$ 1.3&18.8 $\pm$ 0.3&2&3&0.10\\
C15 &17:28:58.722 &-36:36:11.46 &0.5 &2&20.5 $\pm$ 1.5&19.2 $\pm$ 0.4&2&3&0.09\\
C16 &17:29:16.880 &-36:35:47.15 &0.5 &2&17.8 $\pm$ 0.01&24.2 $\pm$ 0.01&3 &1&0.10\\
C17 &17:29:22.812 &-36:34:34.38 &0.5 &2&19.3 $\pm$ 0.2&19.5 $\pm$ 0.03&2 &2&0.09\\
C18 &17:29:06.328 &-36:36:35.52 &0.5 &2&18.8 $\pm$ 0.3&20.0 $\pm$ 0.1&2&1&0.10\\ 

\hline
\end{tabular}
\label{cl18}
\end{center}
\end{table*}

\par The results from the \textit{Herschel Space Observatory} have shown that filaments are all-prevalent in the Galactic plane, particularly at the longer wavelengths \citep{2010A&A...518L.100M}. Such filamentary emission is also seen in the vicinity of this region, IRAS~17256--3631. We are interested in high density clumps characterized by large emission at 350~$\mu$m. The threshold and the contour spacing are given as input parameters in terms of the rms. To locate clumps in the 350~$\mu$m image, the lowest contour level is set at 830~MJy/Sr corresponding to $\sim15\sigma$ with a contour step of 2$\sigma$. Here $\sigma$ is 50~MJy/Sr, the uncertainty determined from a region close by ($\sim 3'$ away). A total of eighteen clumps were obtained by this method. They are listed in Table~\ref{cl18} and are shown in Fig.~\ref{350_aper}. We have also explored the clump structure using the dendrogram algorithm \citep{2008ApJ...679.1338R} and find that the tree structure is the same as that obtained using $\it{2{\rm D}-Clumpfind}$. 

\begin{figure*}
\hspace*{-0.3cm}
\begin{minipage}{0.49 \textwidth}
\includegraphics[angle=270, scale=0.35]{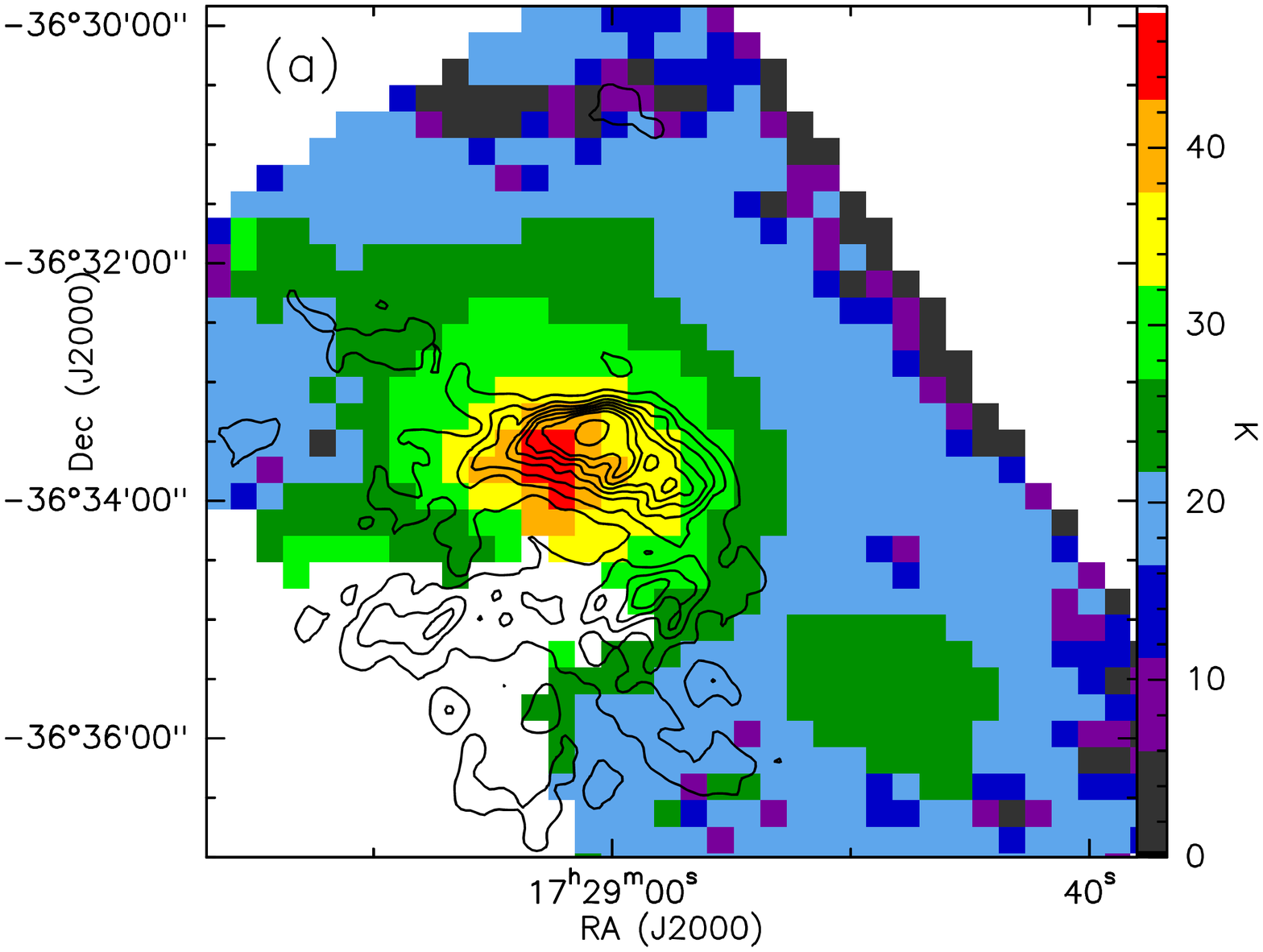}
\label{}
\end{minipage}
\hspace*{0.4cm}
\begin{minipage}{0.49 \textwidth}
\includegraphics[scale=0.37]{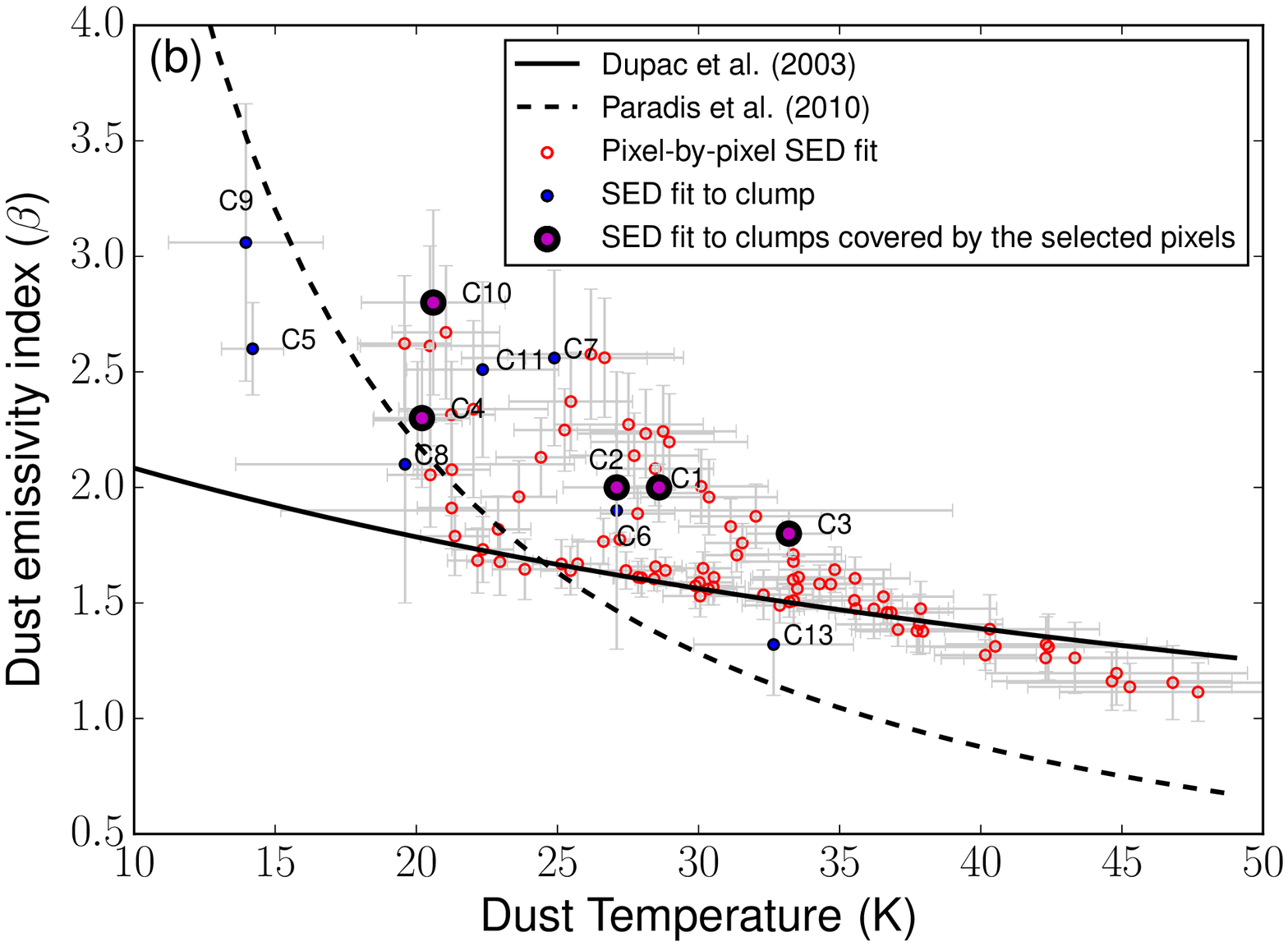} 
\label{}
\end{minipage}
\caption{(a) Dust temperature map overlaid with 1372~MHz radio contours. (b) The plot of T$_d$ versus $\beta$ for pixels within 3-$\sigma$ contour of 1.2~mm map. Also plotted are values for 12 clumps with variable $\beta$. The $\beta$-T$_d$ relations of \citep{2003A&A...404L..11D} and \citep{2010A&A...520L...8P} are also shown in the plot.}
\label{betatd} 
\end{figure*}

\par In order to obtain the flux densities of clumps at other wavelengths (i.e. 70, 160, 250, 500, 870~$\mu$m and 1.2~mm) for clump SED, we adopted arbitrary apertures corresponding to the area covered by each clump in the 350~$\mu$m image.  The apertures used for the clumps are shown in Fig~\ref{350_aper}. A few clumps do not have PACS coverage while others are not detected in 870~$\mu$m or 1.2~mm images. The latter is not surprising  considering the poorer sensitivities of ground-based telescopes that would not sample at low flux levels. A clump is considered to be detected at 870~$\mu$m and/or 1.2~mm if the flux levels of at least 9 pixels (approximately one beam) towards the clump are above the corresponding 3-$\sigma$ contour level. For each clump,  SEDs are constructed using available wavelength bands. The flux densities for each clump within the arbitrary aperture at wavelengths at which it is detected are estimated. Sky emission within an identical aperture is subtracted from the flux densities by considering a region that is $\sim15\arcmin$ away in the Galactic plane that is devoid of bright diffuse emission. The flux densities are fitted with a modified blackbody function of the following form \citep{{1990MNRAS.244..458W}, {2011A&A...535A.128B}}.

\begin{equation}
F_{\nu}=\Omega B_{\nu}(T_d)(1-e^{-\tau_\nu})
\end{equation}

\noindent where

\begin{equation}
\tau_\nu=\mu\  m_H \kappa_{\nu} N(H_2)
\end{equation}

Here, $\Omega$ is the solid angle (aperture size) subtended by the clump, $\kappa_{\nu}$ is the dust opacity, $B_{\nu}(T_d)$ is the blackbody function at a dust temperature $T_d$, $\mu$ is the mean weight of molecular gas which is mostly H$_2$, $m_H$ is mass of hydrogen atom and $N(H_2)$ is the column density. We adopted a value of 2.8 for $\mu$ \citep{2008A&A...487..993K}. The dust opacity is calculated using the expression \citep{2010A&A...518L..92W},

\begin{equation}
\kappa_{\nu}=0.1(\nu/1000\rm{GHz})^{\beta} \rm{cm^2g^{-1}}
\label{opacity}
\end{equation}

In the above expression, $\beta$ is the dust emissivity index, that is set as a free parameter along with dust temperature and column density in our fits. Six clumps: C12, C14, C15, C16, C17 and C18 are neither covered by PACS nor are they detected at 870~$\mu$m and 1.2~mm.  For these clumps, as only three SPIRE wavebands are available, $\beta$ is constrained to 2. The fitting was carried out using the nonlinear least squares Marquardt-Levenberg algorithm. We assumed flux uncertainties of the order $\sim15$\% in all bands \citep{{2013A&A...551A..98L}, {2009A&A...504..415S}, {2004A&A...426...97F}, {2005ApJ...634..436B}}. The clump SEDs are presented in Appendix A. 

\begin{figure*}
\hspace*{-0.3cm}
\begin{minipage}{0.49 \textwidth}
\includegraphics[angle=270, scale=0.35]{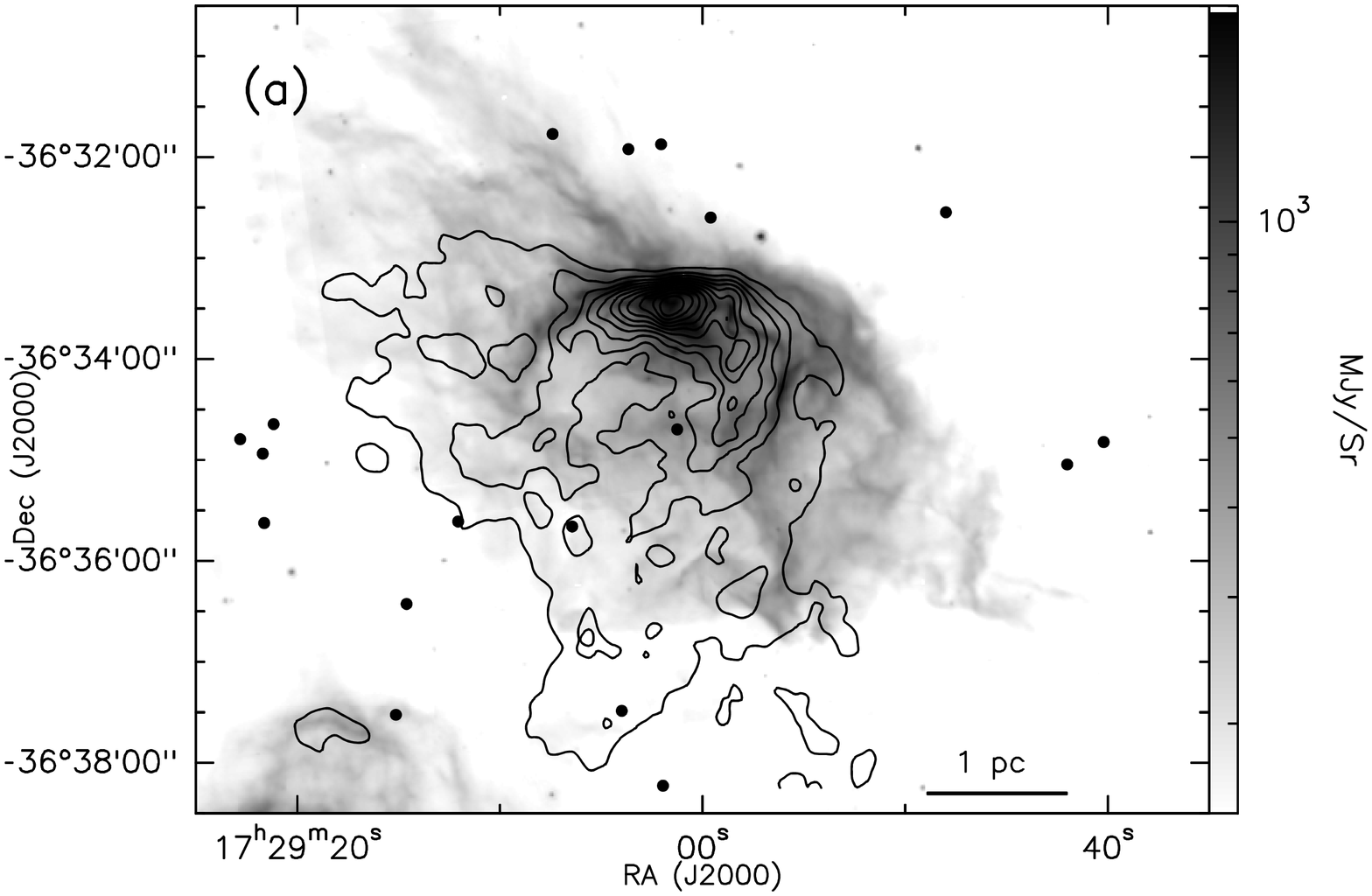}
\label{}
\end{minipage}
\hspace*{0.4cm}
\begin{minipage}{0.49 \textwidth}
\includegraphics[angle=270, scale=0.37]{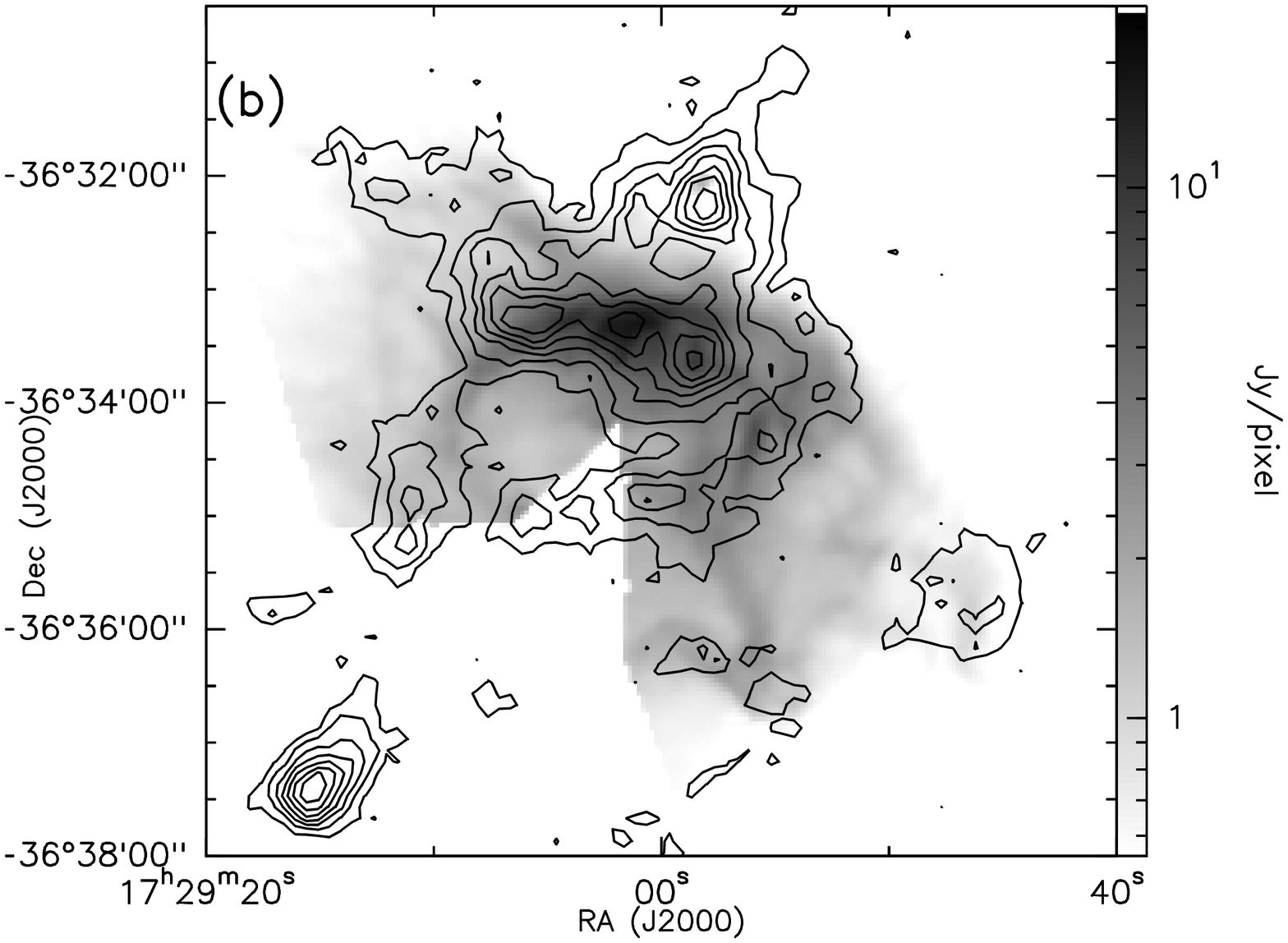} 
\label{}
\end{minipage}
\caption{(a) Overlay of ionized gas contours at 610 MHz on the \textit{Spitzer} IRAC 8~$\mu$m image. The contour levels are at 8, 19.5, 31.0, 42.6, 54.2, 65.7, 77.2, 88.8, 100, 112, 123 and 135~mJy/beam. The positions of 18 YSOs are also marked. (b) Image of PACS 70~$\mu$m emission overlaid with 870~$\mu$m cold dust contours. The contour levels are at 0.24, 0.54, 0.84, 1.14, 1.44, 1.74, 2.04 and 2.34 Jy/beam.}
\label{8m_870} 
\end{figure*}

The dust temperature and column density values obtained using the best-fit are listed in Table~\ref{cl18} for all the eighteen clumps. For twelve clumps, the best-fit $\beta$ values are also tabulated. 
The clump temperatures range from $14-33$~K. The central clump C3 has the highest temperature of 33~K. This is also the location of the radio emission peak. The column densities lie in the range $0.7-8.5\times10^{22}$~cm$^{-2}$. The highest column density is found towards clump C5, that also has the lowest temperature of 14~K. The value of $\beta$ lies between $1.8-2.3$ for the brightest four clumps: C1, C2, C3 and C4. This is close to the normally accepted value of 2 \citep{1983QJRAS..24..267H}. For the other eight clumps fitted with variable $\beta$, the value of $\beta$ ranges from 1.3 to 3.1.
The $\beta$ values obtained show an inverse correlation with the dust temperature. We investigate this further in the next section.
For dense clumps, other authors find that the values of $\beta$ range from $0.9-3.3$ \citep[e.g.,][]{{2010A&A...518L..99A}, {rathborne2010early}}. \citet{1996ApJ...459..619K} obtained a value of 3.7 for Sgr B(2)N region and they have interpreted it as a strong evidence for the presence of ice coated dust mantles in the region. The mean column density value of these clumps, $2.5\times10^{22}$~cm$^{-2}$, is similar to that of other massive star forming regions, eg., mean value of $1\times10^{22} \rm{cm^{-2}}$ for Gemini OB1 molecular cloud complex \citep{2010ApJ...717.1157D} and $5.7\times10^{22}-8.6\times10^{23} \rm{cm^{-2}}$ for 18 young star forming regions \citep{garay2007multiwavelength}. For R$_V=3.1$, using the relation N(H$_2$)/A$_V = 9.4 \times 10^{20}$~cm$^{-2}$mag$^{-1}$ \citep{1978ApJ...224..132B}, we find that A$_V=26$~mag for the mean column density. \textbf{Although this is similar to the value of diffuse interstellar extinction for a distance of 14.7~kpc (Section 1), this is likely to be a coincidence as extinction estimated from the mean column density includes the effect of the dense clump(s) while the latter refers to diffuse interstellar medium alone. }

The column densities are converted to clump masses using the equation

\begin{equation}
M_c= {N(H_2)}\mu\ m_HA
\end{equation}

Here M$_c$ is the clump mass, and $A$ is the physical area of the clump. The area and mass of each clump are listed in Table~\ref{cl18}. The masses of clumps lie between $\sim100-1200$~M$_{\odot}$, and the total mass of the cloud is 6700~M$_{\odot}$. \citet{2006A&A...447..221B} used 1.2~mm continuum data to estimate the total mass of the clumps within this region and they obtained a total mass of 674~M$_{\odot}$. This is a factor of 10 lower than our value, and can be explained on the basis of the fact that the ground based SEST-SIMBA observations at 1.2~mm do not sample a number of clumps as well as the diffuse emission at low flux levels owing to poorer sensitivity. 
%Note that here we have compared the mass from emission corresponding to eight clumps alone. The total mass is expected to be larger than this (discussed later). 
We estimate clump diameters of $0.6$ to $1.2$~pc assuming each clump to be spherical. A clump requires $\sim$30~M$_{\odot}$ to form a high mass star with star formation efficiency of 30\% or greater \citep{2003ARA&A..41...57L}. Similarly, \citet{2007A&A...476.1243M} and \citet{2013ApJ...773..102F} use a lower limit of 40~M${_\odot}$ for identifying potential molecular cores that can form a high mass star. The least massive of our clumps is 100~M$_{\odot}$, and the sizes of our clumps are similar to the sample of these authors: $0.03-0.5$~pc \citep{2007A&A...476.1243M} and $0.4-1.1$~pc \citep{2013ApJ...773..102F}. It is likely few of the clumps in this region either harbor or are capable of forming high mass stars. 

\par Another measure of high mass star formation is the surface density ($\Sigma$) of the molecular cloud \citep{2008Natur.451.1082K}. We divide the mass of each clump by the area and obtain the surface densities of clumps that are listed in Table~\ref{cl18}. The surface densities of the clumps range from 0.03 to 0.4 g/cm$^2$. According to \citet{2008Natur.451.1082K}, these values would indicate that fragmentation could be initiated in these clumps with massive star formation being supressed as a result. However, \citet{2010A&A...517A..66L} and \citet{2013A&A...556A..16G} find lower surface densities even towards massive star forming cores, i.e. those with surface densities of 0.2 g/cm$^2$ or larger. Similarly, \citet{2010A&A...520A.102M} find that it is possible to have lower surface densities of clumps (i.e. lower than 1 g/cm$^2$) for massive star formation, but the core densities could be higher. In addition, \citet{2009ApJS..181..360C} in their analyses comment that lower surface densities of clumps could be attributed to large areas used (i.e large apertures) as they find massive protostars in cores of lower surface densities. 
In our case, the  average surface densities are representative values that could be lower than the surface densities near the core or peak emission of clump as the area of clumps is very large (reaching to $\sim$20\% of peak flux in few cases). Further, there are uncertainities associated with the temperature and column density. Hence, it is possible that few clumps in this region with larger average surface densities such as C5, C9 and C10 have the potential to form massive stars. 

\begin{figure}
\includegraphics[scale=0.4]{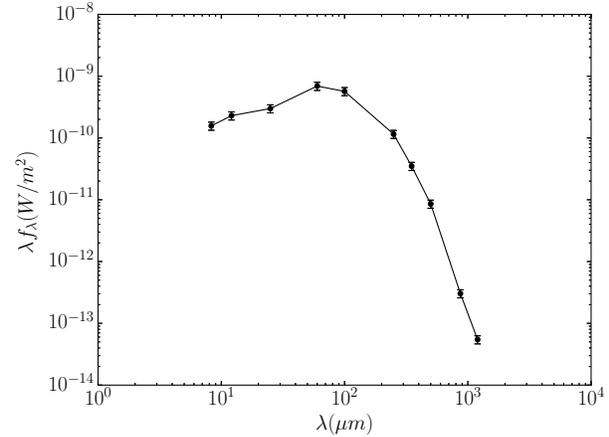}
\caption{Spectral Energy Distribution of this region obtained by integrating fluxes within a circular region of diameter $3.5\arcmin$ centred on the IRAS peak.}
\label{SED}
\end{figure}

\par We have calculated the luminosity of the individual clumps by integrating the flux densities within the best fit curve of 
the modified blackbody. The estimated luminosities are given in Table~\ref{cl18} and range from $10^2 -1.4\times10^4~\rm{L_{\odot}}$. Clump 
luminosities provide an insight into embedded young stellar objects as well as the clump evolutionary stage. According to
 \citet{rathborne2010early}, a bolometric luminosity of 10$^4$~L$_{\odot}$ can be used as a rough threshold to identify clumps 
harboring a high-mass protostar. Using this criteria, we find 4 of the clumps (C1, C2, C3 and C7) have luminosities $>$ 10$^4$~L$_{\odot}$ and may contain a high mass protostar.

\begin{figure*}
\hspace*{-0.5cm}
\begin{minipage}{0.49 \textwidth}
\includegraphics[scale=0.33]{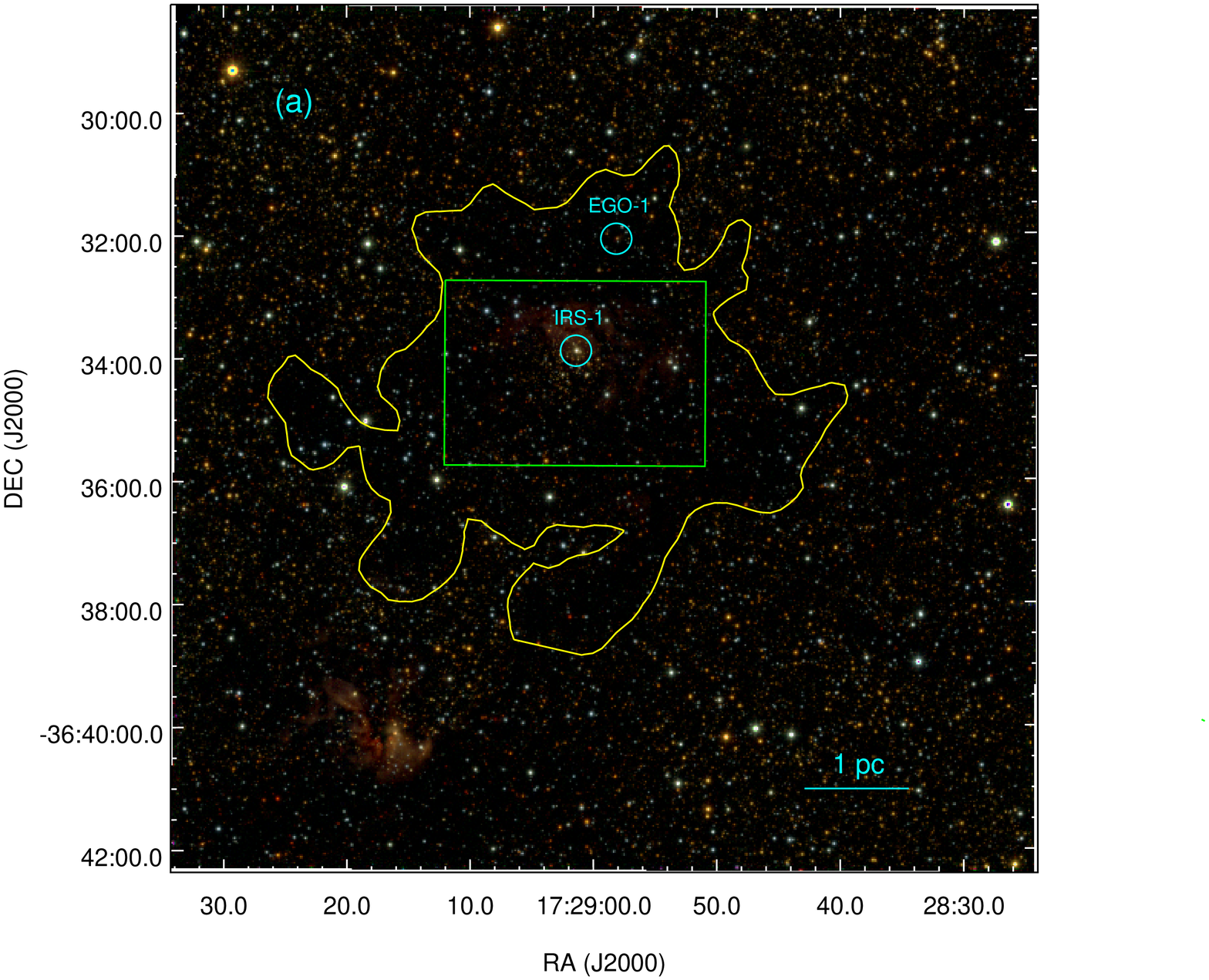} 
\label{}
\end{minipage}
\begin{minipage}{0.49 \textwidth}
\includegraphics[scale=0.32]{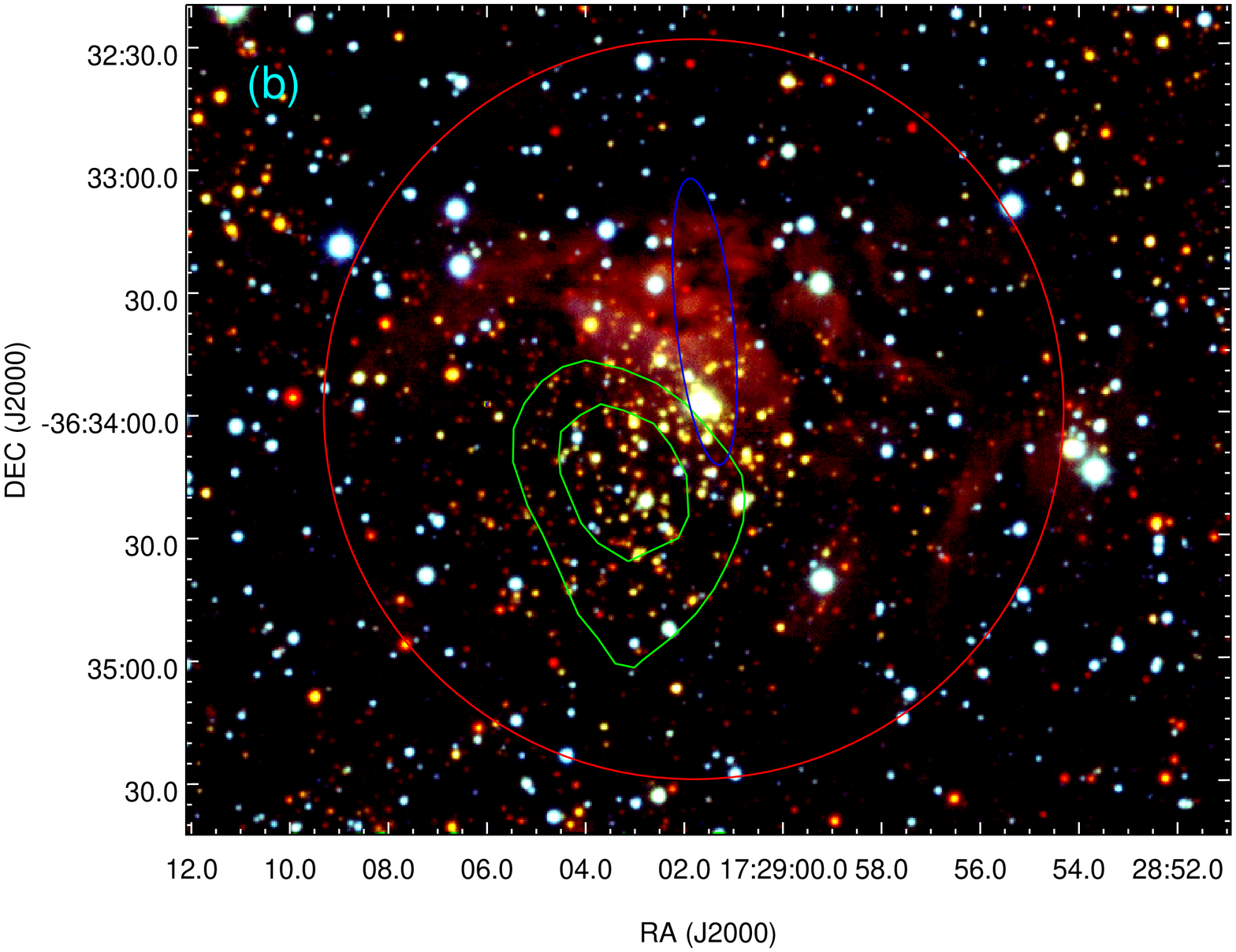}
\label{}
\end{minipage}
\caption{(a) UKIRT color composite image of the IRAS~17256--3631 region in JHK bands. Blue represents J band, green represents H band and red represents K band. Lowest contour of  350~$\mu$m data is marked as yellow contour. The locations of IRS-1 and EGO-1 are also marked in the image. (b) Colour composite image showing the stellar density contours (green) corresponding to the near-infrared cluster. The region of radius 1.5$\arcmin$ centered on IRS-1 is marked as a red circle. The IRAS error ellipse (blue) is also shown in the figure.}
\label{col_com} 
\end{figure*}

\subsubsection{Relation between Temperature and Dust Emissivity Index}  
We observe that clumps with higher dust emissivity index $\beta$ show lower values of dust temperature $T_d$ and vice-versa. Earlier studies have shown an anti-correlation between  these parameters \citep[e.g.][]{{2003A&A...404L..11D}, {2008A&A...481..411D}, {2010A&A...520L...8P}, {2012A&A...542A..10A}}. In order to investigate this further as the $\beta$ and $T_d$ parameters obtained from clump SED fitting represent average values, we have constructed the column density, dust temperature and $\beta$ maps of this region by carrying out pixel-by-pixel fit using a greybody SED (discussed in Section 3.3.1) to 7 images of this region (70 - 1200~$\mu$m). The details are given in Appendix B. However, we obtain only partial maps of this region owing to partial coverage by PACS. The temperature map is shown in Fig.~\ref{betatd}(a). The temperature distribution shows a maximum of 47.7~$\pm$~4.9 K within Clump C3 that also corresponds to the radio emission peak.  For estimating the median values and other statistical analysis such as that of $\beta$-$T_d$ relation, we have selected pixels lying within the 3-sigma contour of 1.2 mm map as it samples regions (clumps) where the greybody SED fits are expected to be fairly good. For these pixels, we find that the reduced chi-square $<$ 5.  The $\beta$, column density and reduced chi-square maps are shown in Appendix B.

\par In order to understand the relation between  $\beta$ and $T_d$, we have plotted  $\beta$ as a function of $T_d$ for these pixels. This is shown in Fig.~\ref{betatd}(b).  We see an anti-correlation between $\beta$ and T$_d$ values for these pixels. Also plotted in the figure are the values of individual 12 clumps that have been fitted with variable $\beta$ (see Table~\ref{cl18}). Among these, the clumps C1, C2, C3, C4 and C10 are covered within the selected pixels. Their dust temperature and $\beta$ values lie close to the region occupied by pixels. The clumps seem to show the anti-correlation albeit with larger uncertainties compared to pixels. Clump C9 shows the largest $\beta$ value of $3.1\pm0.6$. Among the pixels, $\beta$ values range between $1.1-2.7$ in the temperature range $19.5-47.7$~K. The inverse correlation between $\beta$ and $T_d$ observed is similar to that obtained by other studies. \citet{2003A&A...404L..11D} have observed this towards numerous regions in our Galaxy and they attribute this to (i) change in grain sizes in cold dense environments, (ii) differences in chemical composition of grains  or (iii) intrinsic dependence of $\beta$ on $T_d$ due to quantum processes that could be dominant. The best-fit hyperbolic function obtained by them has been overplotted on Fig~\ref{betatd}(b). We have also overplotted the best-fit relations obtained by \citet{2010A&A...520L...8P} towards the region of Galactic longitude $l=30^\circ$ for comparison.  

\par The warmest pixel corresponds to the clump C3 and this matches with the highest average dust temperature obtained for the clump C3. The pixel with the highest temperature $47.7\pm4.9$~K shows the lowest value of $\beta\sim 1.1\pm0.1$. Warm regions show lower dust emissivity index, and vice-versa in the outer envelope.   Assuming this to be true and intrinsic to grain properties,  we can infer that the warmer or active region is comprised of bare silicate grain aggregates or porous graphite grains \citep{{2003A&A...404L..11D},{1989ApJ...341..808M}}, while the higher emissivity index regions could be interpreted as those with icy mantles \citep{1975ApJ...200...30A}. We, however, cannot exclude the possibility of this effect arising due to different molecular clouds along the same line-of-sight \citep{2009ApJ...696..676S} considering that this region lies along the inner Galactic plane.

\subsubsection{Emission from Warm Dust}

\par The emission from warm dust towards IRAS~17256--3631 is seen as extended diffuse emission in the IRAC~8~$\mu$m image. This emission possesses a sharp, arc like edge towards the north. Further, emission in the form of filamentary structures is also seen. The MIR maps in other IRAC bands (3.6 and 4.5~$\mu$m) also show similar emission features as in the 8~$\mu$m map. Diffuse emission from the  warm dust is far more extended than the ionized region in the NE--SW direction as seen in Fig.~\ref{8m_870}a. However, regions of higher flux densities trace the ionized gas morphology towards the north-west. 
\par The 8~$\mu$m emission is dominated by mid-infrared radiation from the heated dust as well as emission from the photodissociation regions (PDRs) \citep{2009PASP..121..213C}. Polycyclic Aromatic Hydrocarbons \citep[PAH;][]{1984A&A...137L...5L} are also significant contributors at these wavelengths. The combination of radio and 8~$\mu$m show that the \hii~region is surrounded by the bright mid-infrared filamentary emission that is probably tracing the PDR associated with this region. The emission morphology at 70~$\mu$m bears a similarity to that at 8~$\mu$m as seen in Fig.~\ref{8m_870}b. 
The better correlation of 70~$\mu$m emission with 8~$\mu$m emission as compared to other Herschel wavelength bands
can be attributed to the fact that emission at 70~$\mu$m is not due to a single dust component. The stochastically heated dust grains as well as grains that are at thermal equilibrium contribute to the 70~$\mu$m emission \citep{2010ApJ...724L..44C}. 
However, there are some prominent differences between 8 and 70~$\mu$m emission that include filamentary emission appearing to emanate from the cometary head. 

\subsubsection{Emission Morphology and Spectral Energy Distribution}

\par The emission from cold dust at far-infrared and submm wavelengths shows a clumpy distribution. The morphology of cold dust emission is different from that of ionized and mid-infrared emission. The three central brightest clumps are along an arc (Clump C1, C2 and C3 in Fig.~\ref{350_aper}). It is along this region where most of the bright mid-infrared emission is located. However, the radio cometary head coincides with the clump peaks and the radio peak is located southwards. There is less near and mid-infrared nebulosity towards the clumps C4, C5 and C8. 
 
We also obtain total SED of this region using flux densities from IRAS-HIRES \citep{1990AJ.....99.1674A}, Midcourse Space Experiment \citep[MSX][]{price2001midcourse}, Herschel-HiGal, ATLASGAL and SEST-SIMBA within a circular region of radius $1.8\arcmin$ centred on the IRAS peak. This is shown in Fig.~\ref{SED}. By integrating under the SED, we find the total luminosity of this region to be $1.6\times10^5$~L$_\odot$. This is marginally larger than the previous estimate, given that we have better wavelength coverage. This bolometric luminosity corresponds to a single ZAMS star of spectral type O6.5.

\subsection{Embedded Cluster and Young Stellar Objects }

We have used the UKIRT and IRAC point source catalogs (discussed in Sections 2.2.1 and 2.3.1, respectively) for studying the associated cluster and the nature of the stellar population in the region around IRAS~17256--3631.

\subsubsection{Infrared Cluster}

\par A JHK colour-composite image of the region around IRAS~17256--3631 is shown in Fig.~\ref{col_com}a. Apart from a 
spatially extended diffuse emission the region also harbours a partially embedded cluster. This cluster has been 
identified by \citet{2003A&A...404..223B} based on 2MASS data.  
Using the source count algorithm outlined in \citet{2010ASInC...1...27S} we revisited this cluster with high 
resolution and deeper UKIRT data. Fig.~\ref{col_com}b shows the stellar surface density contours 
for a grid / bin size of 60$\arcsec$ with separation of 30$\arcsec$. The source count method, that identifies density
enhancements over a background level, clearly identifies only a part of the cluster that is out of the nebulosity. This implies that the cluster is partially embedded.

\begin{figure}
\includegraphics[scale=0.4]{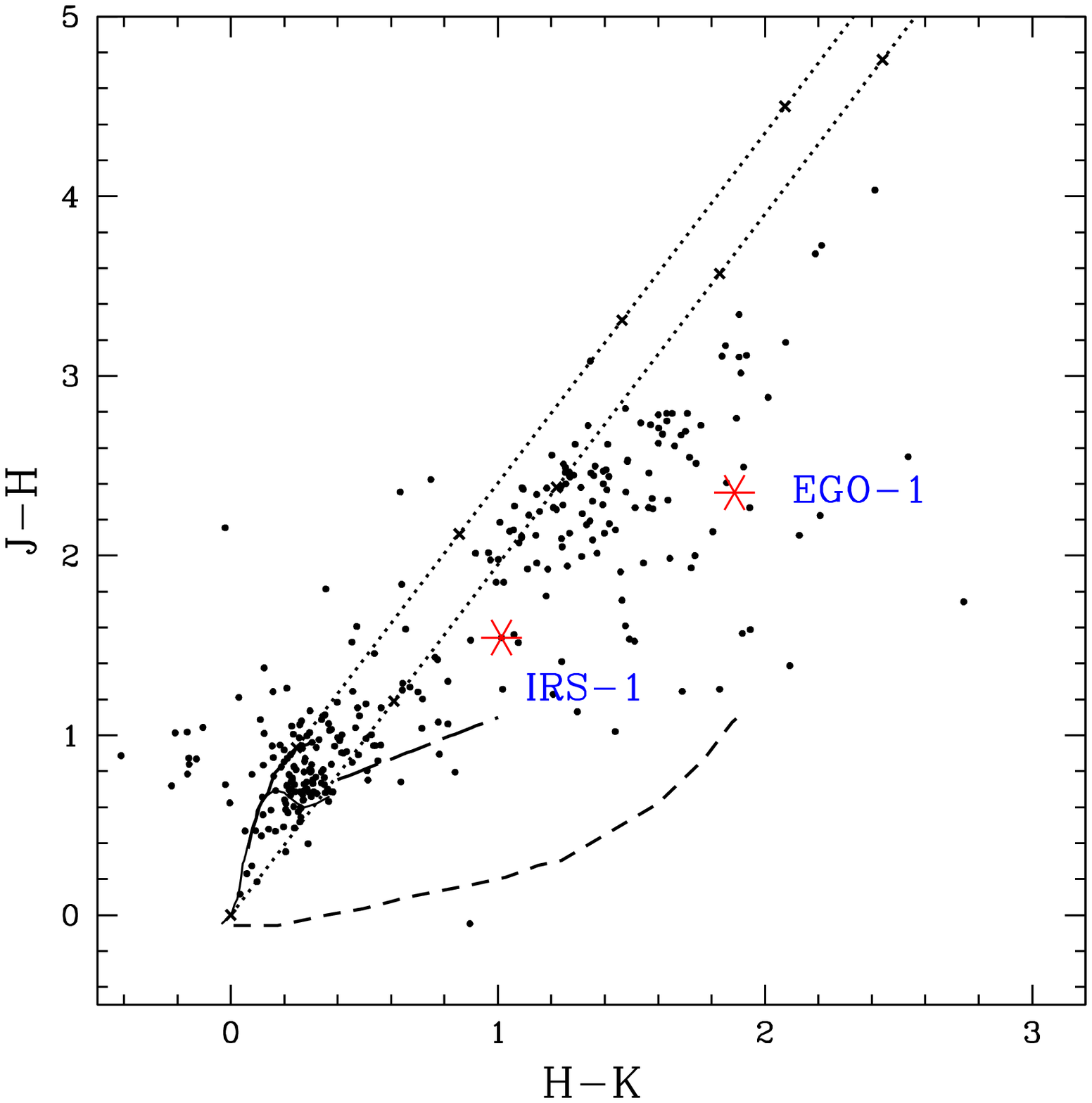}
\caption{The colour-colour diagram of near-infrared sources within 1.5$\arcmin$ radius of IRS-1. Two solid curves represent loci of main sequence (thin line) and giant stars (thick line), respectively, derived from \protect\cite{1988PASP..100.1134B}. Long dashed lines represents T Tauri locus \protect\citep{1997AJ....114..288M}. Short dashed lines represents the loci of Herbig AeBe stars \protect\citep{1992ApJ...393..278L}. Parallel dotted lines represent reddening vectors with cross points placed along intervals corresponding to 10 magnitudes of A$_V$. We have assumed interstellar extinction law of \protect\cite{1985ApJ...288..618R}. The sources IRS-1 and EGO-1 are represented with asterisks.} 
\label{jhk_cc}
\end{figure}

We investigate the nature of the stellar population associated with the cluster using JHK colour-colour diagram 
(CCD) shown in Fig.~\ref{jhk_cc}. For this we have used a region of radius 1.5$\arcmin$ centered on the brightest IR source 
(IRS-1) that lies within the positional uncertainty of the IRAS point source position (see Fig.~\ref{col_com}b). The CCD shows two distinct 
population of objects. The low extinction group ($\rm A_V=0$ -- $5$) is possibly the contamination from field stars and 
the high extinction population (with an average $\rm A_V \sim 15$ and going upto $\rm A_V \sim 30$) are likely to be the cluster members associated with the cluster identified here. As can be seen from the plot, majority of the candidate cluster members are distributed to the right of the reddening 
band of the main sequence stars. This region of the diagram is identified with near-infrared sources (Class II, Class I
and Herbig Ae/Be stars). The near-infrared excess in pre-main sequence Class I and II stars is due to the optically thick circumstellar
disks/envelopes. These disks/envelopes become optically thin with age; hence the fraction of near-infrared excess stars decreases with 
age. For very young ($\rm \leq 1 Myr$) embedded clusters the fraction is $\rm \sim 50 \% $ \citep{{2000AJ....120.3162L}, {2000AAS...197.2402H}} and it decreases to $\rm \sim 20 \%$ for more evolved (2 -- 3 Myr) clusters \citep{{2001ApJ...553L.153H}, {2004A&A...413L...1T}}. Taking the cluster population into consideration, we estimate around 40\% of stars with near-infrared excess. This indicates that
the cluster is a very young and partially embedded cluster, and as the molecular cloud disperses, the cluster members as well 
as the shape of the cluster would be revealed. 

\begin{figure}
\includegraphics[scale= 0.4]{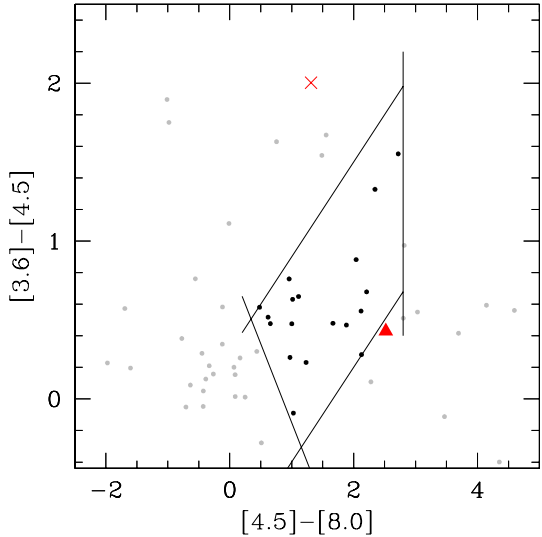}
\caption{IRAC colour-colour diagram of the 58 sources within lowest contour level of 350~$\mu$m. The sources which are not classified as YSOs are represented by gray points. Black dots represent the IRAC YSOs identified using the criteria of \citet{2007ApJ...669..327S}. The triangle and $\times$ represent IRS-1 and EGO-1, respectively. The criteria for YSO identification are shown as solid lines.}
\label{irac_cc}
\end{figure}

\subsubsection{IRAC YSOs} 
To identify the young stellar objects in the vicinity of IRAS~17256--3631, we have used the IRAC photometric data from
the catalog generated (refer Section 2.3.1). We selected point sources within the lowest contour of 350~$\mu$m image (830~MJy/Sr), that 
encompasses most of the diffuse radio, near-infrared, MIR and submm emission. There are 58 sources detected in the three IRAC 
bands of 3.6, 4.5 and 8.0~$\mu$m. Fig.~\ref{irac_cc} shows the ([3.5] - [4.5]) vs  ([4.5] - [8.0]) IRAC CCD. \citet{2007ApJ...669..327S} used the following color cuts to isolate YSOs in a given stellar population based on the SED fits.

\begin{flalign*}
&[3.6]-[4.5] > 0.6 \times ([4.5]-[8.0])-1.0, &\\
&[4.5]-[8.0] < 2.8, &\\
&[3.6]-[4.5] < 0.6 \times ([4.5]-[8.0])+0.3, &\\
&[3.6]-[4.5] > -([4.5]-[8.0])+0.85. &\\
\end{flalign*}

Based on the above criteria, 18 sources were identified as YSOs in our region of interest. These are shown as solid circles in  Fig.~\ref{irac_cc}. The locations of these sources associated with IRAS~17256--3631 are shown in Fig.~\ref{8m_870}a. 
The spatial distribution of infrared objects in different evolutionary stages do not show a morphological segregation. This is because of some foreground and large background contamination due to the location of this region, i.e., towards the Galactic centre. 
 
\par
\begin{figure}
\hspace*{-1.0cm}
\includegraphics[scale=0.30]{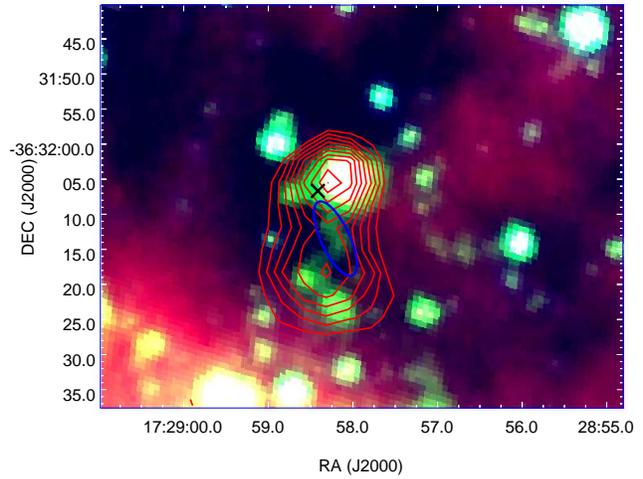}
\caption{IRAC three-color composite image of EGO-1. Blue represents 3.6~$\mu$m, green 4.5~$\mu$m and red 8~$\mu$m. Ellipse marks the extended green emission. The Herschel PACS 160~$\mu$m contours are also shown in the image. The $\times$ locates the position of H$_2$ knot 2 (Fig.~\ref{brg_H2}b).}
\label{ego}
\end{figure}

 We see a point-like source near the clump C4 at 8~$\mu$m and 70~$\mu$m. We looked for this in the all-sky survey from Widefield Infrared Survey Experiment \cite[WISE,][]{wright2010wide} and found that it is present in the 12 and 22~$\mu$m images as well. It is therefore likely to be in an early evolutionary stage. Although this source (represented by an asterisk in the IRAC colour-colour diagram in Fig.~\ref{irac_cc}) does not lie in the region occupied by YSOs, a careful examination of IRAC images reveals that this is likely to be due to the enhanced emission in the 4.5~$\mu$m band, that is shown in Fig.~\ref{ego}. In fact, two green fuzzies associated with this extended green object \cite[EGO,][]{{cyganowski2008catalog}, {2009ApJS..181..360C}} are seen towards the south of this source that are not detected in either 3.6, 5.8 (GLIMPSE) or 8~$\mu$m bands. The locations of these green fuzzies is encircled in Fig.~\ref{ego}. EGOs are characterized by their extended green emission in the IRAC three-color composite images. Their angular sizes vary from a few to 30$\arcsec$ and are believed to be associated with massive young stellar objects \citep{2009ApJ...702.1615C}. We refer to this source as EGO-1. 

\begin{figure}
\includegraphics[scale=0.45]{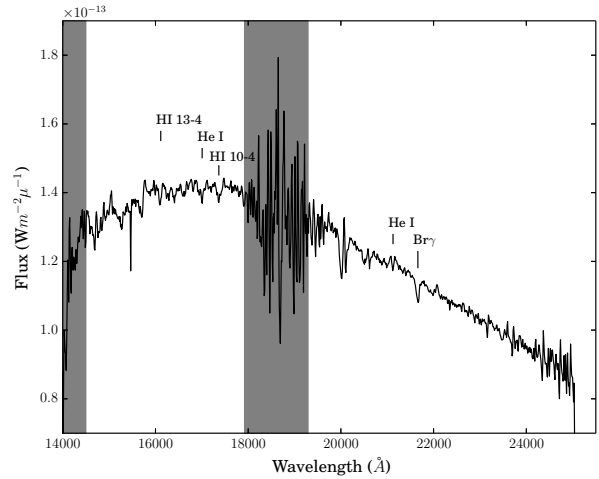}
\caption{The near-infrared HK spectrum of the IRS-1. The shaded area represent the region of poor sky transparency. The near-infrared spectral lines identified are marked over the spectrum.}
\label{spectrum}
\end{figure}

%%%%%%%%%%%%%%  Fig 14 %%%%%%%%%%%%%%%%%%%%%%%%%%
\begin{figure*}
\hspace*{-1.8cm}
\begin{minipage}{0.4 \textwidth}
\includegraphics[scale=0.4]{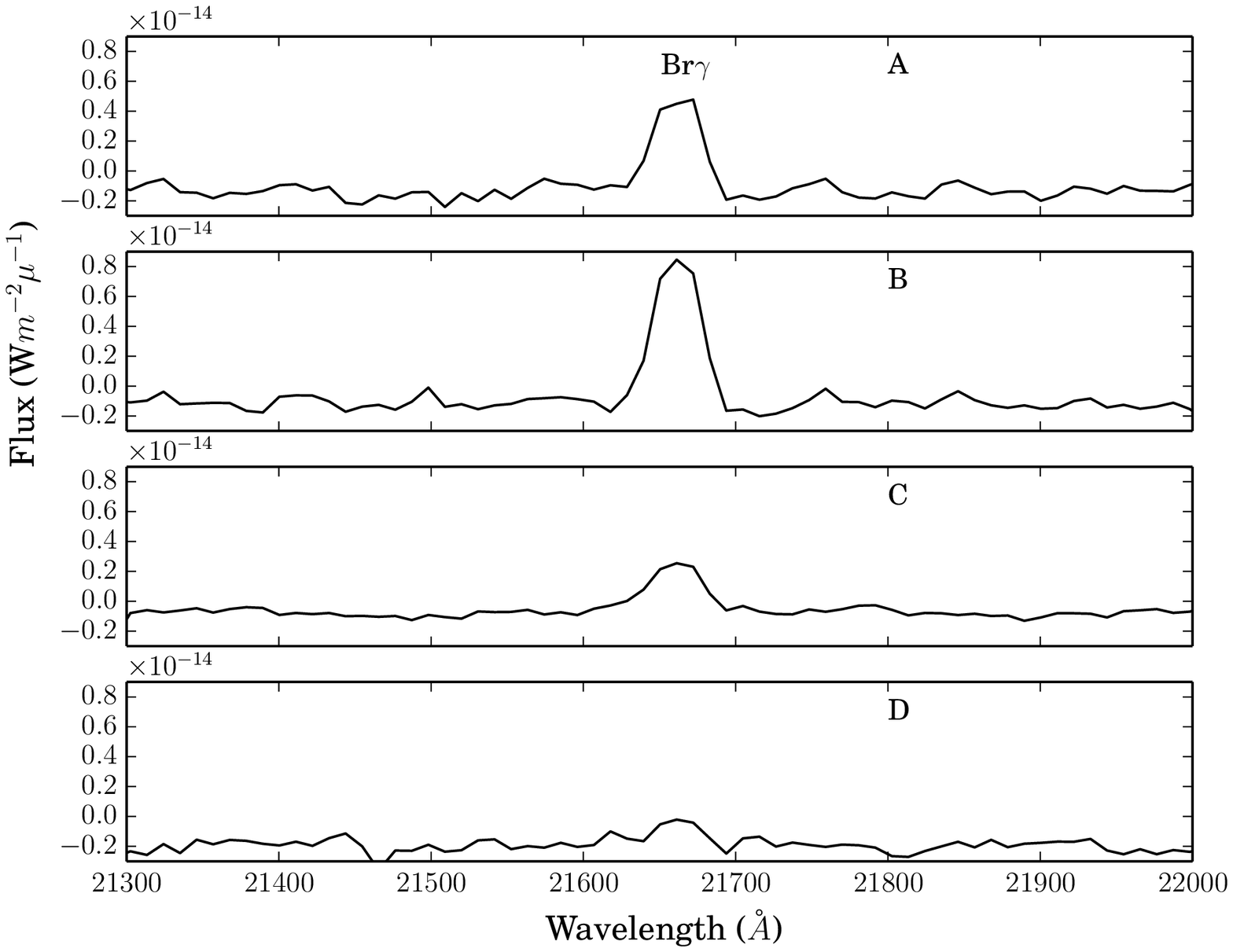}
\label{}
\end{minipage}
\hspace{1.5cm}
\begin{minipage}{0.42 \textwidth}
\includegraphics[scale=0.44]{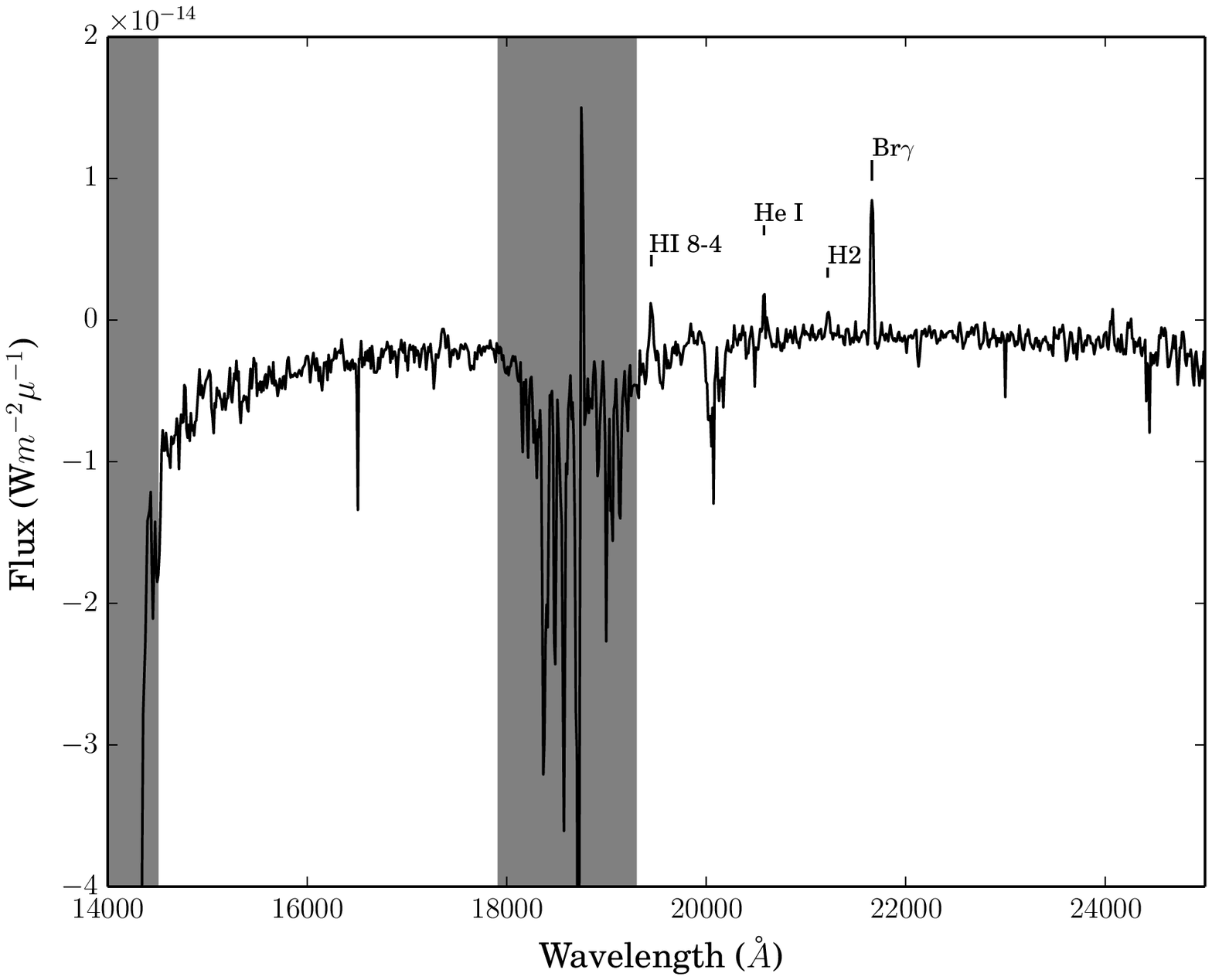}
\label{}
\end{minipage}
\caption{(Left) Br$\gamma$ line across different slit positions labeled as A, B, C and D in Fig.\ref{brg_H2}b. (Right) The near-infrared HK spectrum of nebular region in the vicinity of IRS-1. The shaded region represents region of poor sky transparency.}
\label{neb}
\end{figure*}
%%%%%%%%%%%%%%%%%%%%%%%%%%%%%%%%%%%%%%%%%%%%%

\begin{table}
\footnotesize
\caption{The lines detected in the spectrum of IRS1.}
\begin{center}
%\vskip 0.4cm
%\label{spec}
\hspace*{-0.5cm}
\begin{tabular}{c c c} \hline \hline
Line &Wavelength ($\mu$m) &Comment \\
\hline
Br$\gamma$ &2.165 & IRS-1 (absorption), Nebular (emission) \\
HI (10-4)&1.736&IRS-1 (absorption) \\
HI (13-4)&1.611& IRS-1 (absorption)\\
He I&1.736 & IRS-1 (absorption)\\
He I&1.700 & IRS-1 (absorption) \\
HI (8-4)&1.945& Nebular (emission)\\
He I&2.058 & Nebular (emission)\\
H$_2$ &2.121 & Nebular (emission)\\
\hline
\end{tabular}
\label{spec_lines}
\end{center}
\end{table}

\subsection{Near-infrared Spectroscopy}

\par IRS-1 is the brightest near-infrared source in our region marked in the near-infrared CCD and the location encircled in Fig.~\ref{col_com}a. It is located $\sim\rm{25\arcsec}$ away from the radio peak. As described 
in Section 2.2.1, we carried out spectroscopic observations of this source in the wavelength range of 1.4 -- 2.5 $\mu$m.
The extracted spectrum of IRS-1 is shown in Fig.~\ref{spectrum}. The shaded vertical strips represent the spectral region 
with poor sky transparency. Several hydrogen and helium lines characteristic of massive O and B stars 
\citep{2005A&A...440..121B} are
detected in the spectra of IRS-1. The lines detected above $\rm 3\sigma$ level are listed in Table~\ref{spec_lines}. The spectral 
features seen in IRS-1 are consistent with that of late O type (late than O7.5) and early B type stars as described in the 
K-band spectral atlas of \citet{1996ApJS..107..281H}. 

\par In order to probe the nebulosity associated with IRAS 17256-3631, we oriented the slit along the direction shown
in Fig.~\ref{brg_H2}b. Spectra at positions A, B, C, and D were extracted and presented in the left panel of Fig.~\ref{neb}.
There is a clear variation of the strength of the Br$\gamma$ emission along the slit. It initially increases; from A to B as we move 
southwards, with maximum emission seen in region B that is close to IRS-1. Beyond B it decreases considerably, with very little 
or no emission in region D. The right panel of Fig.~\ref{neb} shows the spectrum of region B. The spectral features are
consistent with that from ultracompact \hii~regions \citep{{2005A&A...440..121B}, {1997ApJ...489..698H}}. The spectrum from region B
shows the presence of weak H$_2$ line at 2.12~$\mu$m. This line can be attributed to two excitation mechanisms, namely shock and UV fluorescence. The signatures of shock excited H$_2$ emission is the absence of transitions from high-$\nu$ levels and a high ratio of fluxes (10:1) in the 2.12 and 2.24~$\mu$m lines \citep{1995A&A...296..789S}. The UV fluorescence, on the other hand, leads to population of both high and low-$\nu$ states. The diagnostic tool for UV fluorescence is a ratio of 2:1 for the 2.12 and 2.24 lines \citep{1976ApJ...203..132B}. The 2.24~$\mu$m line is not detected down to 1.5$\sigma$ level. This appears to favour the shock excitation mechanism although we need better signal-to-noise ratio to ascertain this. It is worth noting that the H$_2$ image shows the presence of a faint H$_2$ nebulosity at B consistent with the location of region B (Fig.~\ref{brg_H2}b).

\begin{figure*}
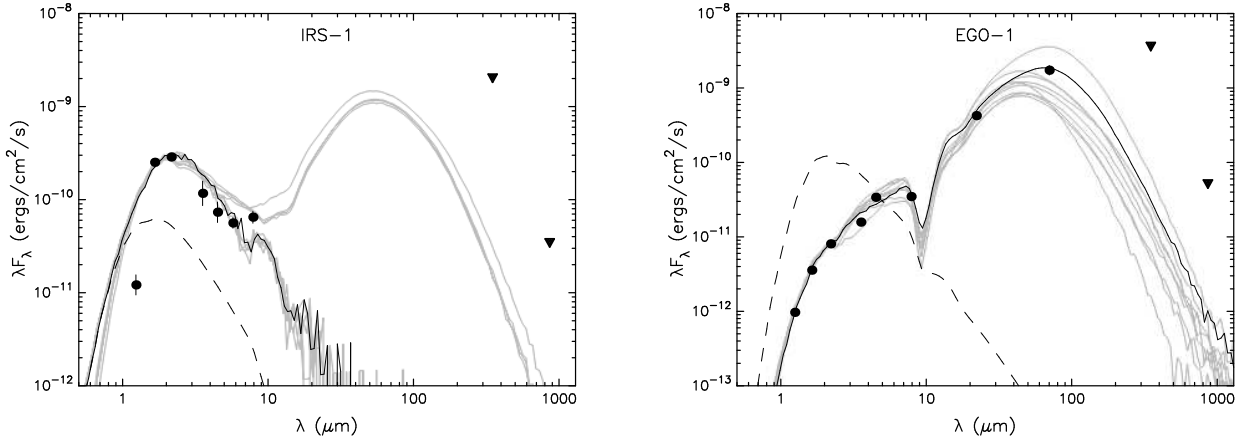

\begin{minipage}{0.49 \textwidth}
\includegraphics[scale=0.65]{Fig15a.eps}
\label{}
\end{minipage}
\begin{minipage}{0.49 \textwidth}
\includegraphics[scale=0.65]{Fig15b.eps} 
\label{}
\end{minipage}
\caption{RWIW best fit models for (left) IRS-1 and (right) EGO-1. The filled circles represent the input fluxes while triangles represent upper limits. The black line shows the best fitting model, and grey lines show the subsequent 9 good fits. The dashed line marks the SED of the stellar photosphere in the best fitting model.}
\label{rwiw}
\end{figure*}

\subsection{SED Models of IRS-1 and EGO-1}
\par In order to get a qualitative idea about the evolutionary stage of IRS-1 and EGO-1, we have fitted their spectral energy distribution (SED) with the models of \citet{2007ApJS..169..328R} (hereafter RWIW). The assumptions include a given gas and circumstellar dust 
geometry as well as a dust grain 
model to carry out radiative transfer modelling. RWIW have pre-computed a large number of radiative transfer models covering a 
wide range of stellar masses and evolutionary stages. We have used the online version of the fitting tool. The SEDs are 
constructed using near and mid-infrared wavelength fluxes shown in Table~\ref{yso_mag}. The associated clump flux densities at 350 and 870~$\mu$m  were given 
as the upper limits for constraining the models. The inputs to the models include distance, visual extinction and source 
flux densities at various wavebands. The SED and the 10 best fitting models for each are shown in Fig.~\ref{rwiw}. Table~\ref{rwiwyso_tb} lists the range of parameters for these models. From the best fit model, the mass of IRS-1 is found to be 14.6~M$_{\odot}$ and 10 best fit models give a mass range of 6.7-- 14.6~M$_{\odot}$ suggesting IRS-1 as a massive star in this region. This is consistent with the spectral type inferred from the near-infrared spectra. The mass of EGO-1 is found to be 6.8~M$_{\odot}$ from the best fit model. The mass ranges from 3.5--6.8~M$_{\odot}$. Although we have attempted to estimate the physical parameters using the RWIW SED fits, we would like to mention that the fitting is based on the assumption that the SEDs of massive YSOs are scaled up versions of their lower mass counterparts. 
\begin{table*}
\footnotesize
\caption{Infrared magnitudes of IRS-1 and EGO-1.}
\begin{center}
%\vskip 0.4cm
%\label{spec}
\hspace*{-0.5cm}
\begin{tabular}{c c c c c c c c c} \hline \hline
Source & $\rm{\alpha_{J2000}}$ & $\rm{\delta_{J2000}}$ & J & H & K & 3.6~$\mu$m &
4.5~$\mu$m &~8.0$\mu$m \\
& ($^{h~m~s}$) & ($^{\degr~\arcmin~\arcsec}$)&(mag) &(mag) &(mag) &(mag) &(mag) &(mag) \\
\hline
IRS-1&17:29:01.39&-36:33:54.21&11.20 $\pm$ 0.01 &9.79 $\pm$ 0.03 &8.86 $\pm$ 0.02 &8.22 $\pm$ 0.01 &7.79 $\pm$ 0.01 &5.27 $\pm$ 0.03 \\
EGO-1&17:28:58.13 &-36:32:05.10 &16.45 $\pm$ 0.01&14.29 $\pm$ 0.01&12.57 $\pm$ 0.01 &10.32 $\pm$ 0.01 &8.32$\pm$ 0.01&7.01 $\pm$ 0.02\\
\hline
\end{tabular}
\label{yso_mag}
\end{center}
\end{table*}

\begin{table*}
\caption{Parameters of the Robitaille et al. (2007) models for IRS-1 and EGO-1. Col. 4 -- 9 gives  Mass, Effective Temperature, Luminosity, Inclination angle, Envelope accretion rate, Envelope mass, Disk Mass, Extinction and Age, respectively. The range of parameters given here are for best fit model and subsequent 9 good fits.}
\scriptsize
\begin{center}
\vskip 0.4cm
\label{spec}
\hspace*{-0.2cm}
\setlength{\tabcolsep}{4pt}
\begin{tabular}{c c c c c c c c c c c c} \hline \hline \\

Source &&$\chi^2$ &Mass  &T$_{eff}$  &Luminosity  &Inc. angle & Env. accretion rate     & Env. mass  &Disk mass  & A$_V$   & Age\\
 & &&(M$_{\odot}$) &(K) &(L$_{\odot}$) & (Deg.) &(M$_{\odot}$/yr) &(M$_{\odot}$) &(M$_{\odot}$) & &(Myr)\\
\hline \\
\multirow{ 2}{*}{IRS-1} & Best fit &103.5 &14.60 & 30955 & 19200 & 31.79 & 0 & 1.39$\times$10$^{-8}$ & 9.04$\times$10$^{-9}$ & 10.0 &1.3 \\
&Range&103.5 - 115.9 &6.66 - 14.60 & 4666 - 30935 & 272 -19200 &18.19 - 75.52 & 0 - 6.03 $\times\rm10^{-5}$ &1.39$\times\rm10^{-8}$ - 22.8& 9.04$\times$10$^{-9}$ - 0.001& 10.0& 0.07 -1.3 \\ \hline

\multirow{ 2}{*}{EGO-1} & Best fit &71.4 & 6.83 & 4530 & 302 & 31.79 &7.48 $\times$ 10$^{-4}$  & 9.19 & 0.35 & 14.09 &0.05 \\ 
&Range &71.4 - 206.5& 6.83 - 3.49 &4223 - 6658 &122 - 497 &117.2 - 286.6 & 2.05 $\times$ $\rm10^{-5}$ - 7.48 $\times$ 10$^{-4}$ &0.3 - 22.6 &0.003 - 0.2 &2.0 - 13.2 &0.005 - 0.2\\
\hline \hline \\
\end{tabular}
\label{rwiwyso_tb}
\end{center}
\end{table*}

\section{Cometary Nature of the \hii~Region}

\subsection{Ionized Gas Distribution}

IRAS~17256--3631 is a cometary \hii~region as seen from the radio continuum images. This morphology is observed in 20\% of all the known \hii~regions \citep{1989ApJS...69..831W}. A cometary \hii~region  comprises of a bright arc-like head and a broad tail of emission. The high resolution 1372 MHz image ($\sim\rm{6\arcsec}$) shows clumpiness in the ionised gas emission. The lower resolution 610 MHz image ($\sim\rm{10\arcsec}$), on the other hand, displays a comparatively smooth cometary structure. The 325 MHz image reveals the presence of large scale diffuse emission at lower flux densities spanning a region of $\sim$7$\arcmin$. The cometary structure is evident in all the frequency bands. The clumpiness of ionized gas can be explained as a result of the local density enhancements in the molecular cloud \citep{2001ApJ...549..979K}. If a single ZAMS star is ionizing the entire region, the spectral type of the ionizing source is O7--O7.5 or earlier. This is consistent with the luminosity of $1.6\times10^5$~L$_\odot$ of this region that corresponds to a single ZAMS O6.5 star (discussed in the next section). Although the radio emission peaks at a location that is $\sim25''$ away from IRS-1, the radio ZAMS spectral type (O7.5) is consistent with that of IRS-1 estimated from near-infrared spectroscopy (O7.5 or later). 
 
Hence, it is likely that IRS-1 is the possible ionizing source of this region. 
\par In the near-infrared, the ionised gas has been probed using Br$\gamma$ line that shows high extinction filamentary features towards the west side of the region. The Br$\gamma$ emission peak coincides with the radio peak. 
The cometary nature is also evident from the radio spectral index maps, where the spectral indices are found to be negative towards the head, indicative of shock-excitation. The spectral index-maps matches well with the H$_2$ emission map as the excited molecular hydrogen is found to lie towards region of negative spectral index (non-thermal) emission. However, not all regions of negative spectral indices show H$_2$ emission, which could be attributed to extinction in the near-infrared. The mechanism for the production of shock excitation across the cometary head is discussed below.

\subsection{Cometary Models}

\par In this subsection, our aim is to understand the origin of the cometary morphology of the \hii~region under investigation here. Cometary \hii~regions vary widely in their appearance and several models have been proposed to explain their shape.  The two prominent ones are (i) the bow-shock model \citep{{1985ApJ...288L..17R}, {1990ApJ...353..570V}, {1991ApJ...369..395M}}, and (ii) champagne flow model \citep{{1978A&A....70..769I},  {1979A&A....71...59T}}. 

In the bow shock model, it has been proposed that cometary regions are the result of relative motion of the star against ambient molecular gas. Here, the \hii~regions are modeled as bow-shocks created by wind blowing massive stars moving supersonically through the molecular clouds. The champagne flow model, on the other hand, explains the cometary shape as a result of density gradient in the molecular cloud, where the ionised gas expands asymmetrically out of a dense clump towards a regions of minimum density. Recently, the effect of strong stellar winds have also been included in these models \citep{{1994ApJ...432..648G}, {2006ApJS..165..283A}}. 

\par From the contours in our radio maps, we see that there is a steep density gradient towards north-west while it is shallow and spread out towards south-east. This implies that the \hii~region is density bounded in north-west and ionization bounded towards the south-east. In order to understand the mechanism responsible for the observed morphology of the \hii~region, we first consider the bow-shock model. We use simple analytic expressions to calculate few shock parameters. According to this model, the stellar wind from a star travelling supersonically through the molecular cloud flows isotropically from the star in all directions, until it encounters a terminal shock. For a star moving in the plane of sky, the bow-shock is expected to trace a parabola. The shock occurs at a stand-off distance $\it{l}$ from the star where the momentum flux in the wind equals the ram pressure of the ambient medium. The stand-off distance $l$ is estimated using the expression below \citep{1990ApJ...353..570V}.

\begin{equation}
\begin{split}
\left[\frac{l}{\rm{cm}}\right]=5.50\times10^{16} \left[\frac{\dot{m}_*}{\rm{10^{-6}M_{\odot}yr^{-1}}}\right]^{1/2}\left[\frac{v_{w}}{\rm{10^8cms^{-1}}}\right]^{1/2}\mu^{-1/2}_H \\
\left[\frac{n_H}{\rm{10^5cm^{-3}}}\right]^{-1/2}\left[\frac{v_*}{\rm{10^6cms^{-1}}}\right]^{-1}
\end{split} 
\end{equation}
\\

where $\dot{m}_{*}$ is the stellar wind mass loss rate, $v_{w}$ is the wind's terminal velocity, $n_{H}$ is the number density of hydrogen nuclei in all forms of the ambient gas and $v_{*}$ is the relative velocity of the star through the medium. The mean mass per particle in the molecular gas is $\it{\mu_H}$=1.4m$_H$ where m$_H$ is taken as one atomic mass unit. The stellar wind mass loss rate ($\dot{m}_{*}$) and wind terminal velocity ($v_{w}$) are calculated using expressions given below \citep{1991ApJ...369..395M}.

\begin{equation}
\left[\frac{\dot{m}_*}{\rm{10^{-6}M_{\odot}yr^{-1}}}\right]=2\times10^{-7}\left[\frac{L}{\rm{L_{\odot}}}\right]^{1.25}
\end{equation}

\begin{equation}
\rm{log}\left[\frac{v_w}{\rm{10^8cms^{-1}}}\right]=-38.2+16.23\ \rm{log}\left[\frac{T_{eff}}{\rm{K}}\right]-1.70\ \left(\rm{log}\left[\frac{T_{eff}}{\rm{K}}\right]\right)^2 
\end{equation} 
  \\
Considering an O7.5 type star, we have assumed a luminosity $L=8.3\times10^4$~$L_\odot$ and effective temperature $T_{eff}=37,500$~K, \citep{1973AJ.....78..929P}. With this, we get $\dot{m}_{*}=0.28\times10^{-6}$~M$_{\odot}$yr$^{-1}$ and $v_{w}=3.0\times10^8$~cm$s^{-1}$. Using these values and taking the number density of hydrogen nuclei to be 1.4$\times$10$^{5}$ cm$^{-3}$ \citep{2005A&A...432..921F} and typical velocity of the star ($v_{*}$) to be 10~km/s \citep{1992ApJ...394..534V}, the stand-off distance is 3.5$\times$10$^{16}$~cm that corresponds to 0.01~pc. The stand-off distance increases by an order of magnitude to 0.12~pc if the velocity of the star is 1~km/s. On the other hand, decreasing the ambient density by an order of magnitude (1.4$\times$10$^4$ cm$^{-2}$) within which the star is moving with a velocity 10~km/s changes the stand-off distance to 0.05~pc. Taking stand-off distance as the distance between the steep density gradient at the cometary head and the radio peak (i.e location of embedded exciting star), we estimate $l$ for IRAS~17256--3631 from our radio images as $\sim0.2$~pc. If we assume IRS-1 to be the ionizing source, the stand-off distance is $\sim$0.44~pc. If we consider a star moving with a moderate velocity of 1~km/s, we get the theoretical stand-off distance as 0.38~pc, close to the observed value only if we consider an ambient medium of density that is lower by an order of magnitude. We note that the viewing angle could play a role in our estimation of $l$. An inclination of $45^\circ$ would lead to a decrease in the theoretical stand-off distance by $\sim70$\%. However, this would enhance the disparity between the theoretical and observed values. 

\par Another parameter that we have estimated in the bow-shock model is the trapping parameter ($\tau_{bow}$) for the ionization front. The supersonically moving star sweeps up dense shells of gas that traps the \hii~region within them and consequently, the ram pressure prevents the \hii~region from expanding dynamically. The shell traps the \hii~region when there are more recombinations in the shell than ionizing photons. The trapping parameter is expected to be much larger than 1 ($\tau_{bow}\gg$~1 ) as computed by \citet{1991ApJ...369..395M} for a number of cometary \textbf{\hii~regions}. They find $\tau_{bow}\sim15-900$ for 4 ultracompact \hii~regions. Here, we calculate the trapping parameter and compare with values obtained for other cometary \hii~regions explained using the bow-shock model. The trapping parameter at the stagnation point is given by the following expression \citep{1990ApJ...353..570V} 

\begin{equation}
\begin{aligned}
\tau_{bow}=0.282\left[\frac{\dot{m}_*}{\rm{10^{-6}M_{\odot}yr^{-1}}}\right]^{3/2}\left[\frac{v_w}{\rm{10^8cms^{-1}}}\right]^{3/2}\left[\frac{v_*}{\rm{10^6cms^{-1}}}\right]^{-1}\\
 \left[\frac{n_H}{\rm{10^5cm^{-3}}}\right]^{1/2}\mu_{H}^{-1/2}\left[\frac{T_e}{\rm{10^4K}}\right]^{-1}\\
\left[\frac{N_{Ly}}{\rm{10^{49}s^{-1}}}\right]^{-1}
\gamma^{-1}\left[\frac{\alpha}{\rm{10^{-13}cm^3s^{-1}}}\right]
\end{aligned}
\end{equation} 
\\

Here $T_{e}$ is the electron temperature of the \hii~region assumed to be 7500~K, and $N_{Ly}$ is the ionizing flux of the star taken as 3.12$\times$10$^{48}$ s$^{-1}$, calculated from the 1372~ MHz data. The recombination coefficient for hydrogen atoms is denoted by $\alpha$ and we assume a value of $3.57\times10^{-13}$~cm$^3$~s$^{-1}$ at 7500~K. This value is estimated by linear interpolation of the optically thick Case B from \citet{osterbrock1989astrophysics}. $\gamma$ is the ratio of specific heats and taking this as 5/3, we obtain the value of trapping parameter as $\mathbf{\tau_{bow}=1.8}$. This implies that even if there is a bow shock, it is quite weak and the \hii~region appearance will be amorphous corresponding to ionized quiescent cloud material surrounding the completely ionized bow shock \citep{1990ApJ...353..570V}. However, in our case we see steep density gradient near the cometary head of the \hii~region.

\par From the considerations above, we see that it is unlikely that the cometary appearance of IRAS~17256--3631 can be explained using the bow shock model alone. Further, we note that although the bow-shock model is generally invoked to explain the cometary morphology of \hii~regions, the origin of the supersonic nature of stellar velocities in young embedded clusters has been under speculation. In fact, bow-shocks from run-away O stars in the interstellar medium have been observed \citep{{1997ApJ...475L..37K}, {noriega1997bow}, {comeron2007very}}. However, in these cases the velocities are high, of the order of 10-100 km/s as they move in diffuse interstellar medium \citep{{2001A&A...378..907R}, {2005ApJ...634L.181E}, {gvaramadze2011search}}. In dense molecular clouds, the adiabatic sound speed is $\sim 0.3$ km/s. Spectroscopic observations have indicated that dispersion velocities of stars in young Galactic clusters that are embedded or just emerged from their natal molecular cloud are typically in the range 0.5-3 km/s \citep{{schmeja2008structures}, {2012MNRAS.421.3206S}}. Although these are close to or larger than the sound speed, the velocities of stars moving at supersonic speeds producing bow shocks are found to be much higher (5--20 km/s) in the models. It is required that the supersonic motion of the star be in the plane of sky or close to it for the \hii~region to appear cometary. Therefore, proper motion measurements would be required to determine the stellar velocity, and that is difficult considering that the star is embedded and the distance is large. Another method to confirm the supersonic velocity of star and formation of bow-shock, involves measurement of the relative velocity of ionized gas with respect to the molecular cloud.

\begin{figure}
\hspace*{0.7cm}
\includegraphics[scale=0.25]{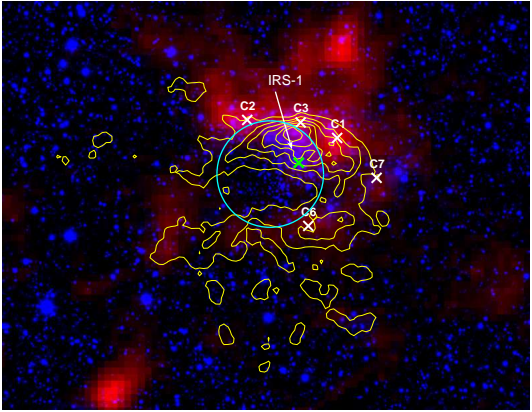}
\caption{Color composite image of the IRAS 17256--3631 region. The size of the field is 6.2$\arcmin$ $\times$ 8.1$\arcmin$. The emission of cold dust appears in red (870$~\mu$m) and near-infrared K-band image (2.2~$\mu$m) is in blue. The yellow contours correspond to radio continuum emission at 1372~MHz. The star cluster is encircled. IRS-1 is highlighted as a green $\times$. The positions of six active/evolved clumps are also shown.} 
\label{clusdis}
\end{figure}

\par An alternative model that explains the cometary morphology is the champagne flow model. A massive star at the center of a molecular clump but close to the edge of the molecular cloud will produce an ionisation front that preferentially expands towards the low density regions \citep{2001ApJ...549..979K}. Here, the density gradients along with hierarchical structure of molecular clouds decide the morphology of the \hii~region. In our case, we propose that the champagne flow plays a prominent role in the morphology of the \hii~region based on the following reasons. (i) If IRS-1 is the driving source, then the high density clumps C1, C2 and C3 are likely to be constraining the flow of ionised gas. (ii) The ionised cometary head matches with the location of three clumps, C1, C2 and C3 (Fig.~\ref{clusdis}). (iii) The radio emission has more than one emission peak that is expected due to high density clumps (a bright peak at north and another at north-west). (iv) The near-infrared H$_2$ emission detected in this region lacks the expected bow shock type morphology and is seen as diffuse filamentary structures (Fig.~\ref{brg_H2}(b)) above the cometary head. This can be attributed to the density gradient in the dense ISM beyond the cometary head.

We propose the following explanation for the observed morphology based on the champagne flow model. If we assume that IRS-1 is the massive star that is the ionising source, then the radio emission shows a steep density gradient towards the north-west of IRS-1.  However, one cannot exclude the possibility of other massive stars also ionising the medium. The star(s) ionize the ambient molecular gas and the expanding ionization front encounters the high density clumps C1, C2, C3 located to the north-west and  distributed in an arc-shaped morphology near the head.  Regions of lower density are towards south-east of the exciting source and a champagne flow develops where the ionized gas escapes through this region producing extended emission towards south-east. The displacement of radio peak with respect to the ionising source is likely to be due to density enhancements or clumpy nature of ambient medium. In fact,  radio emission peak not coinciding with the ionising source is listed as one of the characteristics for champagne flow in \hii~regions, although there could be exceptions because of projection effects \citep{1983A&A...127..313Y}.  The amount of dense molecular gas is less in the southern direction as evident from the cold dust emission map. Another interesting feature is the asymmetry of the cometary tail that is clearly seen in the 325~MHz map. The diffuse emission extends towards the north-east and south-west where there is low flux density due to cold dust. This augurs well for the champagne flow model. We, therefore, strongly believe that the cometary appearance of IRAS~17256--3631 can be explained due to the density gradient as predicted by champagne flow models.

\section{Star Formation Activity}

\par The spatial distribution of dust and ionized material around the infrared cluster is illustrated in Fig.~\ref{clusdis}. This figure shows a colour composite image of the embedded cluster in near infrared K band (blue) and cold dust emission at 870 $\mu$m (red), overlaid with contours of ionised gas emission (yellow; at 1372 MHz). The cold dust emission is absent towards the location of the cluster giving an appearance of a void or cavity here (encircled in the figure). It is likely that the parental molecular cloud has dispersed locally, after the formation of the stellar cluster. Even in the radio emission, we observe low flux levels towards the cluster. IRS-1 is located to the north-west of the infrared cluster and appears within the aperture corresponding to clump C3 although it is nearly $35''$ south of dust emission peak. Most of the high density clumps are also situated north and north-west of the cluster. The void in radio emission near the cluster could be due to stellar winds emanating from stars in the cluster \citep{{smith1973observations},{2005A&A...436..155C}}. Numerical radiaton-hydrodynamical simulations of cometary \hii~regions by \citet{2006ApJS..165..283A} show that champagne flow models with powerful stellar winds from early type stars can produce limb brightened morphologies. Wind blown broken shell/void kind of structures in radio and mid-infrared around clusters of stars have also been observed in other regions, eg. RCW49  \citep{{1997A&A...317..563W},{2013A&A...559A..31B}}.

\par We next attempt to find the relative evolutionary stages of clumps based on the evolutionary sequence proposed by 
\citet{2009ApJS..181..360C} and \citet{2010ApJ...721..222B}. Both the models are similar with the latter being a subtle version of the former. \citet{2009ApJS..181..360C} proposed 
an evolutionary sequence for Infrared Dark Clouds (IRDCs), where the sequence begins with a quiescent clump which evolves later into an active clump (clumps containing an enhanced 4.5~$\mu$m emission and a 24 $\mu$m point source) and finally becomes a red (enhanced 8 $\mu$m emission) clump. \citet{2010ApJ...721..222B} further modified this classification by including radio emission and suggesting that the red clumps are diffuse ones without associated millimeter peaks. They discuss  four important star formation tracers: (1) Quiescent clump (no signs of active star formation), (2) Intermediate clumps that exhibit one or two signs of active star formation (such as shock/outflow signatures or 24 $\mu$m point source), (3) Active clumps that exhibit three or four signs of active star formation (\textquotedouble{green fuzzies}, 24 $\mu$m point source, UC\hii~region or maser emission), and (4) Evolved red clumps with diffuse 8 $\mu$m emission. 

\begin{table}
\footnotesize
\caption{Clump activity and classification.}
\begin{center}

\hspace*{-0.5cm}
\begin{tabular}{c c c c c c} \hline \hline
Clump &Radio &IRAC &WISE &Clump  & Evolutionary  \\
No.&peak& 8~$\mu$m peak&22~$\mu$m source&Activity$^a$&Stage$^b$ \\
\hline
C1 &\checkmark &\checkmark &\text{\sffamily X} &A/E &Type 2\\
C2 &\checkmark &\checkmark &\text{\sffamily X} &A/E &Type 2\\
C3 &\checkmark &\checkmark &\text{\sffamily X} &A/E &Type 2\\
C4 &\text{\sffamily X} &\text{\sffamily X} &\checkmark &I &Type 2\\
C5 &\text{\sffamily X} &\text{\sffamily X} &\text{\sffamily X} &Q &Type 1\\
C6 &\checkmark &\text{\sffamily X} &\text{\sffamily X} &A/E &Type 2\\
C7 &\text{\sffamily X} &\checkmark &\text{\sffamily X} &A/E &Type 2\\
C8 &\text{\sffamily X} &\text{\sffamily X} &\text{\sffamily X} &Q &Type 1\\
C9 &\text{\sffamily X} &\text{\sffamily X} &\text{\sffamily X} &Q &Type 1\\
C10 &\text{\sffamily X} &\text{\sffamily X} &\text{\sffamily X} &Q &Type 1\\
C11 &\text{\sffamily X} &\text{\sffamily X} &\text{\sffamily X} &Q &Type 1\\
C12 &\text{\sffamily X} &\text{\sffamily X} &\text{\sffamily X} &Q &Type 1\\
C13 &\text{\sffamily X} &\text{\sffamily X} &\text{\sffamily X} &Q &Type 1\\
C14 &\text{\sffamily X} &\text{\sffamily X} &\text{\sffamily X} &Q &Type 1\\
C15 &\text{\sffamily X} &\text{\sffamily X} &\text{\sffamily X} &Q &Type 1\\
C16 &\text{\sffamily X} &\text{\sffamily X} &\text{\sffamily X} &Q &Type 1\\
C17 &\text{\sffamily X} &\text{\sffamily X} &\text{\sffamily X} &Q &Type 1\\
C18 &\text{\sffamily X} &\text{\sffamily X} &\text{\sffamily X} &Q &Type 1\\
\hline
\multicolumn{6}{l}{\textsuperscript{a}\footnotesize{Q-quiescent, I-intermediate, A-active, E-evolved based on \citet{2010ApJ...721..222B}}}\\
\multicolumn{6}{l}{\textsuperscript{b}\footnotesize{Classification scheme adopted by \citet{2013A&A...550A..21S}}}

\end{tabular}
\label{clumpact}
\end{center}
\end{table}

 \par In the absence of Spitzer-MIPS 24~$\mu$m data, we searched for WISE 22 $\mu$m point sources associated with the clumps within a search radius of 5$\arcsec$ from the peak position (shown in Fig~\ref{350_aper}) and based on association with radio and mid-infrared emission, we have classified the clumps based on the scheme of \citet{2010ApJ...721..222B}. We also searched for 6.7~GHz methanol and H$_2$O masers, neither being detected here \citep{{1998AJ....116.2936M},{2013A&A...550A..21S}}. There is little information about masers towards all the clumps. The results of clump classification are presented in Table~\ref{clumpact}. The columns in the table list the clump name, association with radio and 8~$\mu$m peaks, and 22~$\mu$m point source. Column 5 gives the clump activity (Q-quiescent, A-active, I-intermediate, E-evolved). Among the eighteen clumps considered, five clumps (C1, C2, C3, C6, C7) are active or evolved clumps. These lie to the north and west of the infrared cluster. The clump C4 is in intermediate stage as this is associated with EGO-1 and no radio emission is detected here. There is a source associated with EGO-1 seen in the PACS~160~$\mu$m image (see Fig.~\ref{ego}). The 160~$\mu$m source exhibits two peaks separated by 13$\arcsec$. The northern peak is located $\sim$2$\arcsec$ away from EGO-1. The southern peak lies near the green fuzzies. This double peaked source is likely to be tracing protostellar outflows. While it is difficult to predict the geometry of this region, the observations indicate that clump C4 harbors very young high mass protostar(s). This is consistent with the RWIW models of EGO-1. The rest of the clumps are quiescent. While few have weak 8~$\mu$m diffuse emission from filamentary structures extending from the active/evolved clumps, it is unlikely that these are associated with the clumps themselves. The clump C2 is resolved into two components in the 870~$\mu$m map. As the 350~$\mu$m map is of lower resolution, there is a chance that some of the clumps actually consists of two or more sources/clumps. 

\par This evolutionary sequence of clumps is consistent with that proposed by \citet{2013A&A...550A..21S}. According to their classification, our clumps can be classified as either Type 1 or Type 2 owing to their detection in far-infrared / millimetre images. We classify a clump as Type 2 if it has associated mid-infrared emission and Type 1 in the absence of mid-infrared emission. Thus, the quiescent clumps are of Type 1 while the intermediate and active clumps are of Type 2. This evolutionary stages of clumps based on this classification is also tabulated in Table~\ref{clumpact}.

\par
\begin{figure}
%\hspace*{0.7cm}
\includegraphics[scale=0.4]{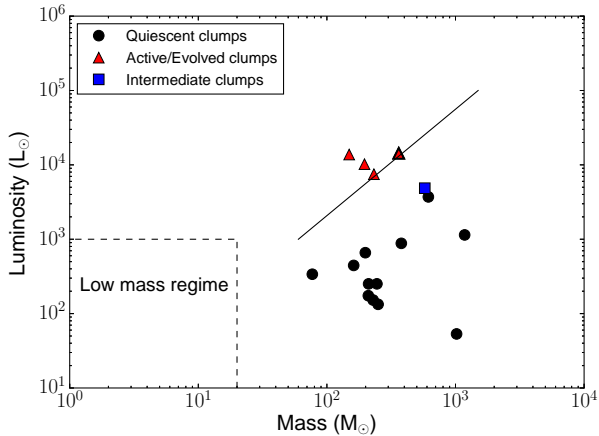}
\caption{Plot of bolometric luminosity versus mass of the clumps. The solid line distinguishes the final-envelope clearing phase (region above) from the accelerating accretion phase (region below) for high mass protostellar objects \citep{2008A&A...481..345M}. The triangle with thick line indicates the location of two active/evolved clumps.} 
\label{Molinari_lm}
\end{figure}

 The differences in evolutionary stages of clumps can also be visualised using a plot of bolometric luminosity versus clump mass \citep{{2002ApJ...566..931S},{2008A&A...481..345M},{rathborne2010early}}. For both the low and high-mass regimes, sources in different evolutionary phases have been shown to lie in distinct regions within this diagram. A plot of the clump masses and luminosities (listed \textbf{in} Table~\ref{cl18}) is shown in Fig.~\ref{Molinari_lm}.  \textbf{At the initial stages of evolution, the massive clumps occupy the lower region in the plot (towards the right). As the protostellar activity increases the sources move almost vertically upwards, towards the higher luminosity side with no significant variation in their masses. Here, the sources are in the accelerating accretion phase. After they reach the ZAMS stage (solid line in Fig.~\ref{Molinari_lm}), they move horizontally towards the lower mass side (i.e towards left in the plot), maintaining nearly constant luminosities.  This phase is known as the envelope clearing phase}. If we assume that each clump forms a single star, then the luminosities and masses would indicate that our clumps lie in \textbf{the} massive star regime. The solid black line distinguishes the accelerating accretion phase from the final envelope clearing phase based on the models shown in Figure 9 of \citet{2008A&A...481..345M}. For this region, we find that clumps classified as A/E (Type 2) lie in the region occupied by the envelope clearing phase. On the other hand, the quiescent clumps (Type 1) lie below in the earlier evolutionary
phase region corresponding to the accelerating accretion phase. The clump classified as intermediate lies between the two groups.  This corroborates with the evolutionary stage  of clumps obtained by us earlier using multi-wavelength tracers.

The evolutionary stages of clumps obtained support the picture of the \hii~region described in the last section. The embedded star cluster has cleared its surroundings of dense gas. and star formation is active towards the central clumps and the clump C4 to the north is in an earlier evolutionary stage.  Since the active clumps surround the infrared cluster towards the north and west where the ionised gas has propagated, it is not unlikely that the ionisation front has triggered star formation in clumps here. The clumps towards the east and south are likely to be quiescent, probably due to lower densities as this is the direction in which the \hii~region is expanding. 

\section{SUMMARY}

From the multiwavelength study of the region IRAS~17256--3631, we arrive at the following conclusions.\\

(i) IRAS~17256--3631 is an \hii~region exhibiting cometary morphology. This is evident from the radio map. The spectral type of ionizing source estimated from radio emission is $\rm{O7-O7.5}$ assuming a single ZAMS star. Narrow band Br$\gamma$ emission from the ionized gas at near-infrared wavelengths is also detected in this region and is consistent with the radio emission.  \\

(ii) The low frequency radio spectral index maps (1372 - 610 and 610 - 325) indicate the presence of thermal and non-thermal emission. The spectral indices are flat near the peak and is negative at the outer edges which points towards shock excitation being a likely mechanism responsible for non-thermal emission. \\

(iii) The narrow band H$_2$ image shows emission in the form of filamentary structures. These indicate the presence of highly excited molecular gas. Several H$_2$ knots are also seen in the region.  \\

(iv) Far-infrared and submm data reveal the presence of 18 high density molecular clumps. The modified blackbody SED fitting to eight of these clumps gives the best fit temperatures in the range $14-33$~K. The total mass of the molecular cloud is estimated to be 6700~M$_{\odot}$ and the total luminosity is $1.6\times10^5~L_{\odot}$. \\

(v) Near-infrared JHK images reveal the presence of a partially embedded cluster as well as high extinction filamentary structures. The spectroscopic study of the brightest star in the cluster (IRS-1) shows that it is a late O or early B type star which is consistent with the spectral type derived from Lyman continuum flux.\\

(vi) Mid-infrared emission traces the ionized gas morphology. Emission in the form of filamentary structures are seen in the region. We also identified 18 young stellar object candidates. An object with extended 4.5~$\mu$m emission, EGO-1 is detected which is likely to be a massive protostellar candidate. \\

(vii) The radio emission is density bounded towards north-west and is ionization bounded towards south-east directions. Several high density clumps are located along the cometary head of ionized emission. The morphology of the \hii~region is explained with the champagne flow model. \\

(viii) The eighteen dust clumps in this region fall into different evolutionary stages. Of these, five active/evolved clumps lie to the north and west of the infrared cluster, one clump is in intermediate phase and the rest are in quiescent phase. The embedded cluster is located towards the void in the cold dust and ionised gas emission indicating that the stellar wind has probably dispersed the parental cloud locally. \\

\bigskip
\noindent \textbf{ACKNOWLEDGEMENT}\\
\par We thank the anonymous referee for a critical reading of the manuscript and highly appreciate the comments and suggestions, that significantly contributed to improving quality of the paper. We thank the staff of GMRT, who made the radio observations possible. GMRT is run by the National Centre for Radio Astrophysics of the Tata Institute of Fundamental Research. We also thank UKIRT staff for their assistance in observations. The UKIRT is supported by NASA and operated under an agreement among the University of Hawaii, the University of Arizona, and Lockheed Martin Advanced Technology Center; operations are enabled through the cooperation of the Joint Astronomy Centre of the Science and Technology Facilities Council of the U.K. When the data reported here were acquired, UKIRT was operated by the Joint Astronomy Centre on behalf of the Science and Technology Facilities Council of the U.K. We thank the Mt.Abu Infrared Observatory staff for assistance in the observations.  We thank M. T. Beltran for providing the 1.2~mm map of this region. We thank A. Sanchez-Monge for providing the ATCA maps. We thank Varsha Ramachandran and Govind Nandakumar for their help in Python programming.
\par  This research made use of NASA/IPAC Infrared Science Archive, which is operated by the Jet Propulsion Laboratory, Caltech under contract with NASA. This publication also made use of data products from $Herschel$ (ESA space observatory). We use ATLASGAL (collaboration between the Max-Planck-Gesellschaft, the European Southern Observatory (ESO) and the Universidad de Chile) data products for our research.

\newpage
\onecolumn  
\appendix
\section{Clump SED Fits}
\begin{figure*}
\centering
\includegraphics[width=4cm]{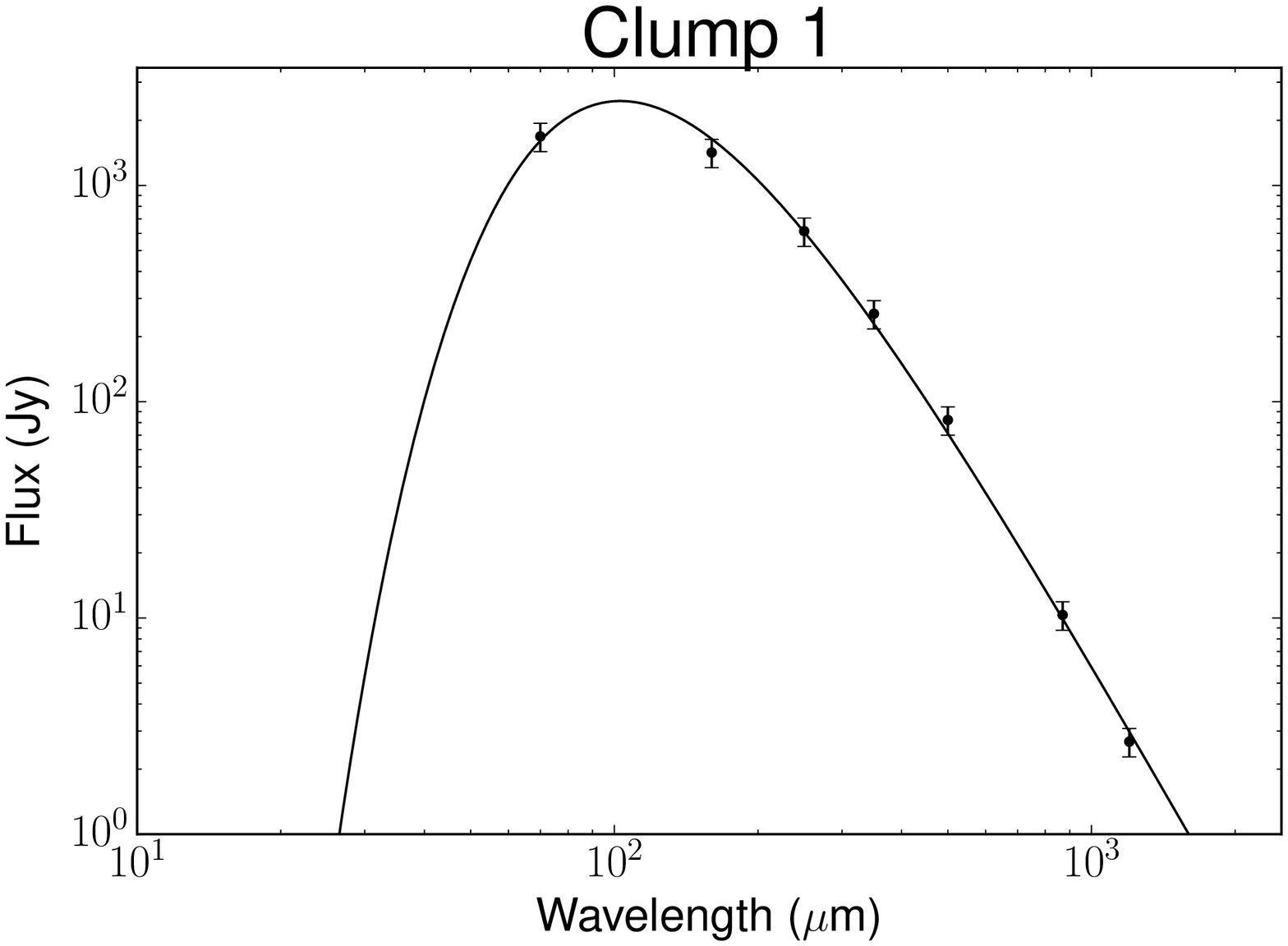}\quad \includegraphics[width=4cm]{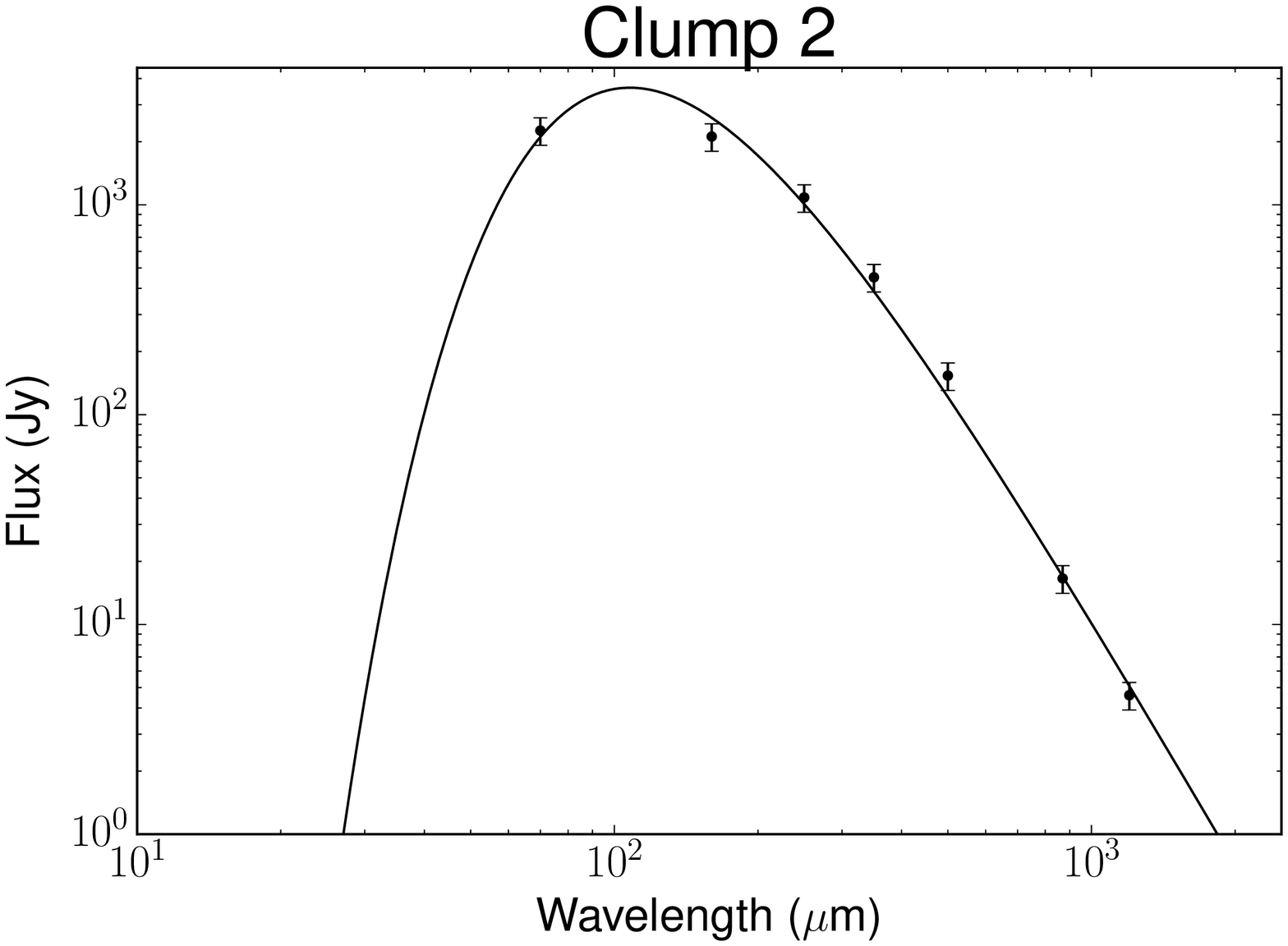}\quad\includegraphics[width=4cm]{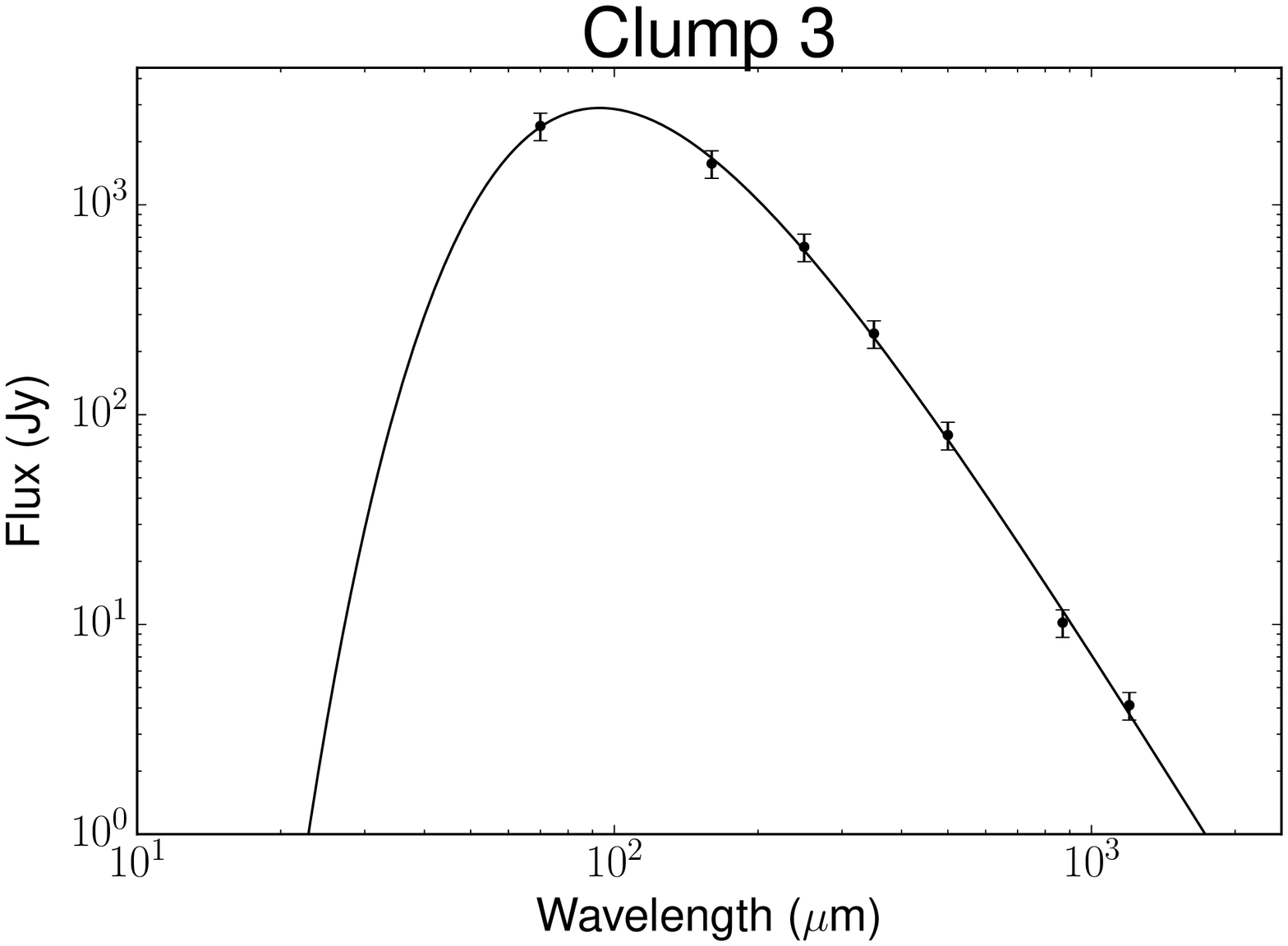}\quad\includegraphics[width=4cm]{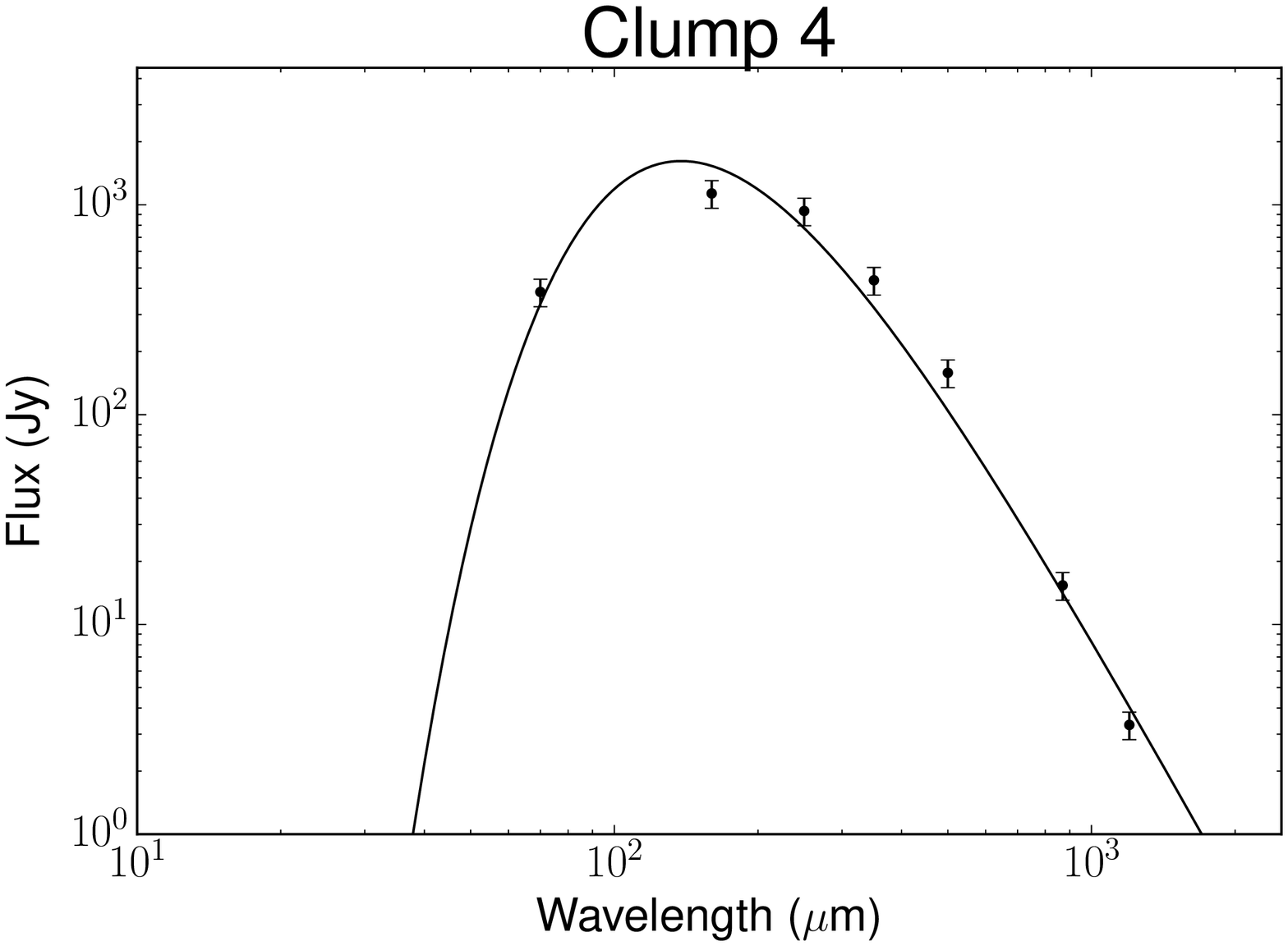}\quad\includegraphics[width=4cm]{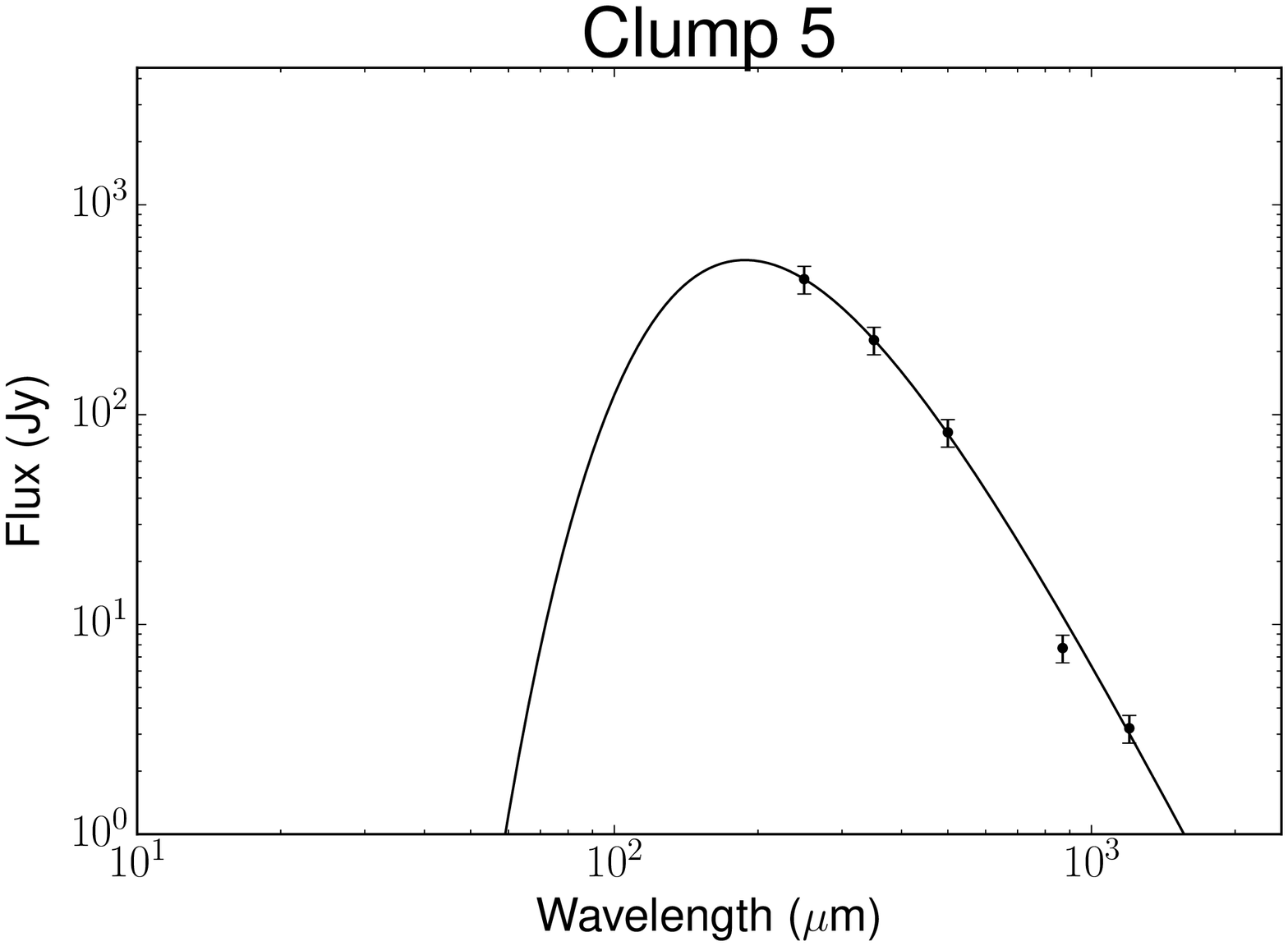}\quad \includegraphics[width=4cm]{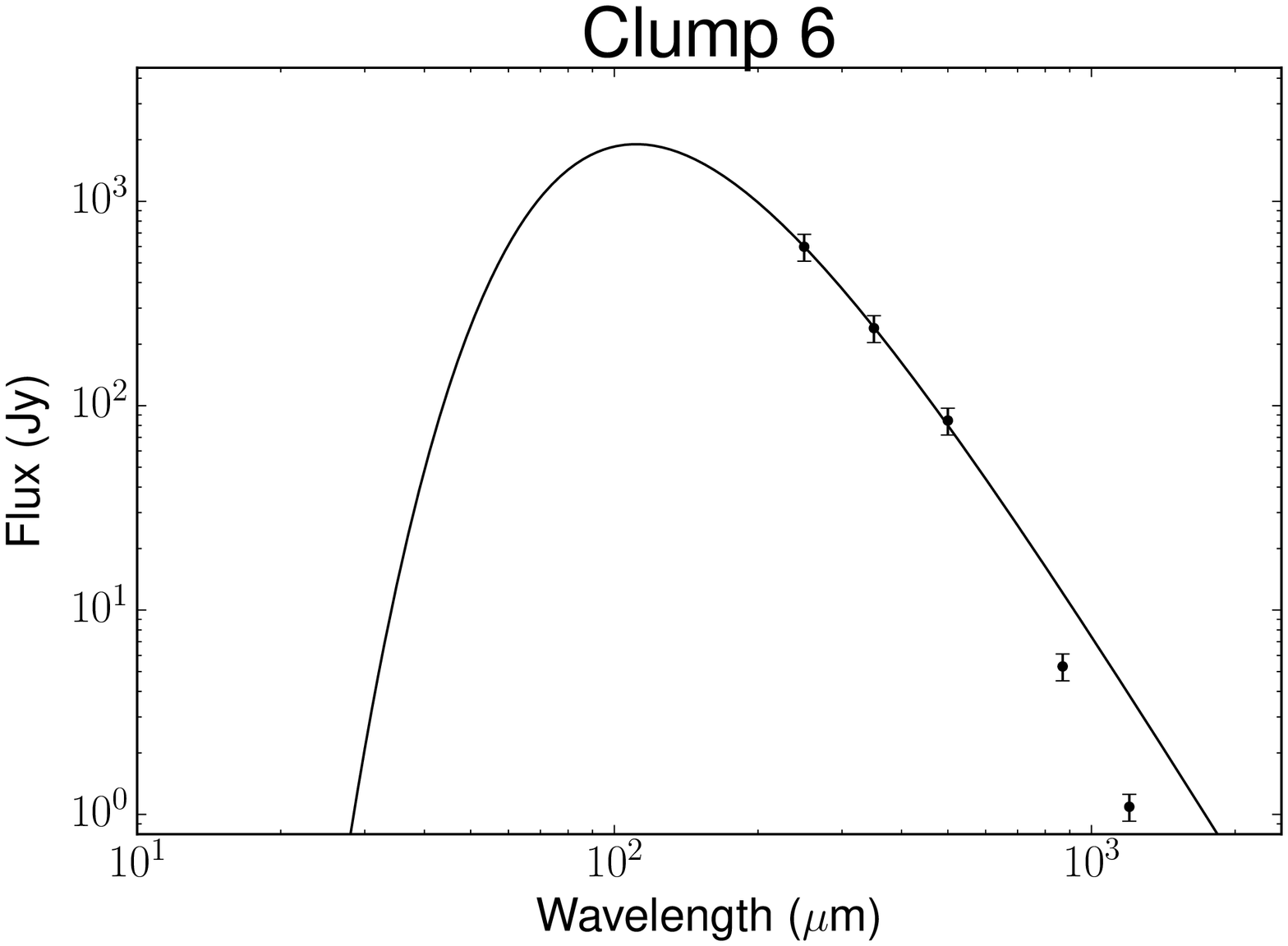}\quad\includegraphics[width=4cm]{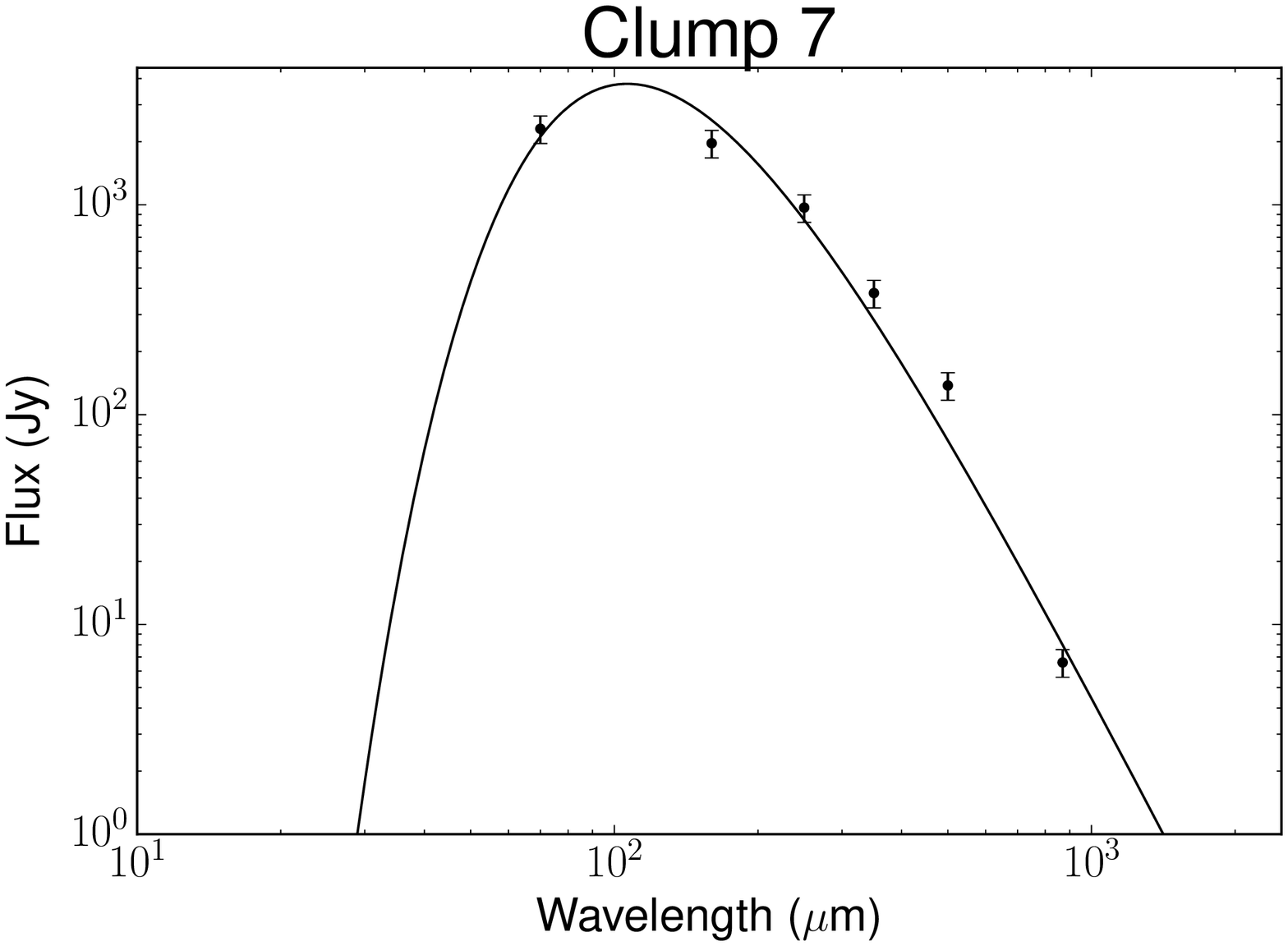}\quad\includegraphics[width=4cm]{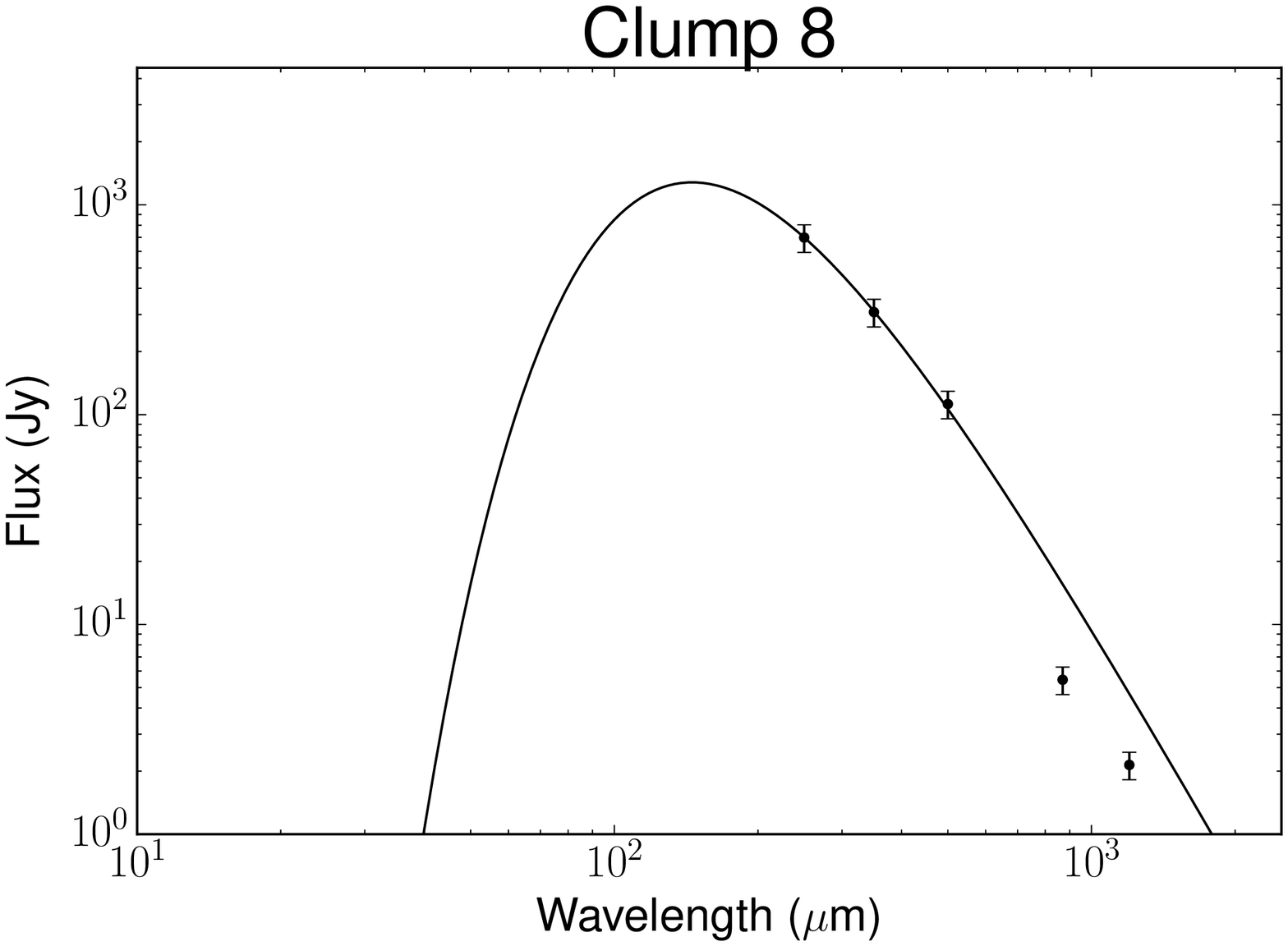}\quad\includegraphics[width=4cm]{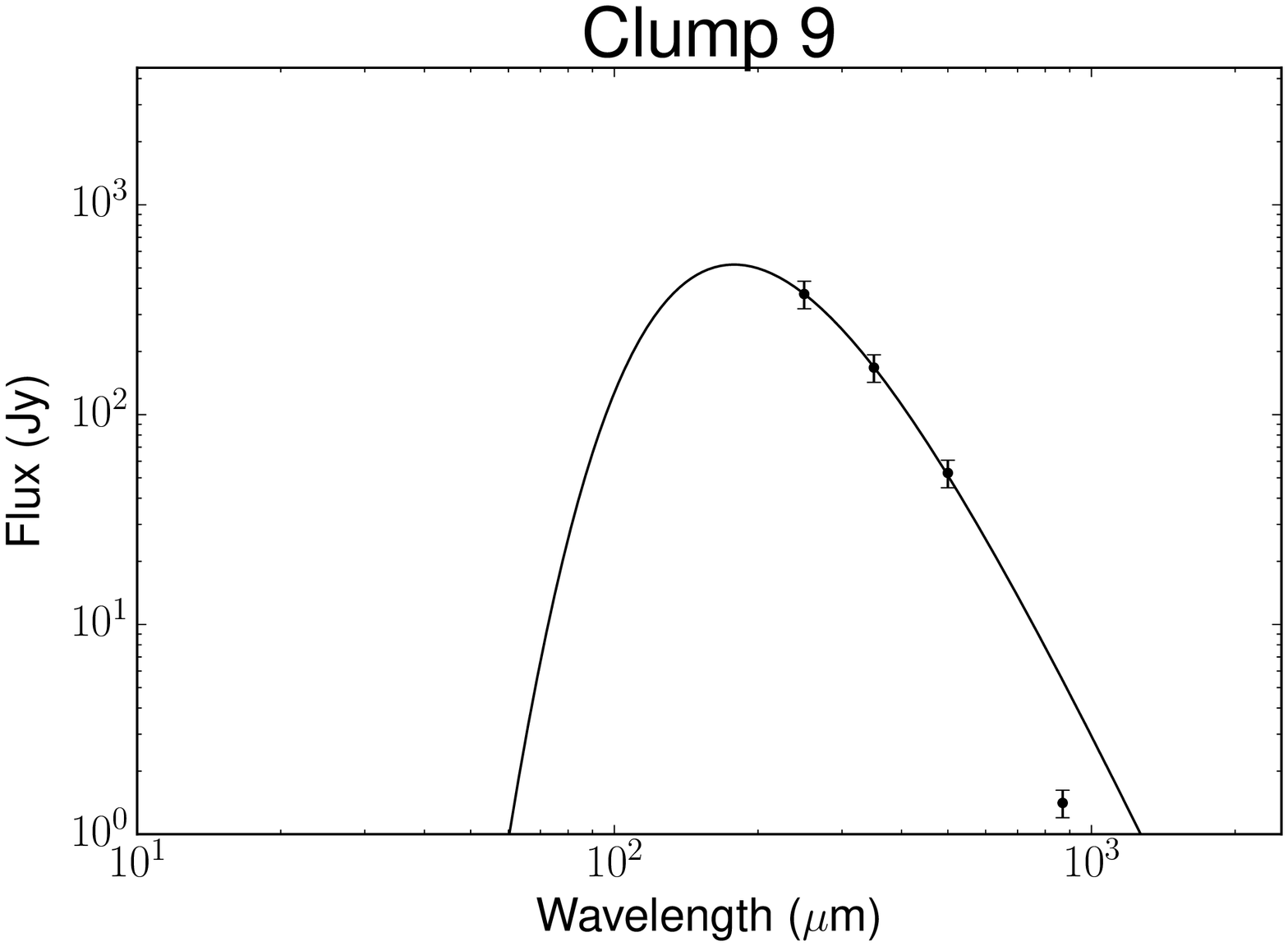}\quad \includegraphics[width=4cm]{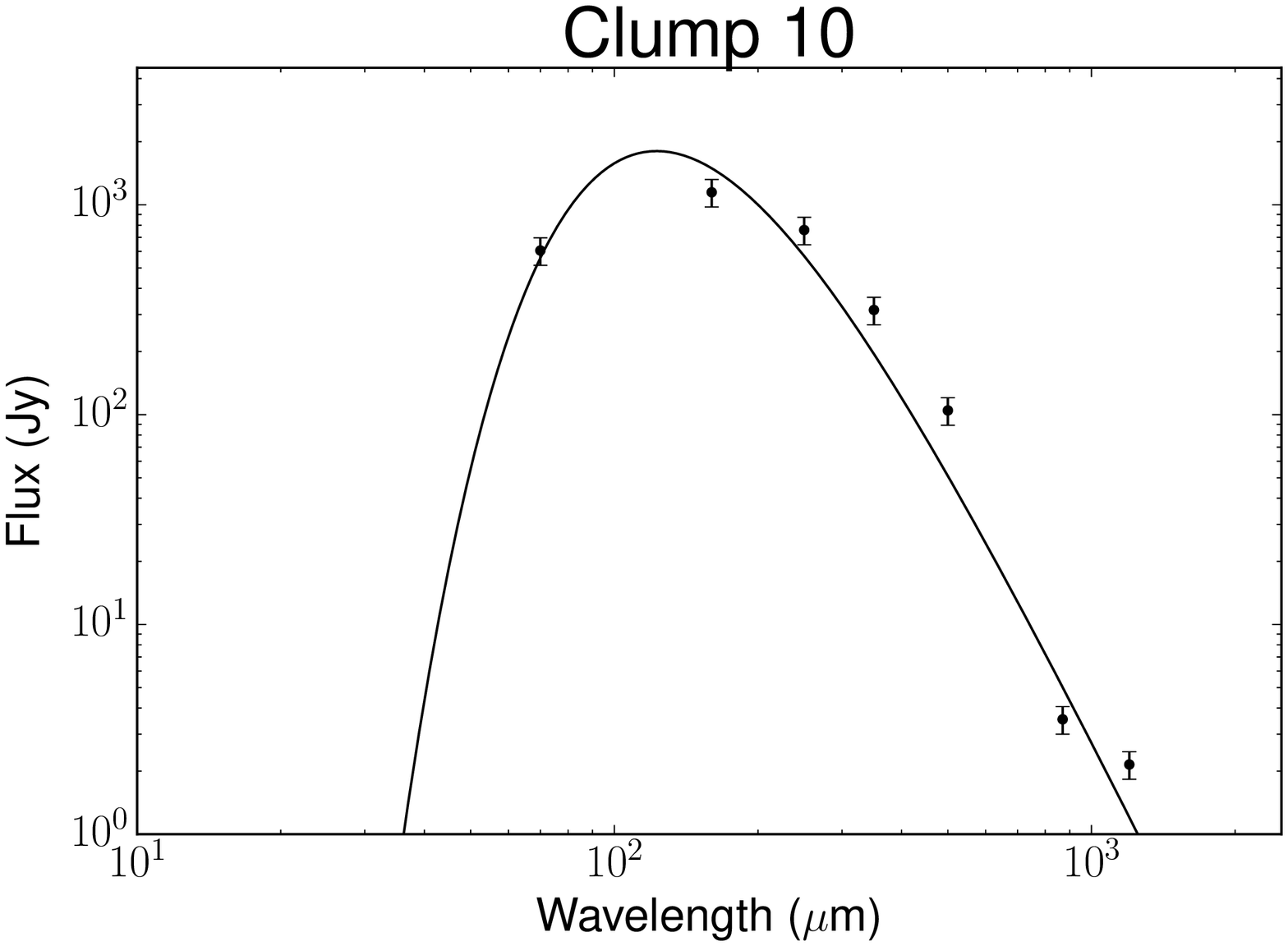}\quad\includegraphics[width=4cm]{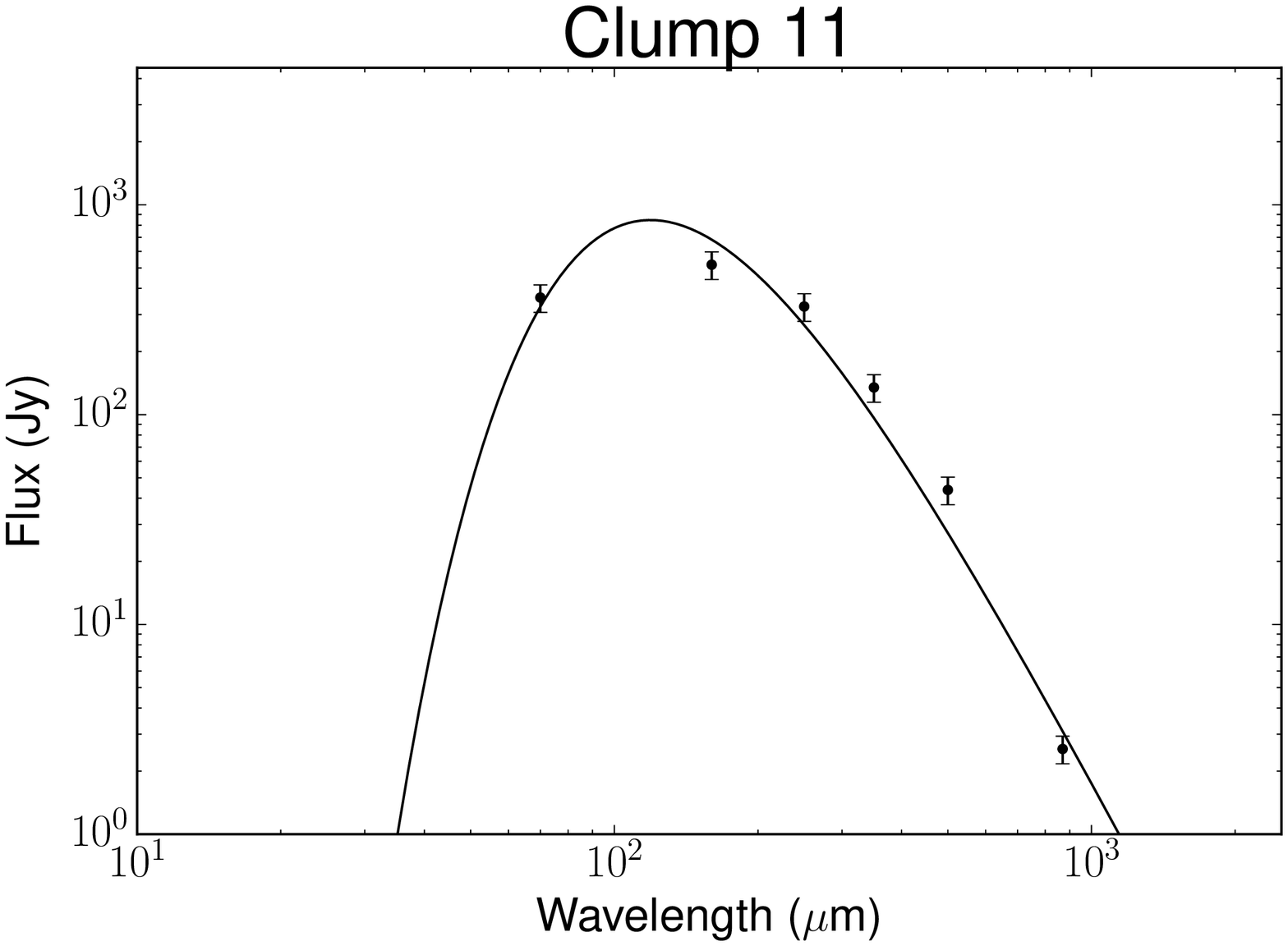}\quad\includegraphics[width=4cm]{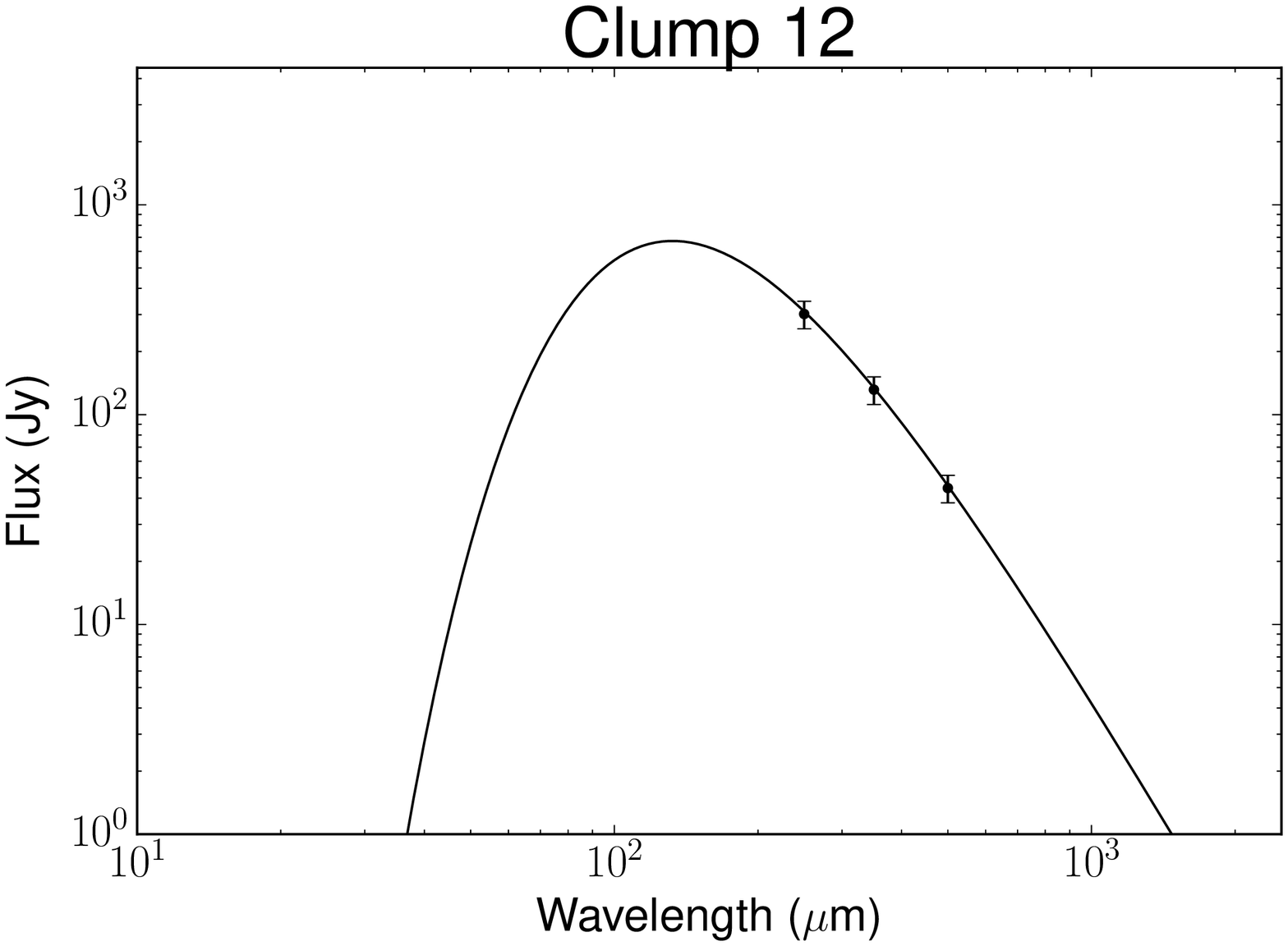}\quad\includegraphics[width=4cm]{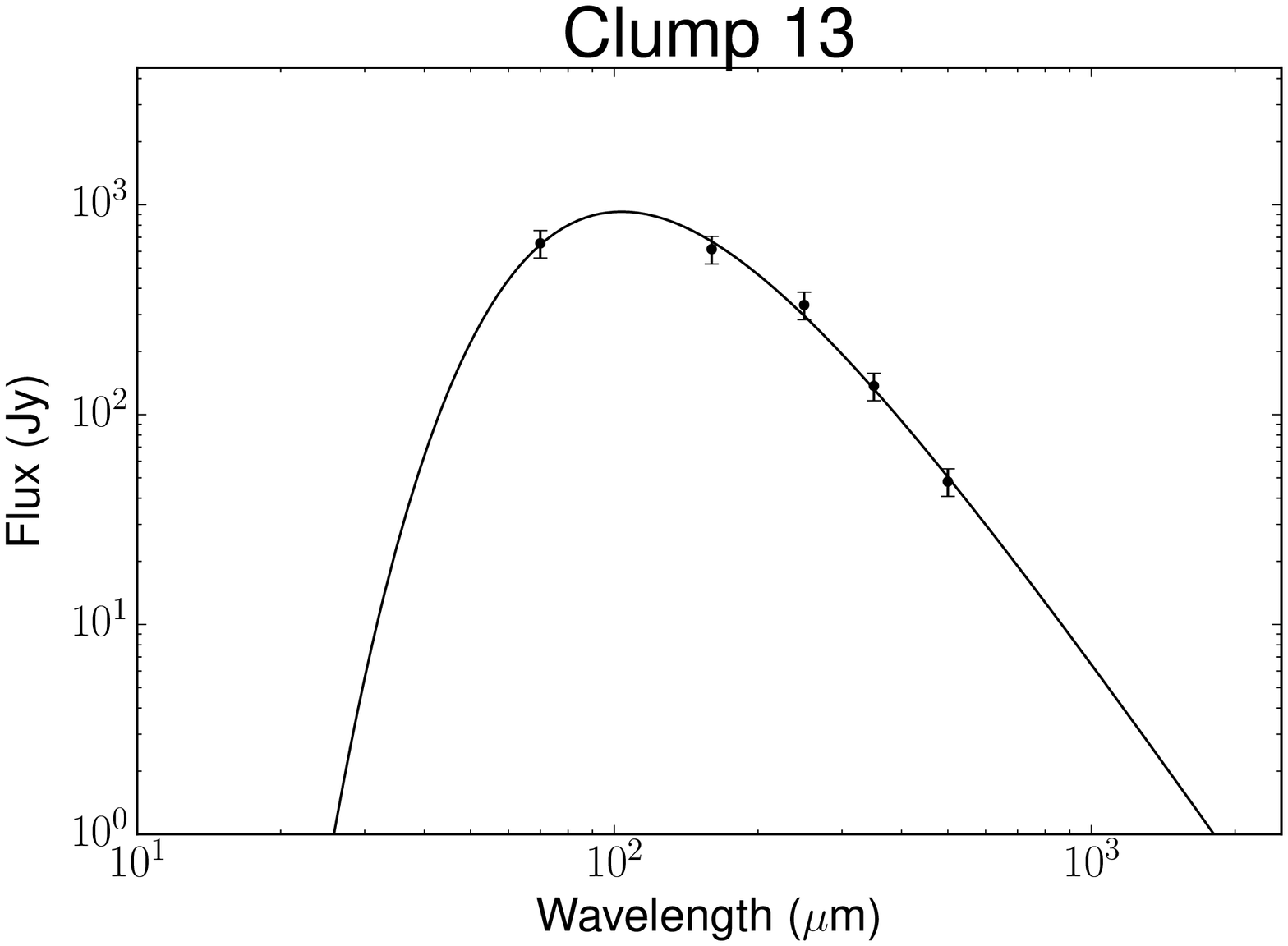}\quad \includegraphics[width=4cm]{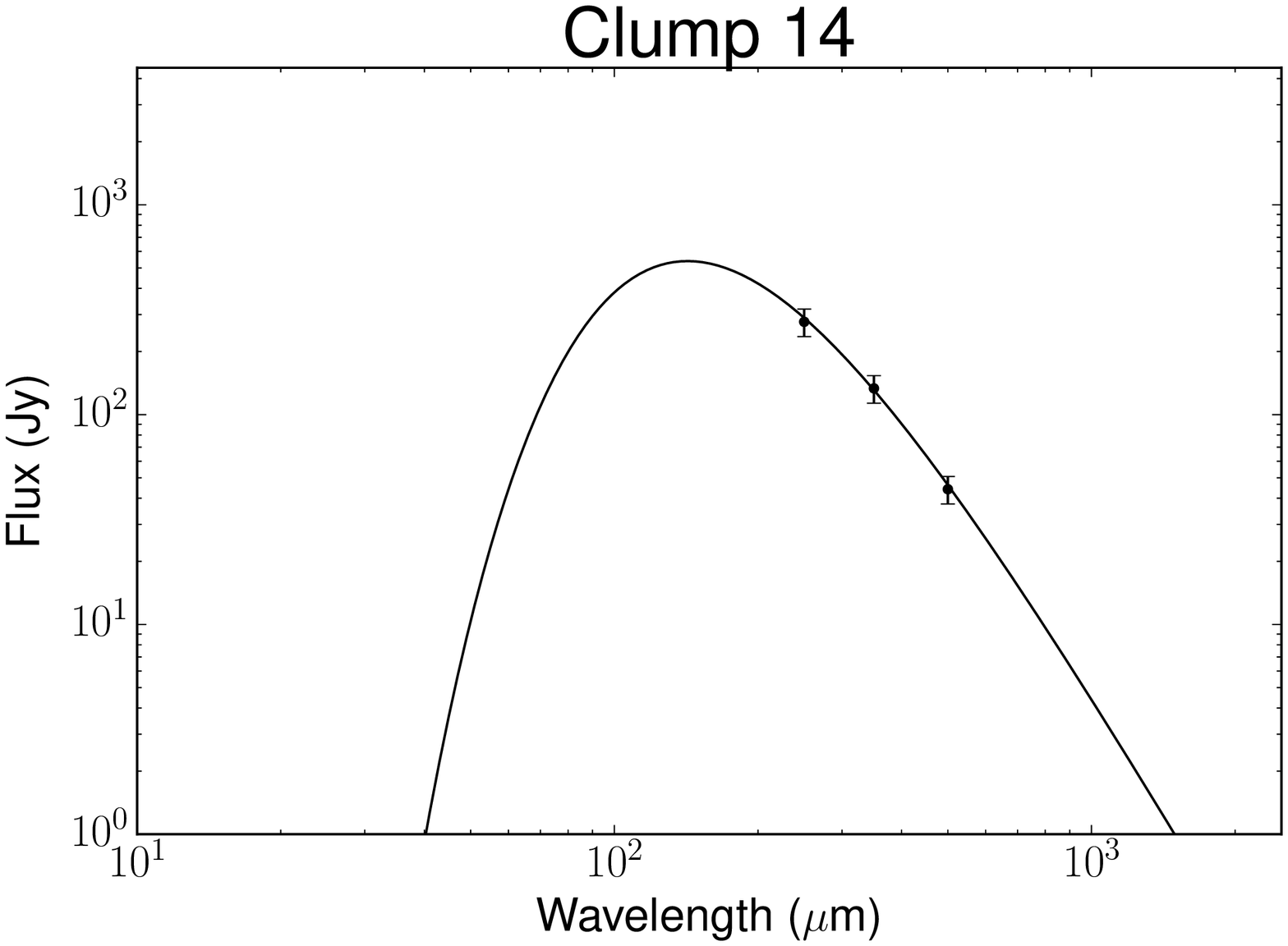}\quad\includegraphics[width=4cm]{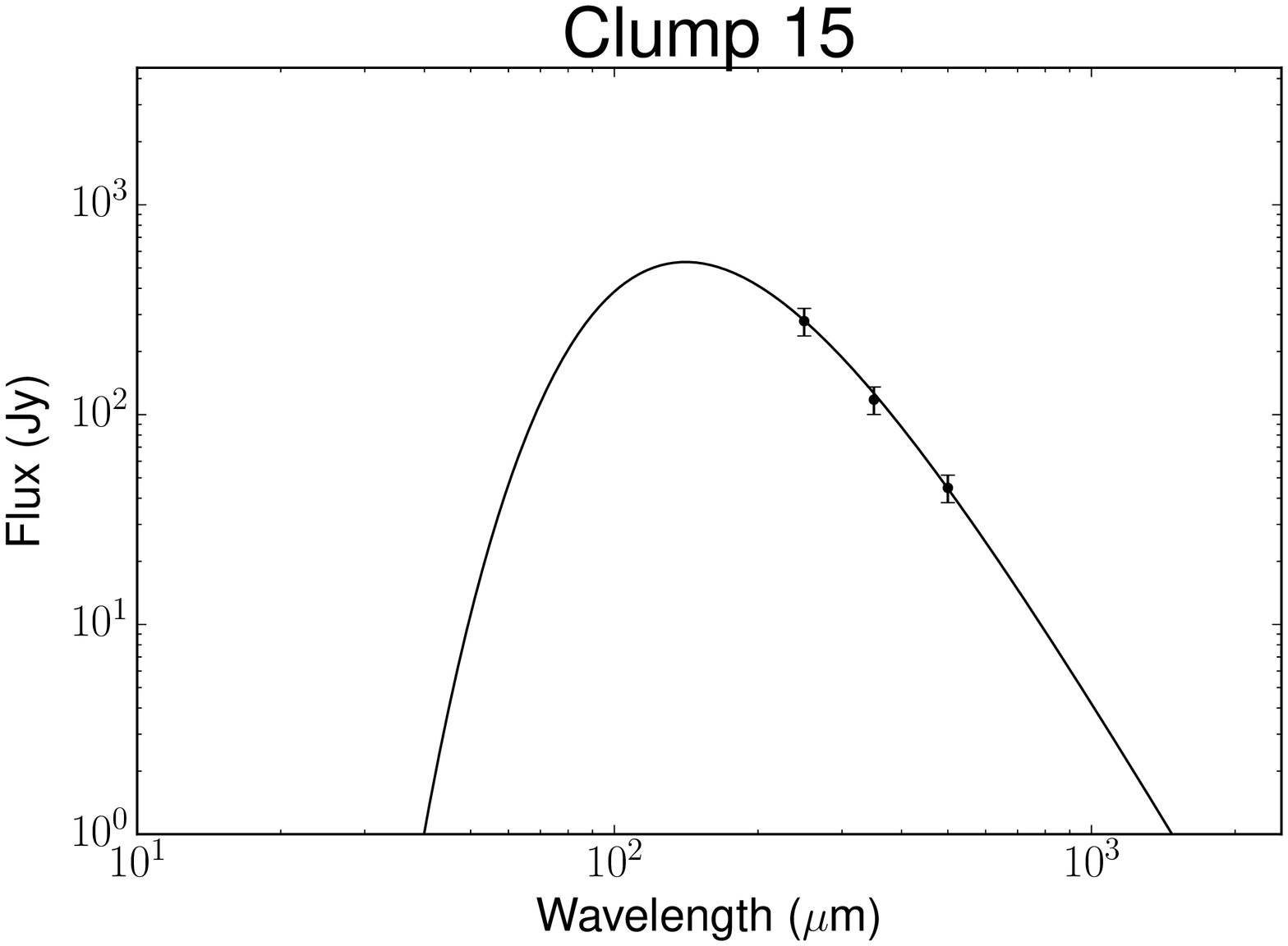}\quad\includegraphics[width=4cm]{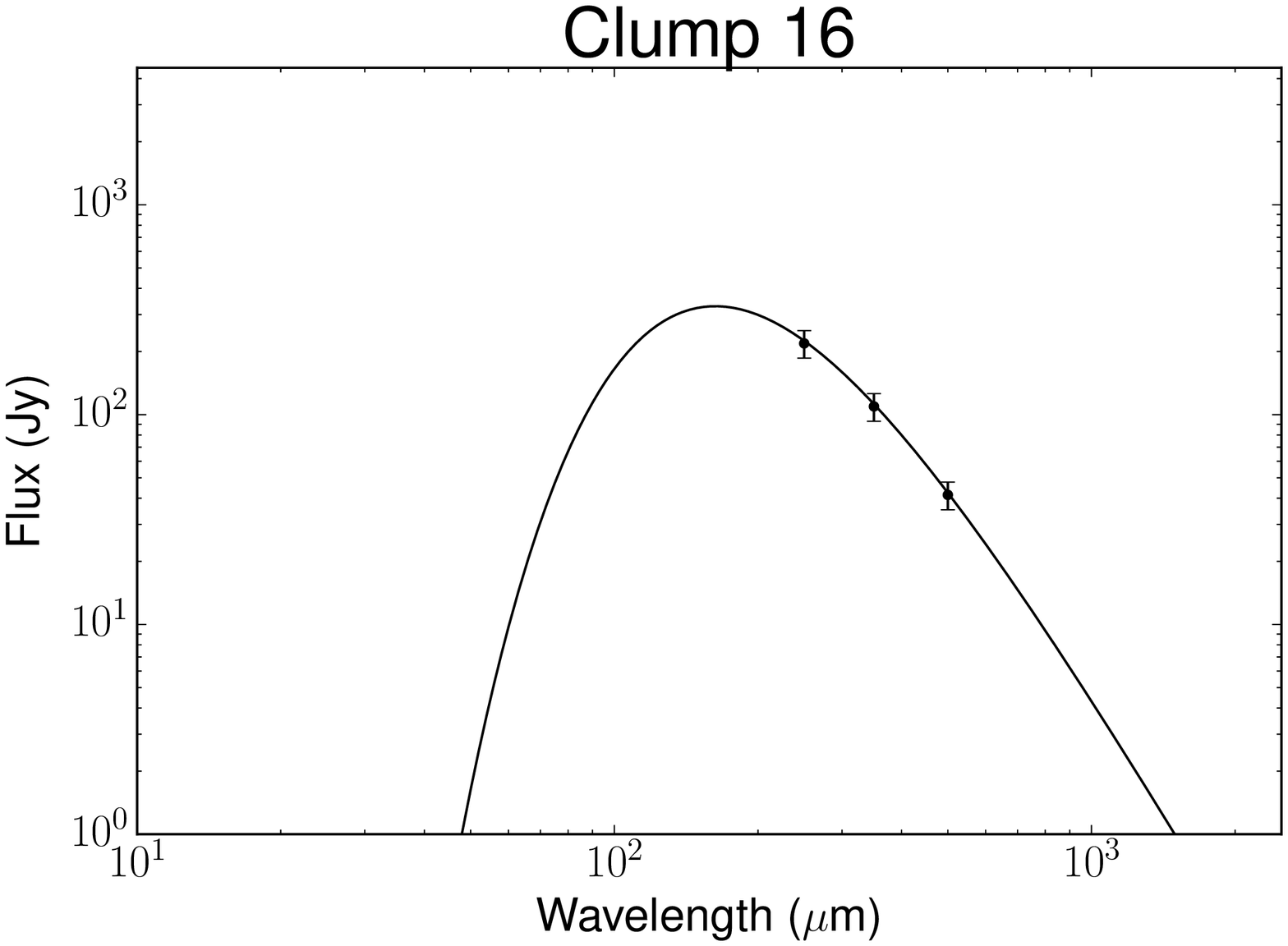}\quad\includegraphics[width=4cm]{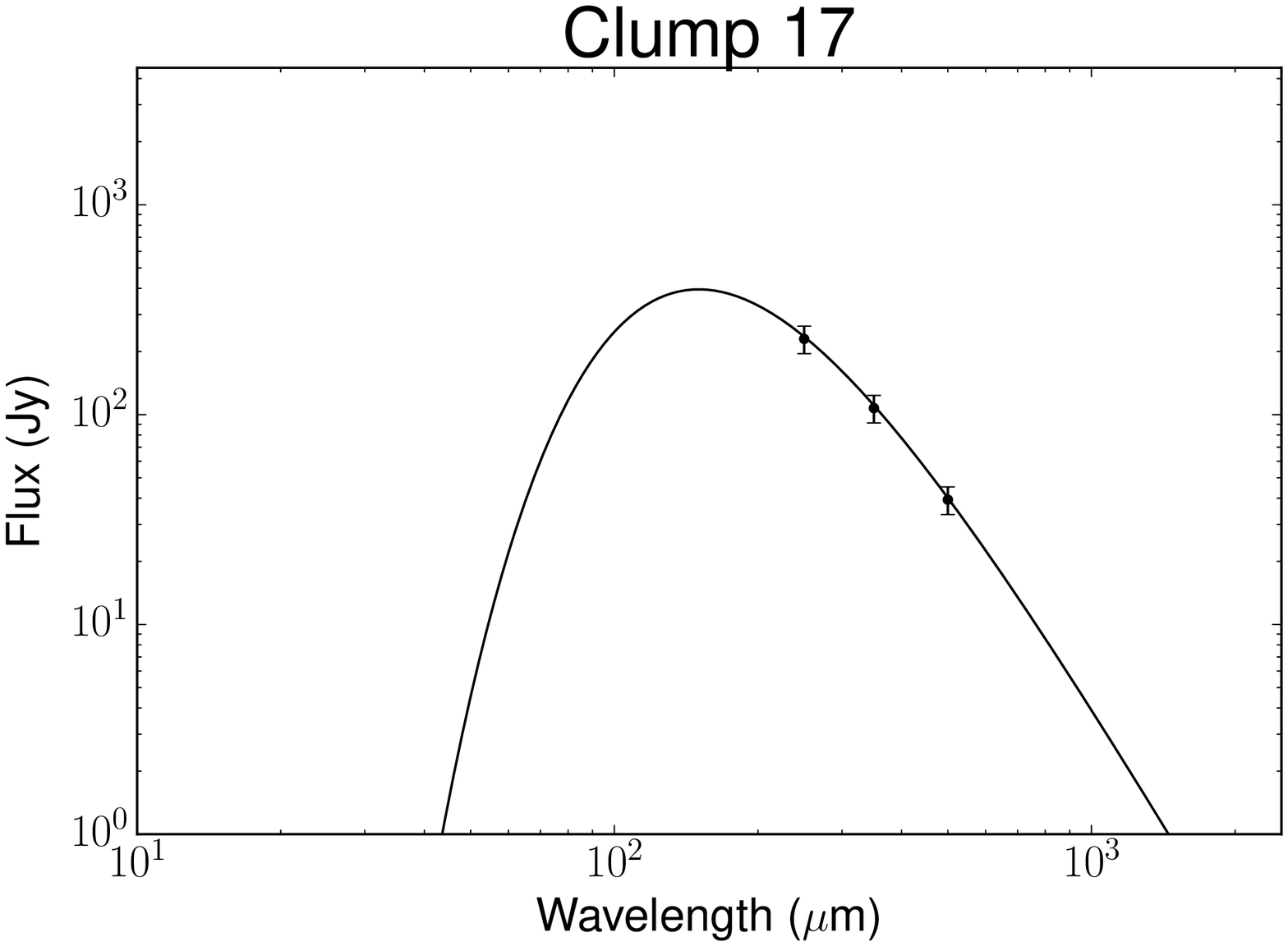}\quad\includegraphics[width=4cm]{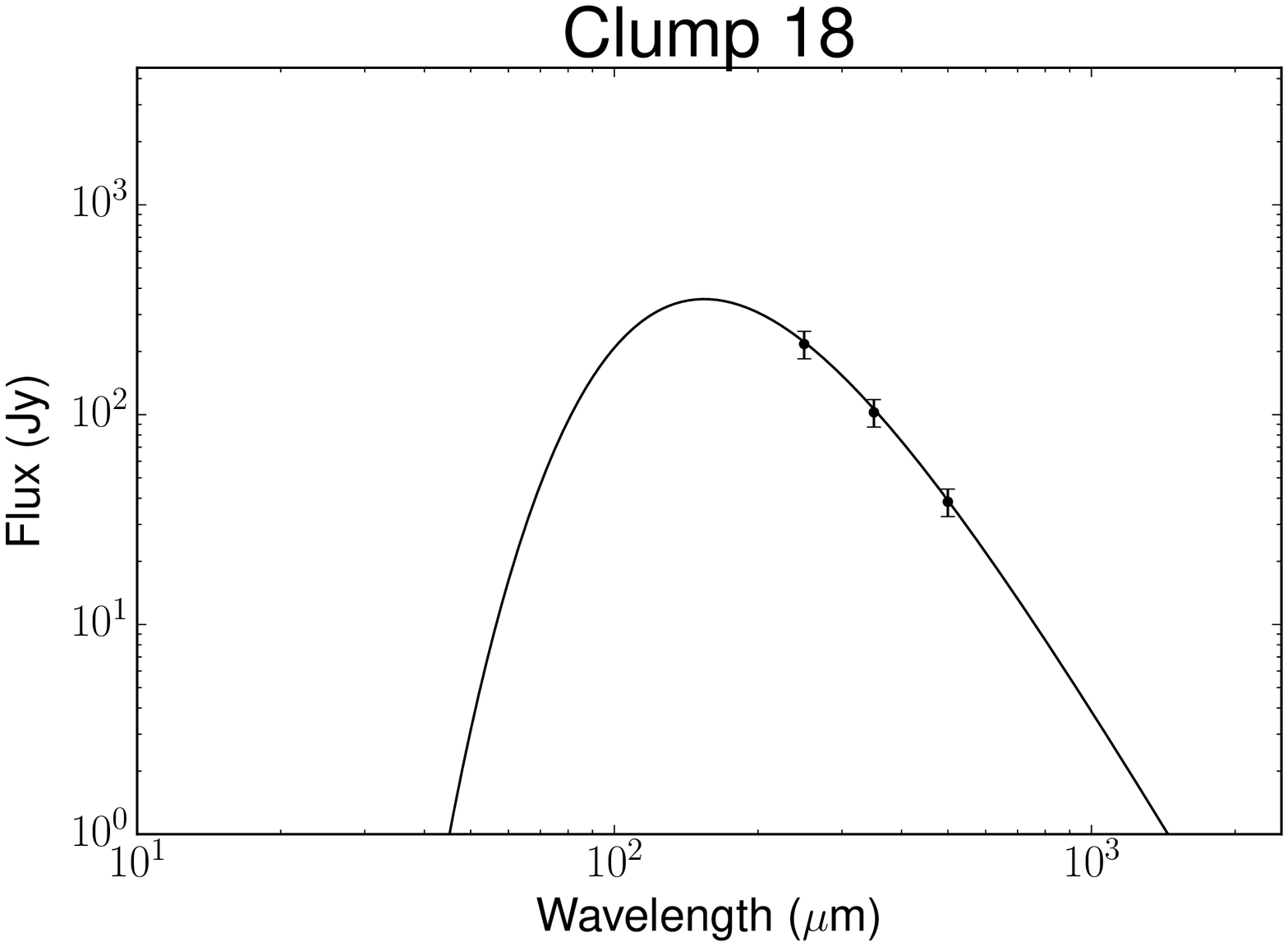}

\caption{Spectral energy distributions of Herschel SPIRE 350 $\mu$m clumps. The integrated flux density within each clump aperture is shown on a log-log plot of wavelength versus flux. The data are shown with 15\% uncertainty error-bars.}
\label{clsed}
\end{figure*}
\newpage
\section{Column Density and Dust Emissivity Index Maps}
We have constructed column density N(H2), dust temperature and $\beta$ maps by carrying out a pixel-by-pixel fit with a greybody SED (Equations same as that used for clumps in Section 3.3.1). The PACs 70 and 160 um images have only partial coverage and do not cover all the clumps in this region. We do not prefer to exclude the 70 and 160 um images, as we will not be sampling the greybody function near the peak. Further, the images are convolved to the poorest resolution ($37\arcsec$) and regridded to a pixel size of $14\arcsec$  (corresponding to the 500~$\mu$m image). The convolution and regridding is carried out in HIPE \citep{2010ASPC..434..139O}. For the Herschel wavebands, we have used the kernels of \citet{2011PASP..123.1218A}, while for the longest two wavelengths (870 um and 1.2 mm), we have used Gaussian kernels for convolution.

\begin{figure}

\centering
\includegraphics[angle=270,scale=0.25]{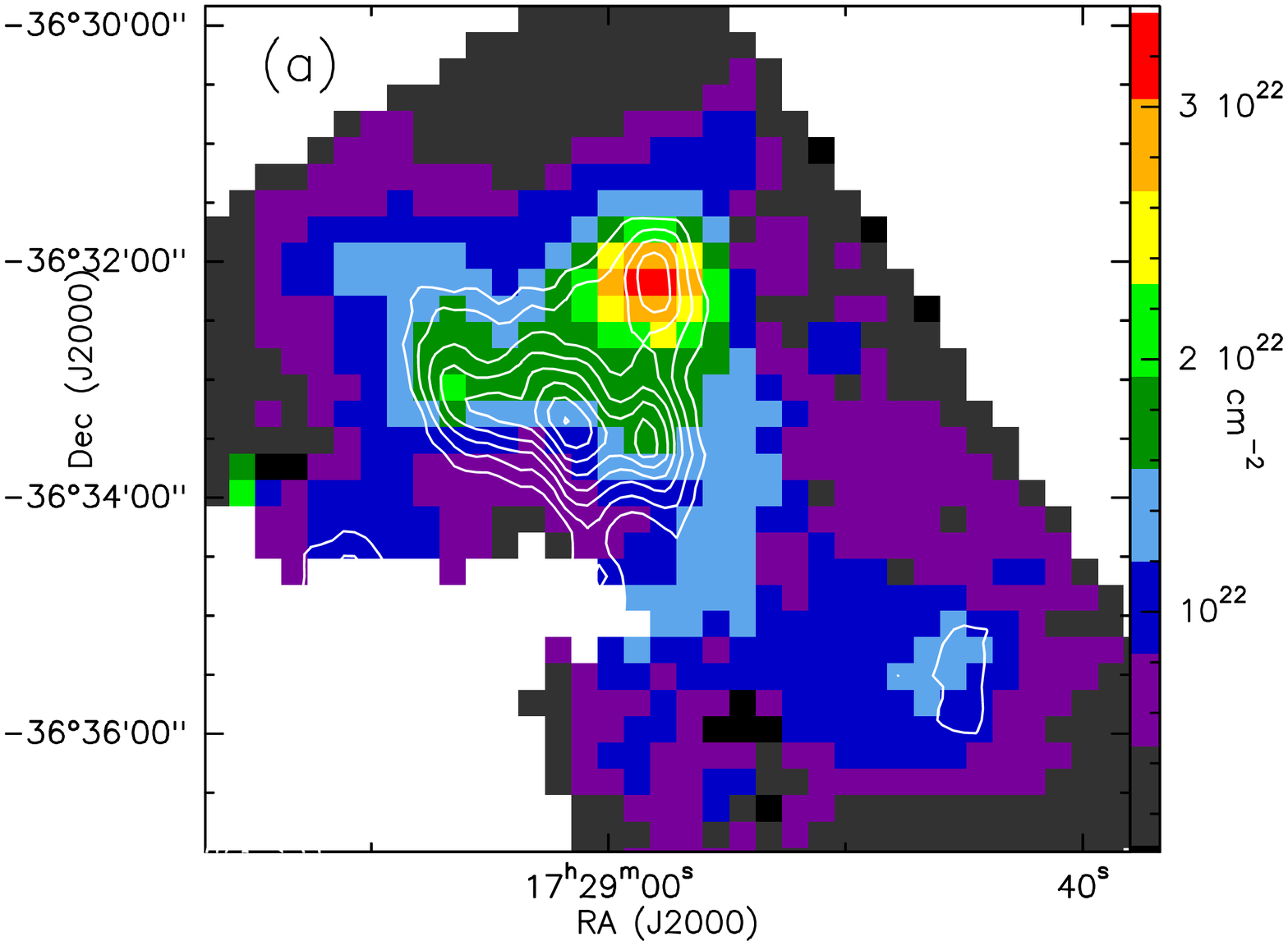} \quad \includegraphics[angle=270, scale=0.25]{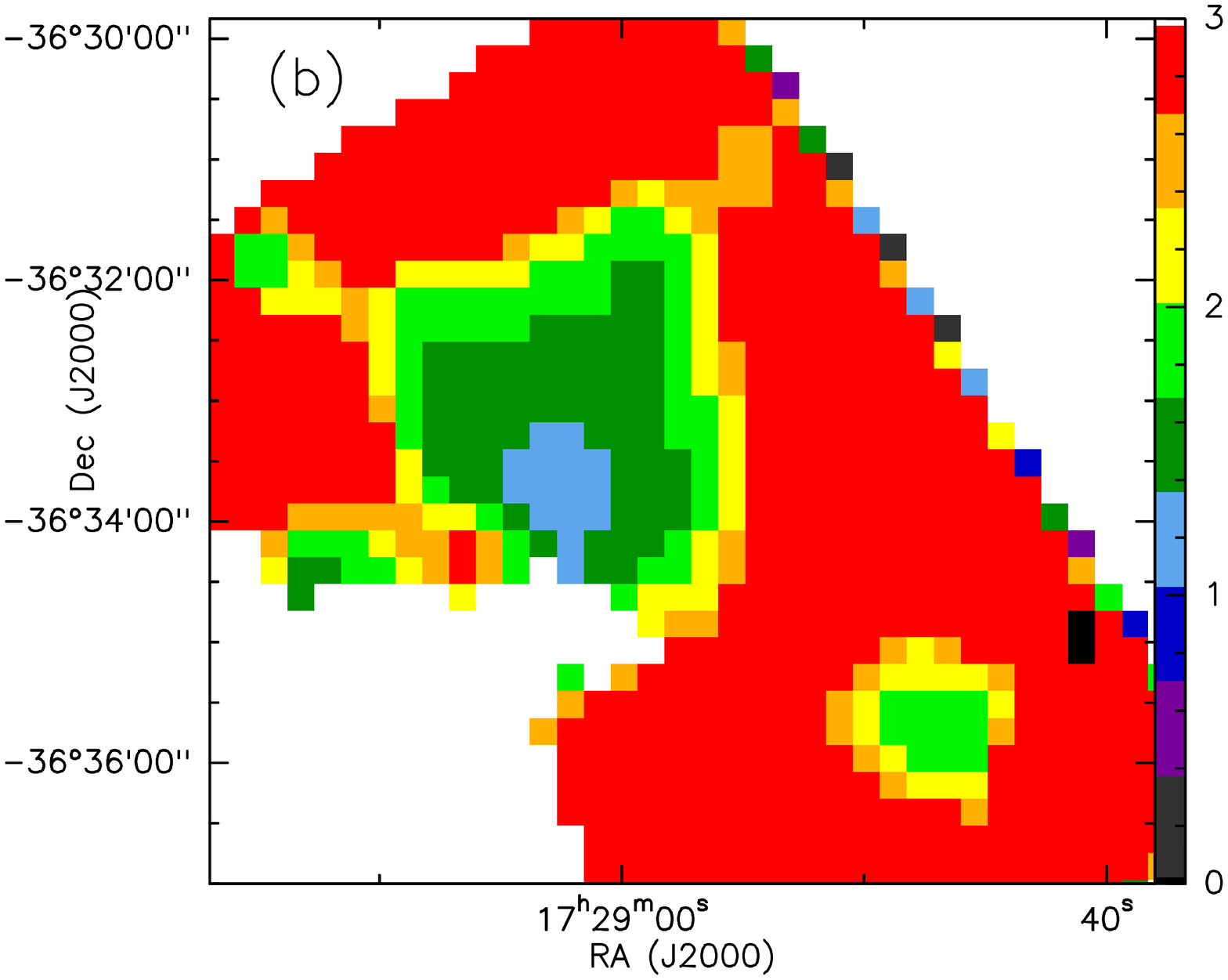} \quad \includegraphics[angle=270, scale=0.25]{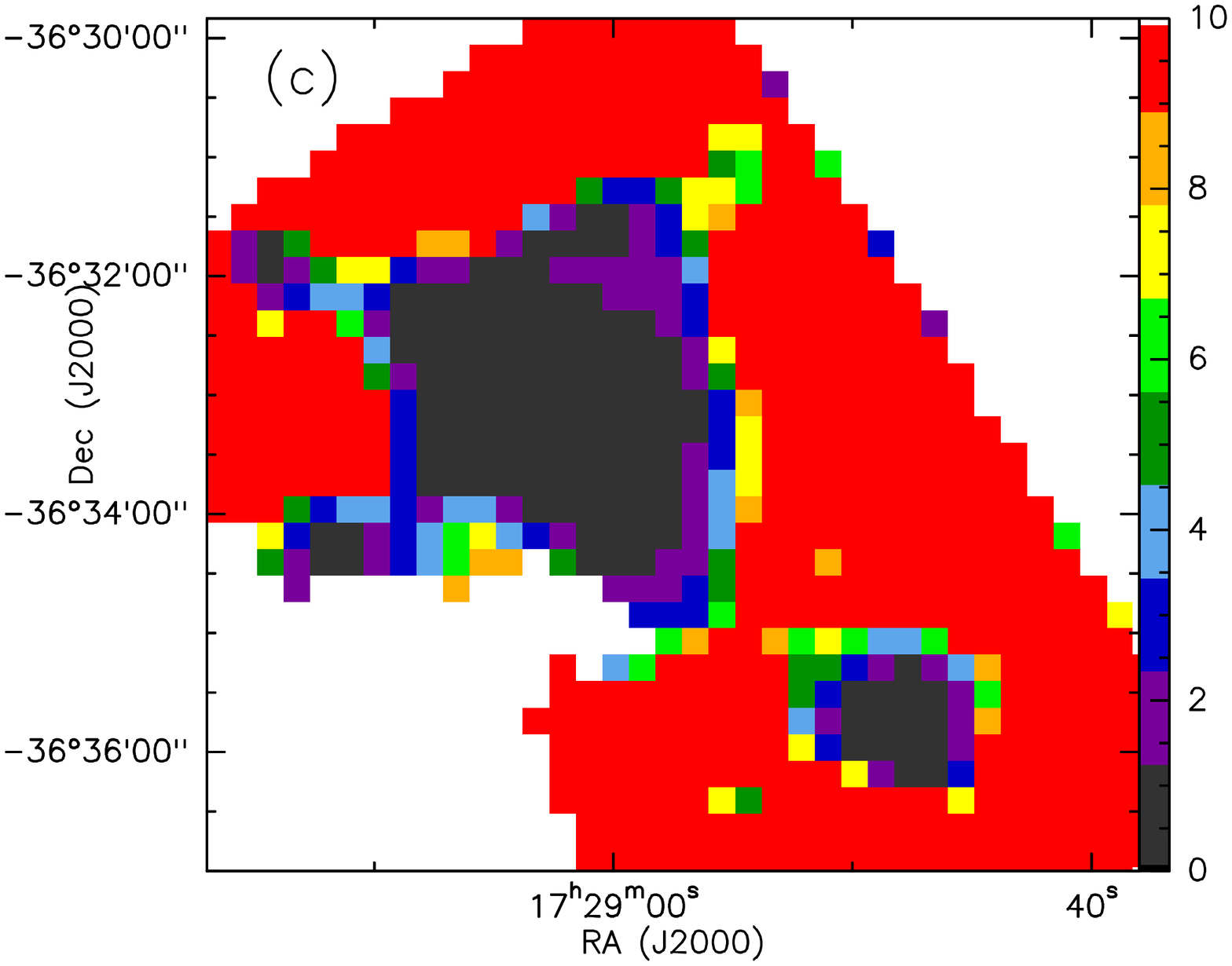}
\caption{(a) Column density map overlaid with 1.2~mm cold dust contours. The contour levels are 250, 380, 510, 640, 770, 900, 1030, 1160, 1290 and 1420 mJy/beam. (b) Dust emissivity index ($\beta$) map. (c) Reduced chi-square ($\chi_{red}$) map.}
\label{appendix_maps}

\end{figure}

The column density, $\beta$, and reduced chi-square maps are shown in Fig.~\ref{appendix_maps}. As the resolution of these maps are poorer compared to the 350~$\mu$m map, we are unable to resolve the central clumps. The peak column density is $3.4\times10^{22}$~cm$^{-2}$ located near the clump C4. The median column density is found to be 1.6$\times$10$^{22}$ cm$^{-2}$. The $\beta$ map is similar in morphology to the dust temperature map. The median value of $\beta$ is 1.7. The reduced-chi square map shows that $\chi_{red}$ falls below 1.0 towards the central clumps while it is larger in the envelope due to lower fluxes in the two longest wavelength bands.

\newpage
\twocolumn  
\bibliography{reference}
\end{document}